%% file: abmp16.tex
\documentclass[prd,aps,preprint,tightenlines,superscriptaddress,nofootinbib,showpacs,showkeys,floatfix]{revtex4}
\usepackage{mathrsfs}
\usepackage{amsfonts}
\usepackage{array}
\usepackage{epsfig}
\usepackage{rotating}
\usepackage{multirow}

\usepackage{amsmath}    
\usepackage{verbatim}   
\usepackage{color}      
\usepackage{colordvi}

\usepackage{subfigure}  
\usepackage{hyperref}   

\usepackage{pstricks,rotating, graphicx}
\usepackage{cancel}

\usepackage{tablefootnote}  

\newcommand{\nn}{\nonumber}

\newcommand{\GeV}{\ensuremath{\,\mathrm{GeV}}}

\newcommand{\msbar}{$\overline{\mathrm{MS}}\, $}

\graphicspath{{../pics/}}
\DeclareGraphicsExtensions{.eps,.ps}

\usepackage{pslatex}
\usepackage{txfonts}
\usepackage[latin1]{inputenc}
\usepackage[T1]{fontenc}

\begin{document}

\begin{titlepage}
\thispagestyle{empty}
\noindent
DESY 16-179\\
DO-TH 16/13\\
\hfill
January 2017 \\
\vspace{1.0cm}

\begin{center}
  {\bf \Large
    Parton Distribution Functions, $\alpha_s$ and Heavy-Quark Masses 
    \\[0.5ex] for LHC Run II
  }
  \vspace{1.25cm}

 {\large
   S.~Alekhin$^{\, a,b}$,
   J.~Bl\"umlein$^{\, c}$,
   S.~Moch$^{\, a}$
   and
   R.~Pla\v cakyt\. e$^{\, a}$
   \\
 }
 \vspace{1.25cm}
 {\it
   $^a$ II. Institut f\"ur Theoretische Physik, Universit\"at Hamburg \\
   Luruper Chaussee 149, D--22761 Hamburg, Germany \\
   \vspace{0.2cm}
   $^b$Institute for High Energy Physics \\
   142281 Protvino, Moscow region, Russia\\
   \vspace{0.2cm}
   $^c$Deutsches Elektronensynchrotron DESY \\
   Platanenallee 6, D--15738 Zeuthen, Germany \\
 }
  \vspace{1.4cm}
  \large {\bf Abstract}
  \vspace{-0.2cm}
\end{center}
We determine a new set of parton distribution functions (ABMP16), the strong coupling constant $\alpha_s$ and the quark masses $m_c$, $m_b$ and $m_t$ 
in a global fit to next-to-next-to-leading order (NNLO) in QCD.
The analysis uses the \msbar scheme for $\alpha_s$ and all quark masses and is performed in the fixed-flavor number scheme for $n_f=3, 4, 5$.
Essential new elements of the fit are the combined data from HERA for inclusive deep-inelastic scattering (DIS), 
data from the fixed-target experiments NOMAD and CHORUS for neutrino-induced DIS, 
and data from Tevatron and the LHC for the Drell-Yan process and the hadro-production of single-top and top-quark pairs.
The theory predictions include new improved approximations at NNLO for the production of heavy quarks in DIS 
and for the hadro-production of single-top quarks.
The description of higher twist effects relevant beyond 
the leading twist collinear factorization approximation is refined.
At NNLO we obtain the value $\alpha_s^{(n_f=5)}(M_Z) = 0.1147 \pm 0.0008$.
\end{titlepage}

\newpage
\setcounter{footnote}{0}
\setcounter{page}{1}

\section{Introduction}

Parton distribution functions (PDFs) are indispensable for theory predictions of scattering processes at hadron colliders.
Within standard factorization in Quantum Chromodynamics (QCD) the PDFs are determined by a comparison of theoretical predictions 
with hard scattering data covering a broad range of kinematics in the Bjorken variable $x$ and the momentum scale $Q^2$.
Steady progress both in the accumulation and in the analysis of
hard-scattering data by experiments at HERA, Tevatron and the Large Hadron Collider (LHC) 
as well as improvements of the relevant theoretical predictions to next-to-next-to-leading order (NNLO) 
in perturbative QCD allows for an accurate description of the parton content of the proton in global fits.
Such fits provide the proton composition in terms of the gluon and the individual light-quark flavors $u$, $d$ and $s$ with a good precision.
Simultaneously, they are also capable to determine the strong coupling constant $\alpha_s$ 
and the heavy-quark masses $m_c$, $m_b$ and $m_t$ to NNLO in QCD.
These results serve as input to high precision predictions for benchmark processes in the Standard Model (SM) 
and cross sections for scattering reactions beyond the SM, measured or being
searched for in run II of the LHC.

PDF extractions have been carried out by us in the past, with ABM12~\cite{Alekhin:2013nda} being our previous global fit.
The present analysis has evolved out of these efforts and results in the new ABMP16 set. 
It incorporates a number of intermediate updates~\cite{Alekhin:2014sya,Alekhin:2015cza}, in particular the ABMP15~\cite{Alekhin:2015cza} fit.
Moreover, it makes use of improvements in the theoretical description of the hard-scattering processes for the production 
of heavy quarks in deep-inelastic scattering (DIS) and for the hadro-production of single-top quarks.
However, the primary motivation for ABMP16 comes from the wealth of the recently published new data 
for the measurements of electron-induced DIS from HERA~\cite{Abramowicz:2015mha} 
as well as $W$- and $Z$-boson production at the Tevatron and the LHC.
These data have great potential to further constrain light-quark PDFs at large and small values of $x$, 
to pin down the gluon PDF and to consolidate determinations of $\alpha_s$ 
using various sets of DIS data published during the last three decades.

In our analysis the PDFs and all QCD parameters which are often correlated with the PDFs, 
i.e., $\alpha_s(M_Z)$ and the heavy-quark masses $m_c(m_c)$, $m_b(m_b)$ and $m_t(m_t)$, 
are determined in the \msbar scheme with the number of flavors fixed,  
$n_f=3, 4, 5$, see, e.g. \cite{Alekhin:2009ni}.
The theoretical accuracy is strictly NNLO in QCD.
Other PDF sets currently available are 
CJ15~\cite{Accardi:2016qay}, 
CT14~\cite{Dulat:2015mca},
HERAPDF2.0~\cite{Abramowicz:2015mha},
JR14~\cite{Jimenez-Delgado:2014twa},
MMHT14~\cite{Harland-Lang:2014zoa},
and 
NNPDF3.0~\cite{Ball:2014uwa},
all of them accurate to NNLO in QCD except for 
CJ15, which has limited the precision to next-to-leading order (NLO).
None of these PDFs uses all of the latest data considered in the current
ABMP16 analysis.
A recent benchmarking of those PDFs performed in \cite{Accardi:2016ndt} 
has shown that differences in the theoretical predictions obtained 
by using various PDFs are a consequence of specific theory assumptions 
or underlying physics models used in the fits of some of these PDFs.
Therefore, it is essential to provide a detailed account of the theoretical framework used in the PDF analyses.

The paper is organized as follows.
We present in Sec.~\ref{sec:set-up} the set-up of the analysis. 
In particular we discuss the various sets of hard-scattering data and their kinematic range in Sec.~\ref{sec:data}.
The improvements in the theory description are given in Sec.~\ref{sec:theory} 
and include new approximate NNLO QCD predictions for heavy-quark DIS 
and for single-top quark hadro-production as well as a refined treatment of 
the higher twist effects.
The results for the ABMP16 PDFs are discussed in Sec.~\ref{sec:results} 
where the quality of the data description is documented, 
the improvements in the PDFs are discussed 
and a detailed comparison with the ABM12 fit and other sets is provided.
Correlations of the various fit parameters are discussed and 
particular attention is paid to the value of the strong coupling constant $\alpha_s$ 
extracted from the global fit and individually from various sets of DIS data. 
The sensitivity of the value of $\alpha_s$ to higher twist terms for all sets of DIS data is quantified.
Furthermore, we report our results on the \msbar heavy-quark 
masses $m_c(m_c)$, $m_b(m_b)$ and $m_t(m_t)$
and compare with other determinations.
Finally, in Sec.~\ref{sec:applications} we present several applications.
We compare the second Mellin moment of the non-singlet quark PDFs to recent lattice measurements 
and we provide cross section predictions with the ABMP16 PDFs for relevant LHC processes, such as 
Higgs boson production in gluon-gluon-fusion and hadro-production of top-quark pairs.
In addition, with the measured values of the strong coupling constant $\alpha_s$
and the top-quark mass $m_t(m_t)$ as input we can solve the renormalization group equations for all SM couplings 
including the scalar self-coupling $\lambda$ of the SM Higgs boson. 
This allows us to study the running of $\lambda$ and to assess 
whether new physics needs to be invoked in order to stabilize the electroweak
vacuum at high scales~\cite{Bezrukov:2012sa,Degrassi:2012ry}.
We also discuss the features of the data grids for the fit results for use with 
the {\tt LHAPDF} library (version 6)~\cite{Buckley:2014ana} and conclude in Sec.~\ref{sec:concl}.

\section{Set-up of the analysis}
\label{sec:set-up}

\subsection{Data}
\label{sec:data}

The data used in present analysis have been updated in an essential manner 
with respect to the ones used in our earlier fits ABM12~\cite{Alekhin:2013nda} and ABMP15~\cite{Alekhin:2015cza}.
The changes concern inclusive DIS data 
as well as data on DIS charm- and bottom-quark production, on the Drell-Yan
(DY) process, and on the top-quark hadro-production 
as follows:
\begin{itemize}
    \item The HERA run I inclusive cross section data on the neutral-current (NC) and charged-current (CC)
    $e^{\pm}p$ DIS have been replaced with the final combination of the 
    run I+II results~\cite{Abramowicz:2015mha}. This input
    provides improved constraints on the small-$x$ gluon and sea-quark PDFs 
    and significant benefits for the 
    separation of the up- and down-quark PDFs by virtue of the precise CC data.

\item The data on production of charm-quarks in $e^{\pm}p$ DIS obtained by the H1~\cite{Aaron:2009af}
    and ZEUS~\cite{Abramowicz:2014zub} collaborations are added. These data are particularly   
   useful for determination of the bottom-quark mass and are also sensitive to the small-$x$ PDFs.
    
\item New data on the charm-quark production in CC neutrino-nucleon DIS collected by the 
    NOMAD~\cite{Samoylov:2013xoa} and CHORUS~\cite{KayisTopaksu:2011mx} experiments
    are added in order to improve the strange sea determination,
    cf. Ref.~\cite{Alekhin:2015cza} for details.
    
\item The latest data on $W^{\pm}$- and $Z$-boson production from LHC and Tevatron are added  
    in order to provide an improved determination of the light-quark PDFs 
    over a wide range of the parton momentum fractions $x$ and to disentangle 
    distributions for quarks and anti-quarks. 
    The data include rapidity distributions for $W^{\pm}$- and $Z$-boson
    production in the forward region at the collision energies of $\sqrt{s}=7$ and 8~TeV  
    obtained by LHCb~\cite{Aaij:2015gna,Aaij:2015zlq,Aaij:2015vua}, 
    D{\O} data on the electron charge asymmetry, which also probes forward kinematics~\cite{D0:2014kma},
    D{\O} data on the muon charge asymmetry in the central region~\cite{Abazov:2013rja},
    new CMS rapidity distributions for the $W^{\pm}$-boson recorded at $\sqrt{s}=7$ and 8~TeV  
    using the muon decay channel~\cite{Chatrchyan:2013mza,Khachatryan:2016pev}
    and the cross section of $W^{\pm}$- and $Z$-boson production at $\sqrt{s}=13$~TeV  
    in the fiducial volume obtained by ATLAS~\cite{Aad:2016naf}.
    
\item A collection of the recent $t$-quark data from the LHC~\cite{Aad:2014fwa,Tepel:2014kna,Aaboud:2016ymp,Chatrchyan:2012ep,
    Khachatryan:2014iya,Sirunyan:2016cdg,Aad:2014jra,Khachatryan:2016mqs,Khachatryan:2015uqb,
    Aad:2014kva,Aaboud:2016pbd,Khachatryan:2016kzg,Khachatryan:2016yzq,Aad:2015pga,CMS:2015toa,ATLAS:2012gpa,
    Aad:2015dya,Chatrchyan:2012vs,Khachatryan:2015fwh,Aad:2012vip,Chatrchyan:2013kff,ATLAS-CONF-2012-031,
    Chatrchyan:2013ual,CMS:2016rtp,Khachatryan:2014loa,CMS:2016pqu}
    and Tevatron~\cite{Aaltonen:2015cra} added to the present analysis 
    provides additional constraints on the gluon PDF and allows to perform a consistent
    determination of the top-quark mass with full account of its correlations with the gluon PDF 
    and the strong coupling $\alpha_s$.
\end{itemize}

With the new measurements included in the present study, the theoretical framework has been
updated correspondingly to account for the best possible precision and
consistency of the PDF fit as discussed in Sec~\ref{sec:theory}. 
In the following, the DIS, DY, and heavy-quark production data sets used in our fit are described in detail.

\subsubsection{Inclusive DIS}

The recent HERA inclusive NC and CC DIS data set~\cite{Abramowicz:2015mha} 
includes a combination of all published H1 and ZEUS measurements performed 
in the runs I and II of the HERA collider. 
The data were collected at the proton beam energies $E_p$ = 920, 820, 
575 and 460~GeV
which correspond to the center-of-mass energies $\sqrt{s}$ = 320, 300, 251 and 225~GeV.
The combined HERA data~\cite{Abramowicz:2015mha} 
cover the range of momentum transfer squared $Q^2$ up to 50000~GeV$^2$ and are the
most precise measurements of $ep$ DIS over that wide kinematic range. 
The high-statistics HERA II data used in the new combination 
improve the accuracy at high $Q^2$, as compared to the HERA I inclusive 
combination, in particular for the NC $e^{-}p$ and 
the CC $e^{+}p$ ($e^{-}p$) data. The latter impose improved 
constraints on the valence down-(up-)quark distributions in the proton and
in combination with the new DY collider data added to our fit 
they allow to avoid using the fixed-target 
DIS data~\cite{Bodek:1979rx,Atwood:1976ys,Mestayer:1982ba,Gomez:1993ri,Dasu:1993vk,Benvenuti:1989fm,Arneodo:1996qe}
collected by the SLAC, BCDMS, and NMC experiments with a deuteron target.
Previously, those samples have been employed in the ABM12 fit and our 
earlier analyses
in order to constrain the down-quark distributions at the expense 
of having to deal also with nuclear effects. 
Now, with the extended DIS and DY input
the experimental uncertainties in the down-quark PDFs do not 
deteriorate as compared to the ABM12 PDFs even in the absence of deuteron 
data, while any additional uncertainty caused by the modeling of 
nuclear effects has been eliminated.

The unprecedented precision achieved for the HERA run I+II data 
facilitates an accurate calibration 
of the earlier fixed-target DIS experiments' normalization. 
Therefore, we introduce a normalization factor for each 
remaining fixed-target data set, SLAC, NMC and BCDMS,  
and fit these factors simultaneously with the PDF parameters. 
The fitted values of normalization factors are determined with an 
uncertainty of ${\cal O}(1\%)$, cf. Tab.~\ref{tab:SLAC}. 
Such a re-evaluation of the normalization is entirely
justified for the SLAC and NMC experiments, 
as it was determined in those experiments in a similar way, using, however, 
less accurate data sets for calibration.
It is also relevant for the BCDMS data~\cite{Benvenuti:1989rh}, which 
were not subject to an additional re-normalization 
in our earlier ABM12 and ABMP15 fits based on the HERA I data. 
Indeed, the BCDMS normalization uncertainty determined in present analysis 
is much smaller than the one of 3\% provided by BCDMS itself.
In general, the normalization factors obtained in the present analysis are 
comparable to unity within the normalization uncertainties quoted by respective experiment.  
However, for the SLAC-E89a experiment~\cite{Atwood:1976ys} the normalization factor deviates from unity by $\sim 5\%$. 
Besides, the data description quality achieved for the SLAC-E89a data 
is significantly worse than for other SLAC experiments.  
It is also worth noting that the SLAC-89a experiment is kinematically separated 
from other SLAC measurements. 
Therefore, having no possibility to clarify the issue of its normalization,
we do not use SLAC-E89a data in the final version of the present analysis. 

\input{table-slac-data1.tex}

\subsubsection{DIS charm- and bottom-quark production}

In addition to the HERA inclusive combination~\cite{Abramowicz:2015mha},
we include into the fit the semi-inclusive HERA data on 
NC DIS charm-quark production 
obtained by a combination of the corresponding H1 and ZEUS results~\cite{Abramowicz:1900rp}.
Those data provide a complementary constraint on the low-$x$ gluon and sea-quark distributions, cf. Ref.~\cite{Abramowicz:1900rp},
and have already been employed in our earlier ABM12 and ABMP15 analyses. 

The CC DIS charm-quark production, which is mostly relevant for disentangling 
the strange sea distribution, is routinely measured by detecting 
di-muons produced in neutrino-nucleon interaction. Two data sets of 
such kind, obtained by the CCFR and NuTeV 
experiments~\cite{Goncharov:2001qe}, were used in our earlier ABM12 
and ABMP15 fits. For the present analysis we add the 
recent precision measurement of di-muon production in $\nu$-Fe DIS 
performed by the NOMAD experiment~\cite{Samoylov:2013xoa}, which  
allows to improve the strange sea determination at large $x$,
cf. Ref.~\cite{Alekhin:2014sya}. One more new measurement of the CC
charm-quark production was performed by the CHORUS 
collaboration~\cite{KayisTopaksu:2011mx} using an emulsion target. 
As a benefit of this technique the charmed-hadrons are detected
directly by their hadronic decays, therefore the CHORUS data
are less sensitive to the details of the charm fragmentation modeling.
Likewise, the data on the charmed-hadron production rates from the emulsion
experiment FNAL-532~\cite{Ushida:1988rt} help to constrain the  
charmed-hadron semi-leptonic branching ratio, which is
required for the analysis of the CCFR, NuTeV, and NOMAD di-muon data, 
cf. Ref.~\cite{Alekhin:2014sya}.

Finally, the bottom-quark DIS production cross sections measured by the 
H1~\cite{Aaron:2009af} and ZEUS~\cite{Abramowicz:2014zub} collaborations 
are also included into the present analysis. 
This allows to determine the value of the bottom-quark mass.

\input{table-dis-dy.tex}

\input{table-ttbar-data.tex}
\input{table-single-top-data.tex}
\subsubsection{Drell-Yan process}

The data on the hadro-production of $W^{\pm}$- and $Z$-bosons 
and the DIS data sets discussed above are mutually complementary 
in the context of disentangling the light-quark PDFs.
In particular, the high statistics data from LHC and Tevatron 
on $W^{\pm}$-production in the forward region allow to improve 
the determination of the up- and down-quark distributions down to 
$x\sim 10^{-4}$, cf. Ref.~\cite{Alekhin:2015cza}.
For the present analysis we select the most recent and statistically significant 
data sets on the $W^{\pm}$- and $Z$-boson production collected by the ATLAS, CMS, and LHCb
experiments at the LHC and the D{\O} experiment at Fermilab, 
cf. Tab.~\ref{dis_dy_table}.
The updated analysis of ATLAS data~\cite{Aaboud:2016btc} collected at $\sqrt{s}=7$~TeV 
was released after completion of our fit. These data are in a good agreement with the 
predictions based on the ABM12 PDFs and therefore should be smoothly
accommodated into a future release.

The data on $W^{\pm}$-production in Tab.~\ref{dis_dy_table} are given 
in form of pseudo-rapidity distributions for the decay electron or muon.
The D{\O} data on $W^{\pm}$-distributions obtained by
unfolding the charged-lepton ones are also available~\cite{Abazov:2013dsa}.
Since those data are sensitive to the details of the 
modeling of the $W^{\pm}$-decay and, in particular, to the PDFs used, 
they are not included into our analysis in order to avoid a bias
due to a mismatch between the PDFs used in the D{\O} analysis and ours,
see also the discussion in Ref.~\cite{Alekhin:2015cza}. 

When available~\cite{Aad:2011dm,Khachatryan:2016pev,Aaij:2015gna,Aaij:2015vua,Aaij:2015zlq}, the absolute measurements of the lepton pseudo-rapidity 
distributions are used. 
In other cases~\cite{Chatrchyan:2013mza,Abazov:2013rja,D0:2014kma} we employ the lepton charge asymmetries. 
However, as a cross-check we also compare our predictions with the LHCb~\cite{Aaij:2015gna,Aaij:2015zlq} 
and CMS~\cite{Khachatryan:2016pev} data on the lepton charge asymmetry, 
although the absolute measurements are used in the fit, cf. Sec.~\ref{sec:dataq}.
The recent ATLAS measurements of the $W^{\pm}$- and $Z$-boson cross sections 
in the fiducial volume at $\sqrt{s}=13$~TeV~\cite{Aad:2016naf} used in our analysis are separated for the electron- and muon-decay channels taking into account 
correlations between these measurements. This gives six data points in total 
for our $\sqrt{s}=13$~TeV ATLAS data set. 

The fixed-target Drell-Yan data provide information on the quark PDFs in the 
high-$x$ region and allow to separate the sea and valence quark distributions. 
In the present analysis two data sets of this kind are employed: 
the ratio of the  
proton-proton and proton-deuterium cross sections from the 
FNAL-866 experiment~\cite{Towell:2001nh} and the proton-copper data 
from the FNAL-605 experiment~\cite{Moreno:1990sf}. 
Both sets have been used in our earlier fits, cf. Ref.~\cite{Alekhin:2012ig}.

\subsubsection{Top-quark production}

Measurements of top-quark production at the LHC and Tevatron
provide a powerful tool for the study of the gluon distribution 
at large $x$ and of $\alpha_s$ at large renormalization scales. 
However, due the strong sensitivity to the value of $m_t$, the accuracy achieved in 
such a study is essentially limited by the uncertainty in $m_t$. 
To take into account this interplay we fit the value of $m_t$ simultaneously with 
the PDF parameters and $\alpha_s$, cf. Sec.~\ref{sec:results}. 
The $t$-quark data included into the present analysis comprise 
the $t\bar t$-production cross sections  measured with various analysis techniques 
and for different decay modes at the center-of-mass energies $\sqrt{s}=5$, 7, 8, and 13 TeV by ATLAS and CMS, cf. Tab.~\ref{tab:data-tt}, 
and those at $\sqrt{s}=1.96$~TeV
obtained at Tevatron~\cite{Aaltonen:2013wca}. 
In addition, single-top production data in the $s$- and $t$-channel from
Tevatron 
and in the $t$-channel from the LHC are considered, cf. Tab.~\ref{tab:data-inp}. 
Single-top production is mediated by the 
electroweak interactions at leading order 
and thus not particularly sensitive to $\alpha_s$ and the gluon distribution. 
Therefore, the latter input dampens the correlation between the gluon PDF,
$\alpha_s$ and $m_t$, which emerges in the analysis of the $t\bar t$-data.  

Due to specifics of the experimental analyses for the $t$-quark detection as
well as necessary extrapolations in phase space, 
the $t$-quark production cross sections usually depend on the value of $m_t$ 
which is taken for the experimental modeling. 
For the Tevatron $t\bar t$-data~\cite{Aaltonen:2013wca} this effect leads to a change  
of ${\cal O}(\pm 1\sigma)$ in the measured cross section when $m_t$ is varied 
by $\pm 2.5~{\rm GeV}$. 
To take this dependence into account we have selected for 
the analysis the value of the cross section of Ref.~\cite{Aaltonen:2013wca}
corresponding to $m_t=170~{\rm Gev}$, 
which is close to our result for $m_t^{pole}$, cf. Sec.~\ref{sec:hq}.  
The sensitivity of the other $t$-quark cross section measurements
used in our analysis  
to the value of $m_t$ is much smaller or not documented. 
Therefore it is not taken into account. 

\subsection{Theory}
\label{sec:theory}

The theoretical description of the hard-scattering processes follows our previous work 
ABM12 and the subsequent updates~\cite{Alekhin:2013nda,Alekhin:2014sya,Alekhin:2015cza}.
We only consider data in the fit, which can be confronted with QCD predictions
at least to NNLO accuracy. 
This allows the analysis to be based on three major types of scattering reactions: 
DIS, the Drell-Yan process and the hadro-production of top-quarks in various channels, 
see also the recent review~\cite{Accardi:2016ndt}.
In this Section we briefly summarize the theory foundations.
Special emphasis is given to the DIS heavy-quark production, where we improve the
approximation of the NNLO Wilson coefficients, to the single-top production in the 
$s$-channel, as well as to the role of power corrections in DIS, where we refine the
treatment of higher twist contributions.

\subsubsection{DIS}

Electron- and neutrino-induced DIS data for NC and CC exchange form the
backbone of basically all PDF analyses.
The theoretical description of these processes uses the operator product expansion (OPE)
on the light-cone~\cite{Wilson:1969zs,Zimmermann:1970,Brandt:1970kg,Frishman:1971qn,Christ:1972ms} 
for fixed values of the Bjorken variable $x$ and the (space-like)
momentum-transfer between the scattered lepton and the nucleon $Q^2 \to \infty$.
The cross sections can be expanded in terms of the well-known (unpolarized) 
DIS structure functions $F_i$, $i=1,2,3$.\footnote{
The alternative definitions $F_T=2xF_1$ and $F_L=F_2-F_T$ are also used
throughout the paper.} 
By virtue of the OPE the latter can be expressed as a product of (Mellin moments of) the Wilson
coefficients $c_{k,i}$ and operator matrix elements (OMEs) of leading twist $\tau = 2$. 
The local OMEs for forward scattering are determined by two states of equal momentum $p$
as~\cite{Politzer:1974fr,Reya:1979zk}
\begin{eqnarray}
  \label{eq:tw2OME}
  \langle p| O_{\mu_1,...\mu_N}|p\rangle
  \, .
\end{eqnarray}
In the three flavor non-singlet cases the renormalization group equation is a scalar differential equation,
while in the singlet case, the quark singlet and the gluon OMEs mix and a $2 \times 2$ system has to be solved. 
In Mellin-$N$ space, for all values of $N \geq N_0$ and $N, N_0 \in \mathbb{N}$, 
the form of these equations is the same and the corresponding anomalous dimensions 
for all $N$ \cite{Moch:2004pa,Vogt:2004mw} are known to NNLO 
and even to N$^{3}$LO for some low moments~\cite{Baikov:2015tea,Velizhanin:2011es,Velizhanin:2014fua,Ruijl:2016pkm} 
and in the large $n_f$-limit~\cite{Davies:2016jie}.
By means of an inverse Mellin transform one obtains the OMEs as a function of Bjorken variable $x$. 
In latter form, i.e. as a function of $x$, the OMEs can also be obtained with
the help of the standard QCD factorization theorems \cite{Politzer:1977fi,
Amati:1978wx,Amati:1978by,Libby:1978qf,Libby:1978bx,Mueller:1978xu,Collins:1981ta,Collins:1985ue,Bodwin:1984hc}, 
due to the one-particle notion of the twist $\tau = 2$ OME in Eq.~(\ref{eq:tw2OME}). 

The QCD corrections to coefficient functions of the hard scattering for NC and CC DIS 
and including mass effects of heavy quarks have been calculated to sufficient accuracy. 
The massless coefficient functions for the NC longitudinal structure function $F_L$ are known 
to NNLO \cite{Moch:2004xu} and for $F_2$ to next-to-next-to-next-to-leading order (N$^{3}$LO) in QCD. 
The corresponding massive ones have been determined to good approximation at NNLO in \cite{Kawamura:2012cr} 
and further improvements will be presented below.

For the neutrino-induced DIS as described by the structure functions $F_{i}^{\,\nu p + \bar\nu p}$, $i=1,2,3$ 
when considering the sum of $\nu$ and $\bar \nu$, exact results 
for the massless coefficient functions are known as well~\cite{Vermaseren:2005qc,Moch:2008fj} to N$^{3}$LO, 
while mass effects have been computed exactly to NLO~\cite{Gottschalk:1980rv,Gluck:1996ve,Blumlein:2011zu} 
and to NNLO in~\cite{Buza:1997mg,moch:2013cc,Blumlein:2014fqa} in the asymptotic region for $Q^2 \gg m_c^2$ 
and in~\cite{Berger:2016inr} completely, i.e., including also those terms beyond the limit $Q^2 \gg m_c^2$.
Details on the treatment of CC DIS heavy-quark production in the PDF analysis
have been given in \cite{Alekhin:2014sya} and will be summarized below.

Note that DIS heavy-quark production is treated entirely in fixed-flavor number scheme 
using $n_f=3$, see, e.g. \cite{Alekhin:2009ni}, and the running-mass
definition~\cite{Alekhin:2010sv} for $m_c$ and $m_b$ is used.

\subsubsection{Higher twist}
\label{sec:hts}

In the twist-expansion 
\cite{Wilson:1969zs,Zimmermann:1970,Brandt:1970kg,Frishman:1971qn,Christ:1972ms}
the unpolarized structure functions $F_{i}(x,Q^2)$ 
for $i=2,L,T$ take the form
\begin{eqnarray}
\label{eq:twexp}
F_{i}(x,Q^2) = F_{i}^{\tau = 2}(x,Q^2) 
+ \sum_{k=1}^\infty \left(\frac{Q_0^2}{Q^2}\right)^k H_{i}^{\tau =2(k+1)}(x,Q^2)
\, ,
\end{eqnarray}
with contributions $H_{i}^{\tau}$ of higher (dynamical) twist 
and $Q_0^2 \simeq 1$\GeV$^2$ denotes a typical reference scale.
Unlike the case of polarized DIS, there are no twist $\tau = 3$ contributions
in Eq.~(\ref{eq:twexp}) but dynamical higher twist terms for $\tau = 2n, n \in \mathbb{N}, n \geq 2$. 
These terms are largely suppressed in the limit of high virtualities $Q^2$.

However, the experimental data often exhibit a correlation between $x$ and $Q^2$ due to similar values of the 
center-of-mass energy $\sqrt{s}$. 
Furthermore, in the NC DIS the largest statistics is localized in the region of lower values of $Q^2$. 
It is often difficult to decide from which scale $Q^2$ onwards a data sample is widely 
free of higher twist contributions. 
It has been proposed \cite{Martin:2002dr,Blumlein:2006be} that a cut 
on the invariant mass of the hadronic system
\begin{eqnarray}
  \label{eq:wdef}
  W^2 = M_P^2 + Q^2 (1-x)/x
  \, ,
\end{eqnarray}
where $M_P$ is the proton mass, might eliminate the higher twist terms. 
Specifically, the ranges $W^2 = 12.5 \div 15 \GeV^2$ for $Q^2 > 4~\GeV^2$ in the non-singlet case at current experimental resolutions have been suggested.
In the singlet case \cite{Alekhin:2012ig} an additional cut of $Q^2 \gtrsim 10~\GeV^2$ 
is necessary to effectively remove the higher twist terms in the current DIS world data. 
These cuts might change as soon as more precise experiments will be performed~\cite{Boer:2011fh,AbelleiraFernandez:2012cc}.
Moreover, in any attempt to determine dynamical higher twist contributions from DIS precision data it is necessary also to include all other mass
effects, i.e., those due to target masses \cite{Georgi:1951sr} and 
due to heavy quarks~\cite{Ablinger:2010ty,Ablinger:2014nga,Ablinger:2014lka,moch:2013cc,Blumlein:2014fqa,Behring:2015roa,Behring:2014eya,Bierenbaum:2009mv}.

The higher twist terms $H_{i}^{\tau}(x,Q^2)$ for $i=2,L,T$ can be decomposed like the leading twist ones 
into process dependent coefficient functions and process independent higher twist OMEs. 
In the massless case the connection between both quantities is given by a series of integration variables $x_j$
\begin{eqnarray}
  \label{eq:twexp1}
  H_{i}^{\tau = 2k}(x,Q^2) \,=\, 
  \sum_n \int_0^1 dx_1 ... \int_0^1 dx_{2k}\, 
  \delta\left(x-\sum_{j=1}^{2k-1} x_j\right)\, c_{i;n}^{\tau = 2k}\left(x_j,\frac{Q^2}{\mu^2}\right)\, O^n\left(x_j,\frac{\mu^2}{\mu_0^2}\right)
  \, ,
\end{eqnarray}
where the sum runs over all contributing operators, using the quasi-partonic operator representation, 
see e.g. \cite{Bukhvostov:1985rn}. 
The local operators of higher twist can be constructed systematically near the light-cone. 
They are formed by more external quark and gluon fields than the twist-two operators 
and potential contributions of lower twist operators if mass scales are present. 
An example for a local twist-four operator is given by
\begin{eqnarray}
:\bar{\psi}(x) \gamma_{\mu_1} \partial_{\mu_2} ... \partial_{\mu_m} \psi(x)
 \bar{\psi}(y) \gamma_{\nu_1} \partial_{\nu_2} ... \partial_{\nu_n} \psi(y):
\, .
\end{eqnarray}
The OMEs which do not belong to the same representation obey different renormalization group equations. 
This applies as well to the higher twist Wilson coefficients $c_{i;n}^{\tau}$ for $\tau \ge 4$, 
which can be calculated perturbatively. 
It is important to solve these evolution equations individually, since the scaling violations,
through which the respective quantity contributes to the structure functions, turn out to be different. 
As a consequence, the higher twist terms contribute additively to the leading twist
term in Eq.~(\ref{eq:twexp}) and not multiplicatively, which is sometimes assumed in the literature~\cite{Abt:2016vjh}.

Among the $(2k-1)$ contributing momentum fractions in Eq.~(\ref{eq:twexp1})
only the value of $x$ can be accessed experimentally.
In particular, there is a priori no way to determine the functional structure of the OMEs $O^n(x_j,\mu^2/\mu_0^2)$ 
with respect to the other variables by fitting data, contrary to the possibility in case of the twist-two terms. 
In the future one might in principle consider lattice simulations of these terms, 
although at the moment no method is known to obtain precise $x$-space predictions in this way.
Therefore, rigorous $x$-space higher twist QCD analyses of the DIS data are currently impossible.

A more realistic scenario for a consistent QCD analysis of dynamical higher twist contributions 
is encountered when working in Mellin space. 
Here the OMEs form matrix representations whose dimensions are growing 
with growing values of the Mellin variable $N$, which corresponds to the variable $x$. 
The growth in the number of contributing operators  becomes more and more significant 
when going to higher twists. 
Here, the OMEs are given by pure numbers, which can be determined in 
an analysis of precise experimental data, realistically up to a specific value of $N$.
However, to comply with the accuracy of the present twist-two analyses the corresponding terms would have to be calculated at NNLO, 
including the massive corrections to the same order. 
Moreover, to measure the corresponding moments of the OMEs, it is necessary to extrapolate in the small- and large-$x$ regions. 
The small-$x$ region is damped to some extent and the problematic part is the region of large values of $x$. 
Still, such an analysis is possible, see e.g.~\cite{Osipenko:2004xg}.

On the theoretical side, systematic twist decompositions have been performed, cf.~\cite{Geyer:1999uq,Geyer:2000ig}.
One forms OMEs with these operators between nucleon states. 
A `partonic' interpretation assumes, that all external lines can be factorized individually. 
Early theoretical investigations of the structure of higher twist operators  and their
anomalous dimensions for $\phi^3$-theory in $D = 6$ dimensions~\cite{Gottlieb:1978,Gottlieb:1978zj,Lam:1984qr} 
and for QCD \cite{Politzer:1980me,Okawa:1980ei,Okawa:1981uw,Wada:1981hx,Wada:1982my,
  Luttrell:1981ke,Luttrell:1982zu,Shuryak:1981dg,Shuryak:1981kj,
  Shuryak:1981pi,Jaffe:1981td,Jaffe:1982pm,Jaffe:1983hp,
  Ellis:1982wd,Ellis:1982cd,
  Bukhvostov:1985rn,Bukhvostov:1987pr,Qiu:1988dn,Bartels:1999km,Braun:2000kw}
revealed a basic structure of these contributions. 
The lowest order anomalous dimensions have been calculated in Refs.~\cite{Bukhvostov:1985rn,Bukhvostov:1987pr,Braun:2000kw}
as well as the Wilson coefficients, 
in different operator bases, in Refs.~\cite{Jaffe:1981td,Jaffe:1982pm,Jaffe:1983hp,Ellis:1982wd,Ellis:1982cd,Qiu:1988dn}. 
More recently, also gluonic operators were considered \cite{Bartels:1999km}. 
A systematic study of the higher twist light-cone distribution amplitudes was given in Ref.~\cite{Braun:2000kw}. 
The renormalization of these operators has been worked out in Refs.~\cite{Braun:2008ia,Braun:2009vc}.
The evolution of the lowest twist-four moments at leading order has been illustrated, e.g., in Ref.~\cite{Glatzmaier:2012iu}. 
We note that the higher twist anomalous dimensions and Wilson coefficients are presently available 
at low orders in QCD only.
Estimates of the higher twist effects have been obtained also by studying renormalon corrections to sum-rules
and DIS structure functions \cite{Braun:1986ty,Balitsky:1989uj,Balitsky:1989jb,Dasgupta:1996hh}, see also 
Refs.~\cite{Beneke:1998ui,Beneke:2000kc}.

In current $x$-space analyses only an effective determination of higher twist contributions is possible.
In the flavor non-singlet case \cite{Blumlein:2006be} one uses the cuts mentioned above 
and studies at lower values of $Q^2$ the deviations from the twist-two prediction determined 
in the high $Q^2$ region based on the N$^{3}$LO corrections in QCD. 
Here, the higher twist contributions to the structure functions are fitted simply as parameters depending on $x$ and $Q^2$, 
i.e. no assumptions are made on the contributing anomalous dimensions or the Wilson coefficients, 
cf. Refs.~\cite{Blumlein:2006be,Blumlein:2008kz,Blumlein:2012se}. 
Other fits of the dynamical higher twist contributions bin by bin in $x$ and $Q^2$, 
both in the non-singlet and the singlet case, have been performed in Refs.~\cite{Eisele:1981nk,
Varvell:1987qu,Virchaux:1991jc,Kataev:1997nc,Alekhin:1998df,Botje:1999dj,
Alekhin:2000ch,Alekhin:2003qq,Alekhin:2007fh,Blumlein:2008kz,Blumlein:2012se}. 
In this approach, it is also important to control the interplay of the size of higher twist terms with contributions 
to the leading twist Wilson coefficients at higher orders in perturbation theory. 
In the large-$x$ region the latter can be obtained for non-singlet DIS from threshold resummation 
and subsequent expansion, which generates approximate N$^{4}$LO corrections, see e.g. Refs.~\cite{Moch:2005ba,Ravindran:2006cg}. 

In view of these considerations, we use in the current fit 
an entirely phenomenologically motivated ansatz for the DIS structure functions including higher twist,
\begin{eqnarray}
\label{eq:htwist}
  \displaystyle
  F_i^{\rm ht}(x,Q^2) &=&
  F_i^{\rm TMC}(x,Q^2)
  +
  \frac{H_i^{\tau=4}(x)}{Q^2}
  \, ,
\qquad\qquad 
  i \,=\, 2,T
  \, ,
\end{eqnarray}
where $F_i^{\rm TMC}$ is given by the leading twist structure function of Eq.~(\ref{eq:twexp}) 
together with the target mass corrections~\cite{Georgi:1951sr}, see also~\cite{Alekhin:2012ig}. 
The reference scale in Eq.~(\ref{eq:twexp}) has been chosen $Q_0^2 = 1$\GeV$^2$ 
and the higher twist terms $H_i^{\tau}$ are taken to be independent of $Q^2$,
i.e. to correspond to the central value of $Q^2$ in the respective $x$-range being analyzed. 
The results for $H_i^{\tau}$ will be presented below in Sec.~\ref{sec:results}.

\subsubsection{DIS heavy-quark production}
\label{sec:ccbar-dis}

The cross section for heavy-quark production in DIS for NC exchange 
by photons of virtuality $Q^2$ is expressed in terms of the 
heavy-flavor structure functions $F_k(x,Q^2,m^2)$ with $k=2,L$.
Here, $m$ is the mass of the heavy quark with $m^2 \gg \Lambda^2_{\rm QCD\,}$.
The structure functions are given as convolutions of PDFs $f_{i}^{}$ 
and coefficient functions $c_{k, i}^{}$, see for instance~\cite{Riemersma:1994hv},
\begin{equation}
  \label{eq:totalF2c}
  F_k^{}(x,Q^2,m^2) \;=\; {\alpha_s\, e_h^{\:\!2} \over 4\:\! \pi^{\:\!2}\,}\, \xi \,\,
  \sum\limits_{i \,=\, q,{\bar{q}},g} \,\,
  \int_{x}^{\,z^{\,\rm max}}
  {dz \over z} \: f_{i}^{}\left({x \over z},\, \mu_f^2 \right)\,
  c_{k, i}^{}\left(\eta(z),\,\xi,\,\mu_f^2,\,\mu_r^2 \right)
  \; ,
\end{equation}
where $z^{\,\rm max} = 1/(1 + 4\:\! m^2/Q^2)$ and $e_h^{}$ is the heavy-quark charge.
The kinematic variables are
\begin{equation}
  \label{eq:eta-xi-def}
  \eta \;=\; {s \over 4\:\! m^2}\: - 1 \quad ,
  \qquad
  \xi \;=\; {Q^2 \over m^2} \;\; ,
\end{equation}
with the partonic center-of-mass energy $s = Q^2 (1/z-1)$.

The coefficient functions can be expanded in powers of $\alpha_s$ 
\begin{eqnarray}
  \label{eq:coeff-exp}
  c_{k, i}^{}(\eta,\xi,\mu^2)
  \; = \;
  \sum\limits_{j=0}^{\infty}\, (4\:\! \pi\, \alpha_s)^j \, 
  c^{\,(j)}_{k,i}(\eta, \xi, \mu^2)
  \; = \;
  \sum\limits_{j=0}^{\infty}\, (4\:\! \pi\, \alpha_s)^j \,
  \sum\limits_{l=0}^{j} c^{\,(j,\ell)}_{k,i}(\eta, \xi)\: 
      \ln^{\,\ell}\frac{\mu^2}{m^2} \:\; ,\quad
\end{eqnarray}
where we have identified the renormalization and factorization scales 
$\mu\,=\,\mu_f\,=\,\mu_r$.

The complete QCD corrections to the coefficient functions in Eq.~(\ref{eq:coeff-exp}) 
are known at NLO, i.e., $c_{k,i}^{\,(1,0)}$ and $c_{k,i}^{\,(1,1)}$, 
see Refs.~\cite{Laenen:1992zk,Riemersma:1994hv,Harris:1995tu}, 
as well as all scale-dependent terms at NNLO, 
i.e., $c_{k,i}^{\,(2,1)}$ and $c_{k,i}^{\,(2,2)\!}$,
see Refs.~\cite{Laenen:1998kp,Behring:2014eya,Alekhin:2010sv}.
Since not all complete results for the scale independent parts
$c_{k,i}^{\,(2,0)}$ at NNLO were available in 2012, 
approximate predictions for the most important 
gluon and the quark pure-singlet coefficient functions $c_{2,g}^{(2,0)}$ and $c_{2,q}^{(2,0)}\!$
covering a wide kinematic range have been provided in Ref.~\cite{Kawamura:2012cr}.
These NNLO approximations are based on significant partial information 
about the threshold region $s \simeq 4\:\! m^2$,
cf. Refs.~\cite{Laenen:1998kp,Kawamura:2012cr}, 
the high-energy regime $s \gg 4\:\! m^2$, cf. Ref.~\cite{Catani:1990eg}, 
as well as the high-scale region $Q^2 \gg m^2$, 
cf. Refs.~\cite{Bierenbaum:2008yu,Bierenbaum:2009zt,Bierenbaum:2009mv,Ablinger:2010ty}.
Since then, important new results have appeared~\cite{Ablinger:2014nga,Ablinger:2014lka,Behring:2015roa,Behring:2014eya}
which are valid in the limit $Q^2 \gg m^2$ and allow for a substantial improvement of the constructions 
of Ref.~\cite{Kawamura:2012cr} as we will discuss below.

In the limit $Q^2 \gg m^2$ the heavy-quark coefficient functions are subject to 
an exact factorization into the respective coefficient functions with massless quarks 
$c^{\,\rm light}_{k,j}$ and heavy-quark OMEs $A_{ij}$. 
Schematically we have for $Q^2 \gg m^2$, that is large $\xi$,
\begin{equation}
  \label{eq:hqfact}
    c_{k, i}(\eta,\xi,\mu^2) \;\: \to \;\:
    c^{\,\rm asy}_{k, i}\left(x,\frac{Q^2}{\mu^2},\frac{m^2}{\mu^2}\right) \:=\:
    \bigg[
    A_{ji}\left(\frac{m^2}{\mu^2}\right)\,
    \otimes\,
    c^{\,\rm light}_{k, j}\left(\frac{Q^2}{\mu^2}\right)
    \bigg](x)
  + {\cal O}\left(\frac{m^2}{Q^2}\right)
  \, ,
  \qquad
\end{equation}
where we have indicated that the variable $\eta$ in Eq.~(\ref{eq:eta-xi-def}) factorizes as 
\mbox{$\eta \to Q^2/(4 m^2) (1/x-1) + {\cal O}(1)$} and $\otimes$ denotes the standard Mellin convolution, 
cf.~Eq.~(\ref{eq:totalF2c}).

\begin{figure}[t!]
\begin{center}
\includegraphics[width=0.49\textwidth, angle=0]{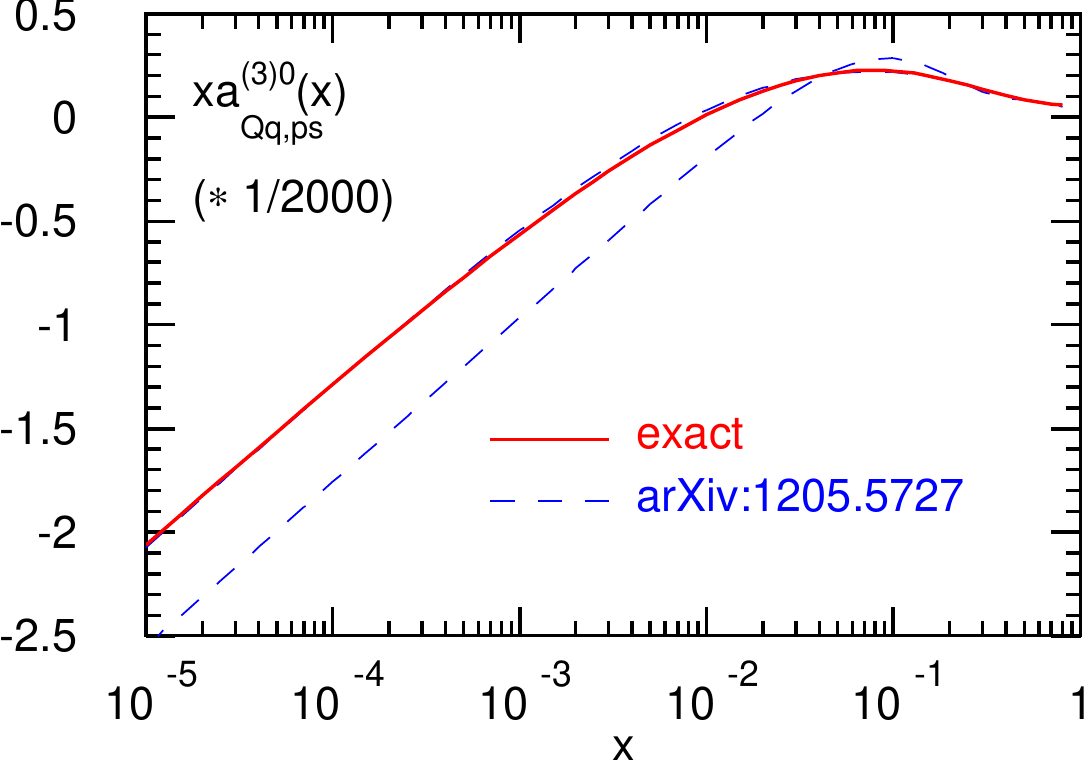}
\hspace*{3mm}
\includegraphics[width=0.47\textwidth, angle=0]{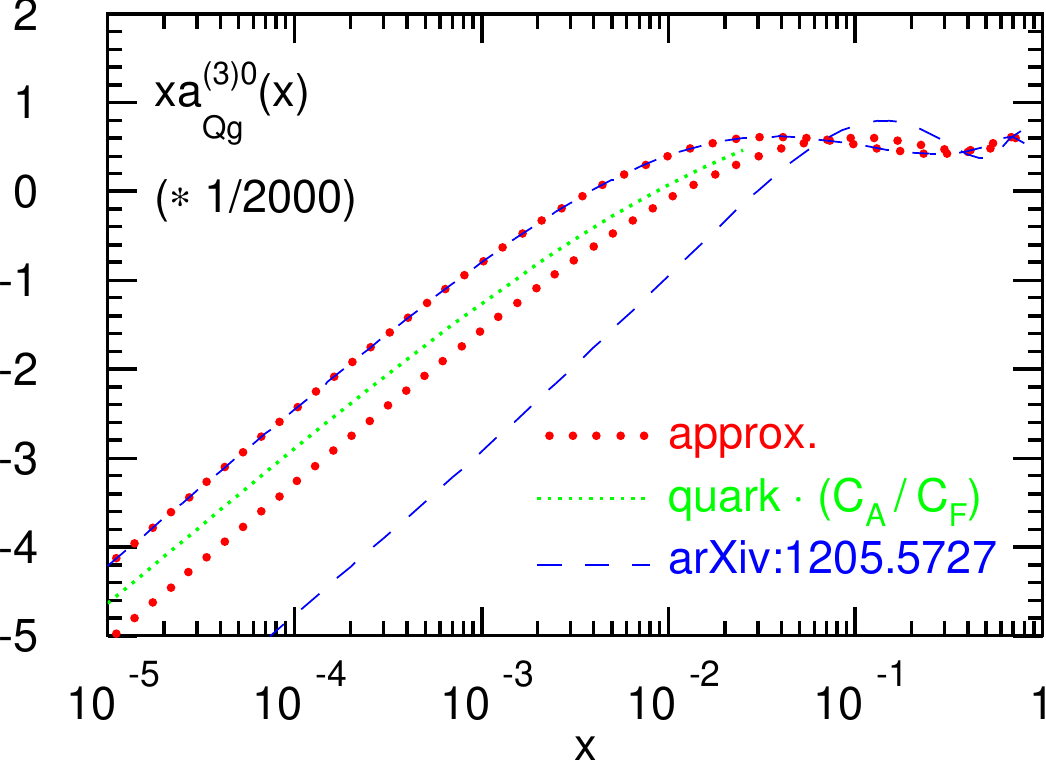}
\caption{\small 
\label{fig:aQj30} 
Left: 
The exact result for the OME $a^{(3)\,0}_{Qq,\,\rm ps}$ 
and comparison to previous approximations of Ref.~\cite{Kawamura:2012cr}. 
Right:
The new approximations for $a_{Qg}^{\,(3)\,0}$ of Eq.~(\ref{eq:fitA}) and (\ref{eq:fitB}) 
based on the `Casimir-scaled' results for $a^{(3)\,0}_{Qq,\,\rm ps}$ indicated by the thin dotted (green) line
and comparison to previous approximations of Ref.~\cite{Kawamura:2012cr}. 
}
\end{center}
\end{figure}

The factorization in Eq.~(\ref{eq:hqfact}) can be used to compute the
asymptotic expressions $c^{\,\rm asy}_{k, i}$ of the heavy-quark coefficient functions 
for $Q^2 \gg m^2$ at NNLO based on knowledge of the anomalous dimensions~\cite{Moch:2004pa,Vogt:2004mw} 
and coefficient functions for DIS with massless quarks~\cite{Zijlstra:1992qd,Moch:1999eb,Vermaseren:2005qc} 
and the results for the massive OMEs~\cite{Bierenbaum:2008yu,Bierenbaum:2009zt,Bierenbaum:2009mv,Ablinger:2010ty} 
up to three loops. 
In analogy to Eq.~(\ref{eq:coeff-exp}) the heavy-quark OMEs in
Eq.~(\ref{eq:hqfact}) can be expanded in 
powers of $\alpha_s$ (note the different normalization convention) as 
\begin{eqnarray}
  \label{eq:OMEexp}
  A_{ij}
  \; = \;
  \delta_{ij} +
  \sum\limits_{k=1}^{\infty}\, 
  \left(\frac{\alpha_s}{4\:\! \pi}\right)^k \, A^{(k)}_{ij}
  \; = \;
  \delta_{ij} +
  \sum\limits_{k=1}^{\infty}\, 
  \left(\frac{\alpha_s}{4\:\! \pi}\right)^k \,
  \sum\limits_{\ell=0}^{k}\, a^{(k,\ell)}_{ij}\, 
  \ln^{\,\ell}\frac{\mu^2}{m^2} 
  \; ,\qquad
\end{eqnarray}
where the genuinely new $k$-th order information resides in the expressions 
$a^{(k,0)}_{ij}$ for which we will use the short-hand $a^{(k)}_{ij} \equiv a^{(k,0)}_{ij}$.
Previous information on $a^{(3)}_{ij}$ at three loops included 
a number of even-integer Mellin moments~\cite{Bierenbaum:2009mv} 
and the complete $n_f$-dependence~\cite{Ablinger:2010ty,Blumlein:2012vq}. 
Thus, for the two important OMEs, 
the heavy-quark gluon $a^{(3)}_{Qg}$ and the heavy-quark pure-singlet one $a^{(3),{\rm ps}}_{Qq}$, 
decomposed in powers of $n_f$ as
\begin{eqnarray}
  \label{eq:aQg30nf-exp}
a^{(3)}_{Qg}\;\;
  &\!=\!&
  a^{(3)\,0}_{Qg\!}
  \;\; +\, n_f\: a^{(3)1}_{Qg}
  \; ,\qquad
\\
  \label{eq:aQqps30nf-exp}
a^{(3),{\rm ps}}_{Qq} \!\!
  &\!=\!&
  a^{(3)\,0}_{Qq,\,\rm ps}
  \,+\, n_f\: a^{(3)1}_{Qq,\,\rm ps}
  \; ,\qquad
\end{eqnarray}
the expressions for $a^{(3)1}_{Qg}$ and $a^{(3)1}_{Qq,\,\rm ps}$ are known exactly~\cite{Ablinger:2010ty},
while approximations for $a^{(3)0}_{Qg}$ and $a^{(3)0}_{Qq,\,\rm ps}$ 
based on some the fixed Mellin moments~\cite{Bierenbaum:2009mv} had been given
in Ref.~\cite{Kawamura:2012cr}.
As a new result the complete exact expression for $a^{(3),{\rm ps}}_{Qq}$ is now available~\cite{Ablinger:2014nga}.
In Fig.~\ref{fig:aQj30} (left) we show $a^{(3)0}_{Qq,\,\rm ps}$ compared
to the previous approximations~\cite{Kawamura:2012cr}. 
The plot demonstrates that the uncertainty band estimates of Ref.~\cite{Kawamura:2012cr} 
have been reasonable, particularly in the small-$x$ region.

\begin{figure}[t!]
\begin{center}
\includegraphics[width=16.2cm]{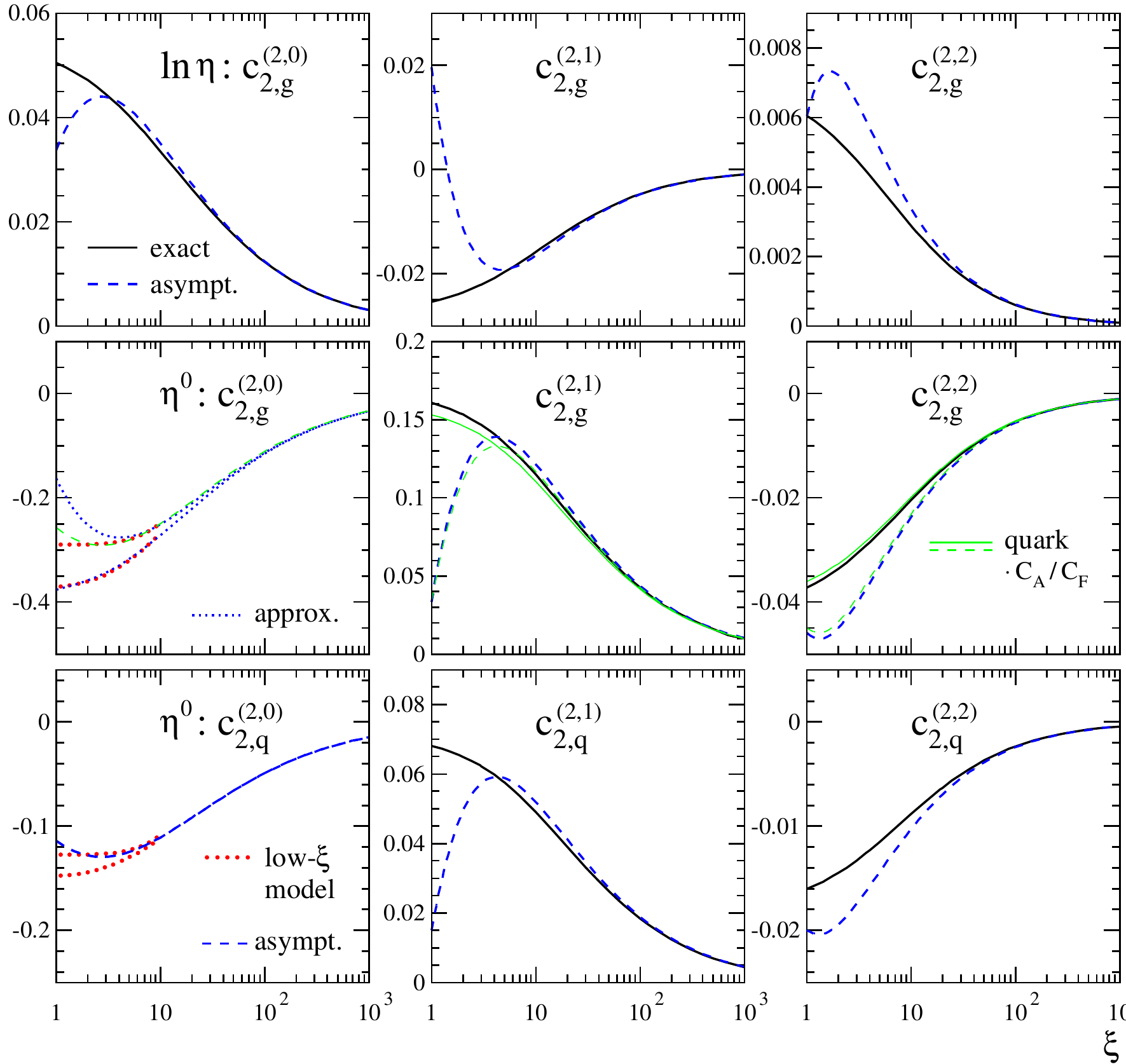}
\caption{\small
  \label{fig:c2i-nllx}
  Top panels: the coefficients of the leading small-$x\,/\,$high-$\eta$
  logarithm for the contributions $c_{2,g}^{(2,\ell)}$, $\ell = 0, 1, 2$,
  to the NNLO gluon coefficient function defined in Eq.~(\ref{eq:coeff-exp}).
  Middle and bottom panels: the respective next-to-leading $\eta^{\,0}$ 
  coefficients for $c_{2,g}^{\,(2)}$ and $c_{2,q}^{\,(2)}$.
  The solid (black) lines are the exact all-$Q^2$ results, the dashed (blue) 
  ones the high-scale asymptotic results $c_{2,g}^{\,(2)\,\rm asy}$ and $c_{2,q}^{\,(2)\,\rm asy}$.
  defined in Eq.~(\ref{eq:hqfact}); 
  the dotted (red) ones the low-scale extrapolations of 
  Eqs.~(\ref{eq:c2g20-nllxA})--(\ref{eq:c2q20-nllxB}).
  Also illustrated, by the thin (green) lines in the bottom panels, is the 
  small next-to-leading high-$\eta$ deviation of $c_{2,g}^{\,(2)}$ from the 
  `Casimir-scaled' results for $c_{2,q}^{\,(2)}$. 
}
\end{center}
\end{figure}

Knowledge of the exact result for $a^{(3)0}_{Qq,\,\rm ps}$ offers a 
possibility for improvement of $a^{(3)0}_{Qg}$ as well, 
since the gluon and pure-singlet quark OME are closely related in the small-$x$ limit.
In fact, the leading small-$x$ terms proportional to $x^{-1}\ln x$ 
(denoted by the superscript LLx in the following)
are identical up to simple scaling with the QCD color factors $C_A/C_F$, 
that is~\cite{Catani:1990eg}, 
\begin{equation}
  \label{eq:casimir-scaling}
 a^{(3)0,\,{\rm LLx}}_{Qq,\,\rm ps} \,=\, \frac{C_A}{C_F}\,\, a^{(3)0,\,{\rm LLx}}_{Qg}
 \, .
\end{equation}
For the sub-leading small-$x$ terms proportional to $x^{-1}$ 
this `Casimir-scaling' is not exact anymore, 
but the deviations are numerically small in known cases, see for instance~\cite{Vogt:2004mw}.
One can, therefore, improve the approximation for $a^{(3)0}_{Qg}$ based on the 
fixed Mellin moments of Ref.~\cite{Bierenbaum:2009mv} by 
fixing the coefficient of the $x^{-1}$ term of $a^{(3)0}_{Qg}$ with the help of the $C_A/C_F$ relation
in Eq.~(\ref{eq:casimir-scaling}) and allowing for an additional variation of $\pm 10\%$ to model the uncertainty.
This leads to the following two approximations shown in Fig.~\ref{fig:aQj30} (right) 
\begin{eqnarray}
  \label{eq:fitA}
   a_{Qg,A}^{(3)\,0}(x) &\!=\!&  
        354.1002 \, \ln^3(1-x)
     \,+\, 479.3838 \, \ln^2(1-x)  
     \,-\, 7856.784 \,(2-x)    
\nn\\&& \mbox{}
     \,-\, 6233.530 \,\ln^2 x
     \,+\, 9416.621 \,x^{-1}
     \,+\, 1548.891 \,\,x^{-1}\,\ln x 
     \; ,
\\[1mm]
  \label{eq:fitB}
   a_{Qg,B}^{(3)\,0}(x) &\!=\!&
        226.3840  \, \ln^3(1-x)
     \,-\, 652.2045 \, \ln^2(1-x)  
     \,-\, 2686.387\, \ln(1-x)  
\nn\\&& \mbox{}
     \,-\, 7714.786 \,(2-x)    
     \,-\, 2841.851 \,\ln^2 x
     \,+\, 7721.120 \,x^{-1}
     \,+\, 1548.891 \,\,x^{-1}\,\ln x 
     \; ,
\end{eqnarray}
which can be considered as two extremes of the approximation procedure
described in Ref.~\cite{Kawamura:2012cr}.
The result for $a_{Qg,A}^{(3)\,0}$ in Eq.~(\ref{eq:fitA}) has been taken over from Ref.~\cite{Kawamura:2012cr} 
since its small-$x$ behavior is close to the result for $a^{(3)0}_{Qq,\,\rm ps}$ rescaled by the factor $C_A/C_F$ 
plus the additionally added shift of $10\%$ as can be seen in Fig.~\ref{fig:aQj30} (right) 
were the previous approximations of Ref.~\cite{Kawamura:2012cr} are shown as well.
It is obvious from the plot that the new information on the small-$x$ behavior~\cite{Ablinger:2014nga}  
helps to reduce the uncertainty on $a_{Qg,A}^{(3)\,0}$ significantly.

\begin{figure}[t!]
\begin{center}
\includegraphics[width=0.47\textwidth, angle=0]{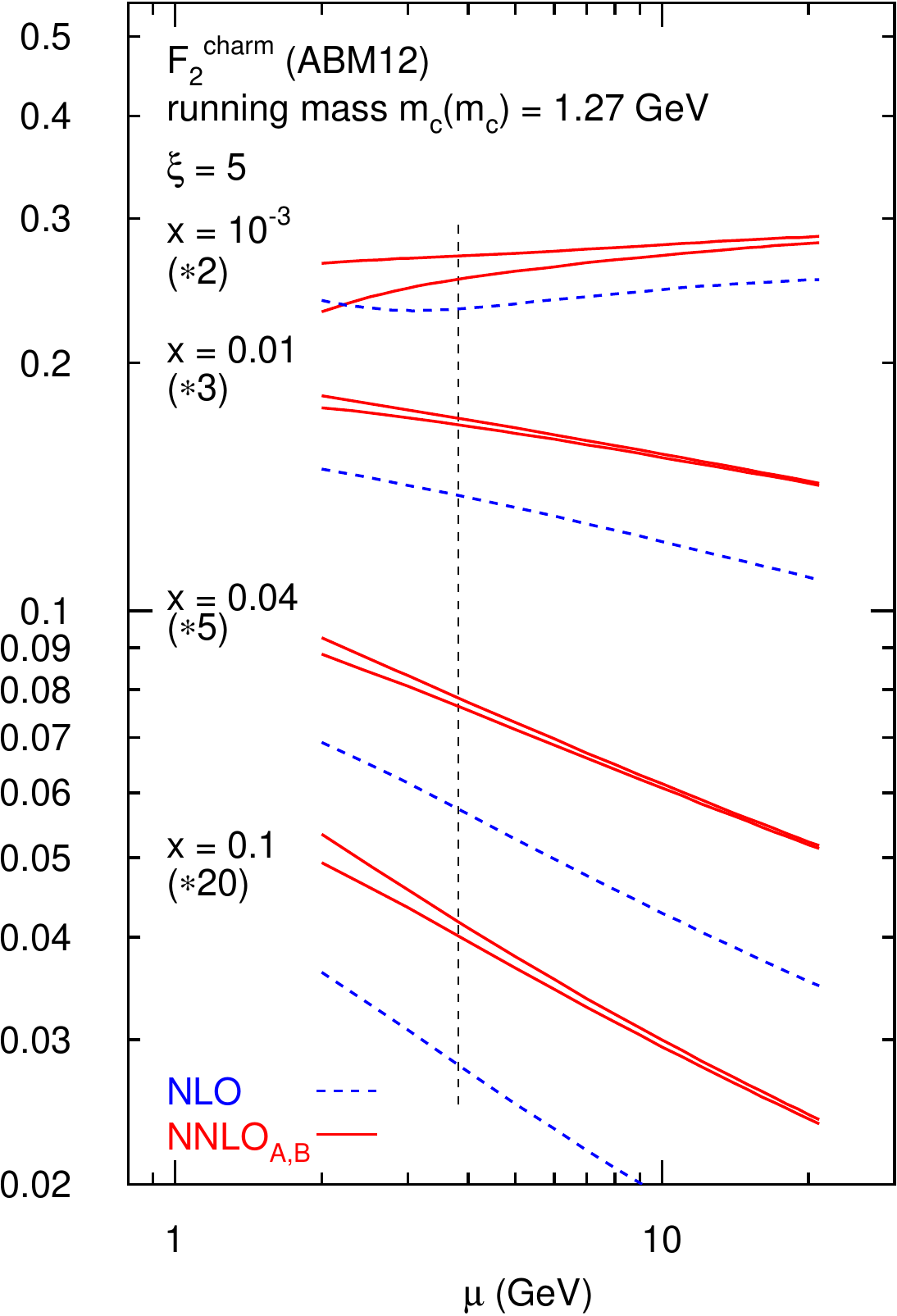}
\hspace*{3mm}
\includegraphics[width=0.471\textwidth, angle=0]{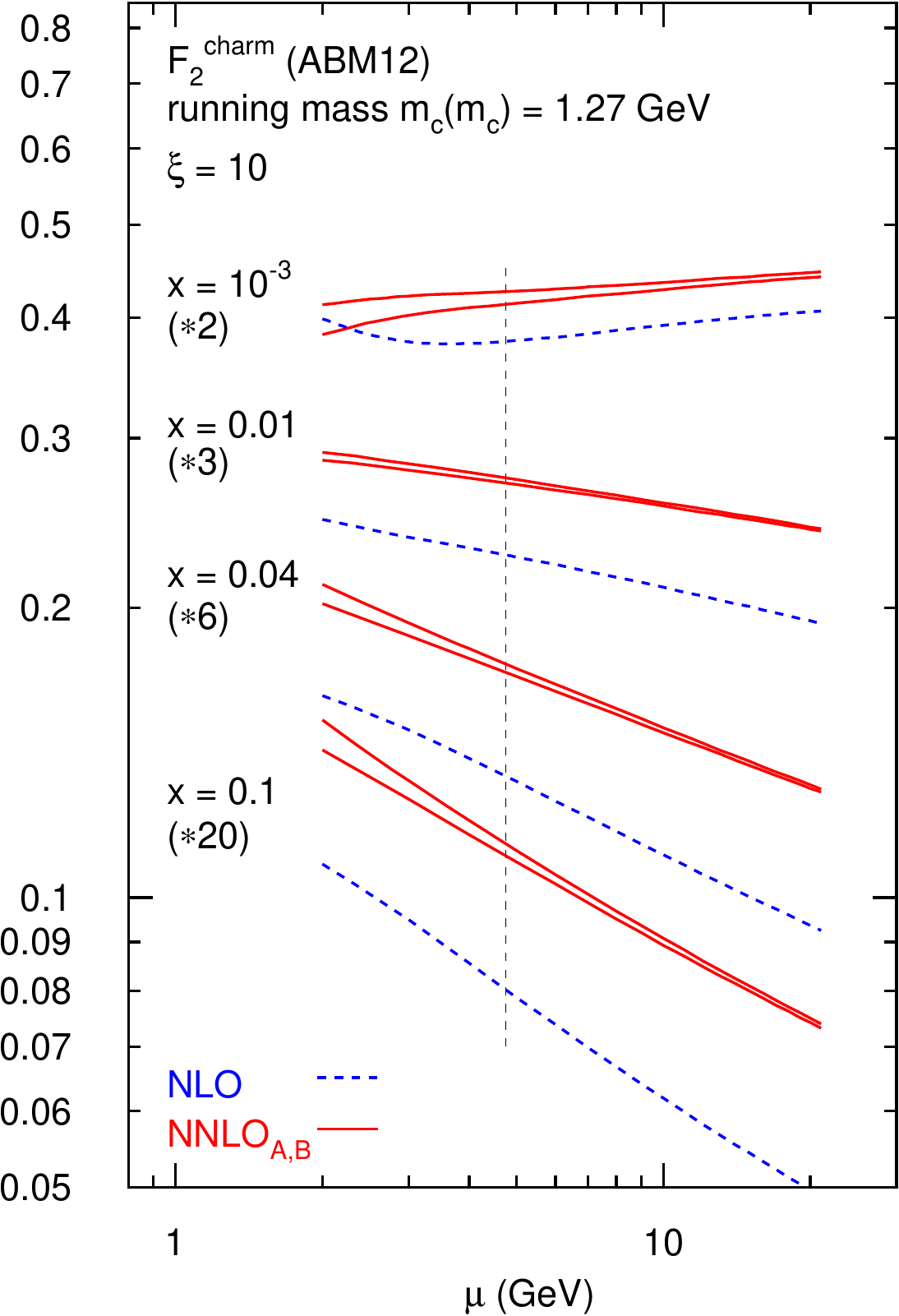}
\caption{\small 
\label{fig:F2c-mu-dep}
Scale dependence of the heavy-quark structure function $F_{2}^{\,c}$ for $\mu=\mu_R=\mu_F$ 
at NLO and at NNLO with the approximations A and B of Eqs.~(\ref{eq:assembly20A}) and (\ref{eq:assembly20B}) 
for two values of $\xi=Q^2/m_c^2$.
The dashed vertical line denotes the choice $\mu=\sqrt{Q^2 + 4m_c^2}$. 
}
\end{center}
\end{figure}

We are now in a position to improve the NNLO approximations of the 
gluon and pure-singlet heavy-quark coefficient functions, $c_{2,g}^{(2,0)}$
and $c_{2,q}^{(2,0)}$.
To that end, we need to combine the results from the various kinematic regions, 
i.e., those near threshold $s \simeq 4\:\! m^2$, at high energies $s \gg 4\:\! m^2$ 
and for large $Q^2 \gg m^2$.
We use the following ansatz for the 
NNLO coefficient functions~\cite{Kawamura:2012cr} 
\begin{eqnarray}
  \label{eq:assembly20A}
  c_{2,i}^{\,(2,0),\,A} &\!\!=\!\!& 
  \delta_{ig}\, \left( c_{2,g}^{\,(2,0)\,{\rm thr}} \right)
  \,+\, \left(1 - f(\xi) \right) \beta\,\,\,\, c_{2,i}^{\,(2,0){\,\rm asy},\,A}
\nn \\ & & \mbox{}
  +\, f(\xi)\, \beta^3
  \bigg( - c_{2,i}^{\,(2,0)\,{\rm LLx}}\; \frac{\ln \eta}{\ln x}
  \:+\: c_{2,i}^{\,(2,0){\,\rm NLL},\,A}
  \,\frac{\eta^{\gamma}}{C + \eta^{\gamma}} 
  \bigg)
  \, ,
\end{eqnarray}
and
\begin{eqnarray}
  \label{eq:assembly20B}
  c_{2,i}^{\,(2,0),\,B} &\!\!=\!\!& 
  \delta_{ig} \left( c_{2,g}^{\,(2,0)\,{\rm thr}} 
  \,+\, f(\xi)\: 2\:\!c_{2,g}^{(2,0)\,{\rm const}} \right)
  \,+\, \left(1 - f(\xi) \right) \beta^3\, c_{2,i}^{\,(2,0){\,\rm asy},\,B}
\nn \\ & & \mbox{}
  +\, f(\xi)\, \beta^3
  \bigg( - c_{2,g}^{\,(2,0)\,{\rm LLx}}\; \frac{\ln \eta}{\ln x}
  \:+\: c_{2,g}^{\,(2,0){\,\rm NLL},\,B}
  \,\frac{\eta^{\delta}}{D + \eta^{\delta}} 
  \bigg) 
  \, ,
\end{eqnarray}
where $\beta \;=\; \sqrt{1-4\:\!m^2/s}$.
The threshold contributions 
$c_{2,g}^{\,(2,0)\,{\rm thr}}$ and $c_{2,g}^{\,(2,0)\,{\rm const}}$ 
for the gluon coefficient function are given by the scale-independent terms 
in Eqs.~(3.18) and (3.19) of Ref.~\cite{Kawamura:2012cr}.
The asymptotic results for the quark and gluon coefficient functions  
$c_{2,i}^{\,(2,0)\,{\rm asy}}$ at large $Q^2 \gg m^2$ 
are given by the scale-independent terms of Eqs.~(B.8) and~(B.10)
of Ref.~\cite{Kawamura:2012cr} where it is understood that 
in the case of the gluon coefficient functions $c_{2,g}^{(2,0)}$ 
the results for $a_{Qg,A}^{(3)\,0}$ and $a_{Qg,B}^{(3)\,0}$ 
in Eqs.~(\ref{eq:fitA}) and (\ref{eq:fitB}) are used to
account for an uncertainty band due to the heavy-quark OME $A_{Qg}$ 
while for the pure-singlet heavy-quark coefficient functions $c_{2,q}^{(2,0)}$
the exact result for $a^{(3)0}_{Qq,\,\rm ps}$ of Ref.~\cite{Ablinger:2014nga}
is to be used in all cases.

The function $f\,(\xi) = (1 + e^{\,2(\xi - 4)})^{-1}$ joins the asymptotic expressions 
for the large-$\xi$ limit with the low-$\xi$ region and 
provides a smooth transition between these two regimes.
Here $c_{2,g}^{\,(2){\rm LLx}}$ is the leading contribution given in Eq.~(3.39) of Ref.~\cite{Kawamura:2012cr}; 
$c_{2,q}^{\,(2){\rm LLx}}$ is given by the same expression rescaled by $C_F/C_A$ according to Eq.~(\ref{eq:casimir-scaling}).
Division of the factor $\ln x$ and substitution $\ln x \to -\ln \eta$ ensures
the correct the slope in $\eta$ at all values of $\xi$, cf. Eq.~(\ref{eq:eta-xi-def}).

The next-to-leading large-$\eta$ terms denoted by $c_{2,i}^{(2){\rm NLL}}$ 
are currently unknown in the low-$\xi$ region, but we can derive constraints 
in the low-$\xi$ region at high-$\eta$ 
\begin{eqnarray}
  \label{eq:c2g20-nllxA}
  c_{2,g}^{\,(2,0)\,{\rm NLL},\:A}(\xi)
  &\!=\!& 
  0.01\,\left(\frac{\ln\xi}{\ln5}\right)^{\!4} \,-\;0.29 \; ,
\\[1mm]
  \label{eq:c2g20-nllxB}
  c_{2,g}^{\,(2,0)\,{\rm NLL},\:B}(\xi)
  &\!=\!& 
  0.05\,\left(\frac{\ln\xi}{\ln5}\right)^{\!2} \,-\,0.37 \; ,
\end{eqnarray}
as well as 
\begin{eqnarray}
  \label{eq:c2q20-nllxA}
  c_{2,q}^{\,(2,0)\,{\rm NLL},\:A}(\xi)
  &\!=\!& 
  0.0045\,\left(\frac{\ln\xi}{\ln5}\right)^{\!4} \,-\;0.1275 \; ,
\\[1mm]
  \label{eq:c2q20-nllxB}
  c_{2,q}^{\,(2,0)\,{\rm NLL},\:B}(\xi)
  &\!=\!& 
  0.0175\,\left(\frac{\ln\xi}{\ln5}\right)^{\!2} \,-\,0.1475 \; .
\end{eqnarray}
These extrapolations are shown in Fig.~\ref{fig:c2i-nllx}
Finally, the matching is performed with the 
factors $\eta^{\gamma}/(C + \eta^{\gamma})$ and $\eta^{\delta}/(D +
\eta^{\delta})$, respectively, 
where the suppression parameters $\gamma,\: C$ and $\delta,\: D$ 
take the values
\begin{equation}
  \label{eq:suppr-param21}
  \gamma = 1.0\: ,\quad C = 20.0\: ,
  \qquad {\rm and}\qquad
  \delta = 0.8\: ,\quad D = 10.7
  \: ,
\end{equation}
as determined in Ref.~\cite{Kawamura:2012cr}.

With these improvements at hand, we can provide new approximate NNLO results for 
the charm-quark structure functions $F_2(x,Q^2,m^2)$ in Eq.~(\ref{eq:totalF2c}).
In Fig.~\ref{fig:F2c-mu-dep} we display $F_2(x,Q^2,m^2)$ at NLO and NNLO 
for charm-quarks at the values of $\xi=5$ and $10$ and for a range of scales $\mu$.
We use the \msbar mass definition with $m_c(m_c)=1.27 \GeV$ 
and the ABM12 PDFs~\cite{Alekhin:2013nda} at NNLO.
It is clear from Fig.~\ref{fig:F2c-mu-dep} that the residual uncertainty 
in $F_2(x,Q^2,m^2)$ at NNLO due to the approximations A and B 
of Eqs.~(\ref{eq:assembly20A}) and (\ref{eq:assembly20B}) is small 
and becomes even negligible in most kinematic regimes compared 
to the residual theoretical uncertainty from truncating the perturbative expansion.
The latter is conventionally estimated by a variation of the scale $\mu= \kappa \sqrt{Q^2 + 4m_c^2}$
in the range $1/2 \le \kappa \le 2$ around the nominal scale choice $\kappa = 1$ 
as indicated by the dashed vertical lines in Fig.~\ref{fig:F2c-mu-dep}.
Except for very small $x$ and low $\xi$ the scale variation always dominates.

Further theoretical improvements for structure functions of heavy-quark DIS 
can be expected from the complete three-loop result for the heavy-quark OME $A_{Qg}$ 
in analogy to Ref.~\cite{Ablinger:2014nga} including $a^{(3)0}_{Qg}$, which is in progress.
This would narrow down the uncertainty band spanned by the approximations A
and B in Eqs.~(\ref{eq:assembly20A}) and (\ref{eq:assembly20B}) in the small-$x$ region.
In addition, a dedicated computation of the next-to-leading terms 
in the high-energy regime $s \gg 4\:\! m^2$ along the lines of Ref.~\cite{Catani:1990eg} 
would eliminate the remaining uncertainties from the low-scale extrapolations of 
Eqs.~(\ref{eq:c2g20-nllxA})--(\ref{eq:c2q20-nllxB}).

The fit of the heavy-quark measurements from HERA 
with the new NNLO approximation for the heavy-quark structure functions 
presented here leads to significantly reduced theoretical uncertainties, in
particular for the charm-quark mass $m_c(m_c)$ in the \msbar scheme.
Results will be discussed below in Sec.~\ref{sec:results}.

\bigskip

For heavy-quark production in CC DIS one considers at parton level in Born approximation the process
\begin{equation}
\label{eq:ccborn}
s(p) + W^*(q) \to  c\, ,
\end{equation}
where the initial $s$-quark is taken massless and the final state charm-quark is heavy.
The coupling to the $W$-boson involves the usual parameters of the Cabibbo-Kobayashi-Maskawa (CKM) matrix.
The cross section for this process including higher order QCD 
corrections is expressed in terms 
of the corresponding heavy-quark CC DIS structure functions $F_k$, $k=1,2,3$, as
\begin{equation}
  \label{eq:totalF2c-CC}
  F_k(x,Q^2,m^2) =
  \sum\limits_{i = q,{\bar{q}},g} \,\,
  \int\limits_{\chi}^1\,
  {dz \over z} \,\, f_{i}\left({x \over z}, \mu_f^2 \right)\,\,
  {\cal C}_{k, i}\left(z,\xi,\mu_r^2,\mu_f^2 \right)
  \, ,
\end{equation}
where ${\cal C}_{k, i}$ denote the heavy-quark coefficient functions of CC DIS 
with kinematical variables as defined in Eq.~(\ref{eq:eta-xi-def}).
The integration over the parton momentum fraction $z$ 
in Eq.~(\ref{eq:totalF2c-CC}) is limited by $\chi = x/\lambda$ and 
$\lambda$ is given by $\lambda = 1/(1 +   m^2/Q^2) =  \xi/(1 + \xi)$.

For the QCD description of the structure functions $F_k$ in Eq.~(\ref{eq:totalF2c-CC}) 
we aim at NNLO accuracy, which implies to keep for the coefficient functions all terms 
${\cal C}_{k, i}^{(l)}$ with $l \le 2$ 
in the perturbative expansion defined in Eq.~(\ref{eq:coeff-exp}).
At NLO exact expressions for ${\cal C}_{k, i}^{(1)}$ are available~\cite{Gottschalk:1980rv,Gluck:1996ve,Blumlein:2011zu}. 
At NNLO, following the approach already used in Ref.~\cite{Alekhin:2014sya}, 
we approximate ${\cal C}_{k, i}^{(2)}$ with the respective results 
in the asymptotic limit $Q^2 \gg m^2$ derived in Refs.~\cite{Buza:1997mg,moch:2013cc,Blumlein:2014fqa}.
Given the relevant kinematic range of the data, the use of these approximate NNLO predictions 
for ${\cal C}_{k, i}^{(2)}$ is well justified and sufficiently accurate
for the HERA data~\cite{Behring:2014eya,Behring:2016hpa}.
In addition, the QCD corrections for CC DIS heavy-quark production are
generally small for scale choices $\mu= \sqrt{Q^2 + m^2}$ and the main effect
of the NNLO correction is a reduction of the theoretical uncertainties due to 
variations of $\mu_r$ and $\mu_f$.
This has been confirmed in a recent computation~\cite{Berger:2016inr} 
of the exact NNLO contributions to Eq.~(\ref{eq:totalF2c-CC}), 
i.e., including the complete result for ${\cal C}_{k, i}^{(2)}$ 
with all terms beyond the asymptotic limit $Q^2 \gg m^2$.

\subsubsection{Drell-Yan process}
\label{sec:dy-theory}

The QCD predictions for the Drell-Yan process are known to NNLO for fully exclusive kinematics.
This is essential, since the differential distributions in the lepton rapidity
from the $W^\pm$- and $Z$-boson decay provide important constraints for the flavor separation 
of the light-quark PDFs.
In addition, due to the detector acceptance being limited at collider
experiments the data are obtained in a restricted phase space and 
the $W^\pm/Z$-boson event selection criteria typically impose a cut on the lepton's transverse momentum $p_T^l$. 

We have used the publicly available code {\tt FEWZ} (version 3.1)~\cite{Li:2012wna,Gavin:2012sy}  
for the computation of the fully differential QCD predictions to NNLO for the lepton rapidity distributions 
used in the present analysis.
{\tt FEWZ} (version 3.1) can estimate PDF uncertainties in the cross sections 
by sampling over all members of a given PDF set simultaneously. 
This allows also for a fast and efficient algorithm to compute the NNLO QCD 
predictions for the current parameters of a new fit 
using the $1\sigma$ variations in the fitted parameters provided by the PDF set members.
This approach has been used in the ABM12~\cite{Alekhin:2013nda} and ABMP15~\cite{Alekhin:2015cza} fits previously.

A new feature in the current analysis already discussed in ABMP15~\cite{Alekhin:2015cza} 
is the change in the parameterization of the light-quark PDFs so that 
the shape of the iso-spin asymmetry $I(x)=x[\bar d(x) - \bar u(x)]$ is now model-independent.
Previously in the ABM12 fit, a constraint $I(x) \sim x^{0.7}$ has been imposed, 
which was motivated by Regge-phenomenology arguments valid in the asymptotic limit for $x\rightarrow 0$. 
Details for the explicit onset of such an asymptotic behavior have thus far not been
specified in the literature, though, and the fit results of ABMP15~\cite{Alekhin:2015cza} 
have returned a non-zero iso-spin asymmetry of the light-quark sea $I(x)$ at small values of Bjorken $x\sim 10^{-4}$. 
A turnover of this trend at even smaller $x$ still allows for a Regge-like shape at $x\sim 10^{-6}$. 
These findings are corroborated in the present analysis and will be discussed below in Sec.~\ref{sec:results}.

\subsubsection{Hadro-production of top-quarks}
\label{sec:top-theory}

Theory predictions including QCD corrections to NNLO are known exactly for the hadro-production of top-quark pairs 
and for single-top production in the $t$- and $s$-channel to good approximation.
The various top-quark production processes determine the top-quark mass and 
help to constrain the gluon PDF ($t{\bar t}$ data) as well as the 
light-quark PDFs in the ratio $d/u$ at large $x$ (single-top $t$-channel data).

In the current analysis we apply the NNLO QCD predictions for inclusive $t{\bar t}$ cross
section~\cite{Baernreuther:2012ws,Czakon:2012zr,Czakon:2012pz,Czakon:2013goa} 
together with the conversion for the top-quark mass $m_t(\mu_r)$ in the \msbar scheme  
as discussed in Refs.~\cite{Langenfeld:2009wd,Aliev:2010zk,Dowling:2013baa}.
The NNLO QCD corrections for single-top production in the $t$-channel have
been obtained in the structure function approximation~\cite{Brucherseifer:2014ama} 
(see also Ref.~\cite{Berger:2016oht}), which neglects color suppressed contributions 
and is sufficient in view of the current experimental precision.
The higher order QCD corrections are small so that we can use the factor $k=0.984$ calculated in ~\cite{Brucherseifer:2014ama} 
for the inclusive cross section for $t$-channel single-top production to rescale the NLO QCD corrections to NNLO accuracy, 
see also \cite{Alekhin:2016jjz} for further discussions.
For the $s$-channel single-top production the QCD corrections are known to NLO~\cite{Smith:1996ij,Harris:2002md}.
In addition, approximations for the NNLO corrections to the inclusive cross section
have been provided in Refs.~\cite{Kidonakis:2006bu,Kidonakis:2007ej,Kidonakis:2010tc} 
based on soft-gluon resummation and have been applied in Ref.~\cite{Alekhin:2016jjz}. 

All necessary theory predictions are computed with the {\tt Hathor} package~\cite{Aliev:2010zk,Kant:2014oha} 
and we always use the top-quark mass $m_t(\mu_r)$ in the \msbar scheme.
In this renormalization scheme for the mass, the cross sections typically 
exhibit very good perturbative convergence and scale stability with respect to
variation of the renormalization and factorization scales $\mu_r$ and $\mu_f$.
The fit results for the top-quark mass $m_t(m_t)$ are reported below in Sec.~\ref{sec:results}.

\section{Results}
\label{sec:results}

\subsection{Quality of data description}
\label{sec:dataq}

The total number of data points ($NDP$) used in the fit is 2860 and the 
value of $\chi^2$ per number of data points $\chi^2/NDP=1.18$ obtained
is comparable to the one of the ABM12 fit. As in our previous studies 
the data do not demonstrate any statistically significant trend with respect to the fit 
and no additional improvements can be achieved by further releasing the 
PDF shape, cf. Sec.~\ref{sec:pdfs}. A detailed breakdown 
of the values of $\chi^2$ for the separate processes and data sets is given in 
Tabs.~\ref{tab:datahq}, \ref{tab:dydata} and discussed in the following.

\input{table-heavyquark1.tex}

\input{table-WandZ-data.tex}

\subsubsection{DIS data}

\begin{figure}[t!]
\centerline{
  \includegraphics[width=16.0cm]{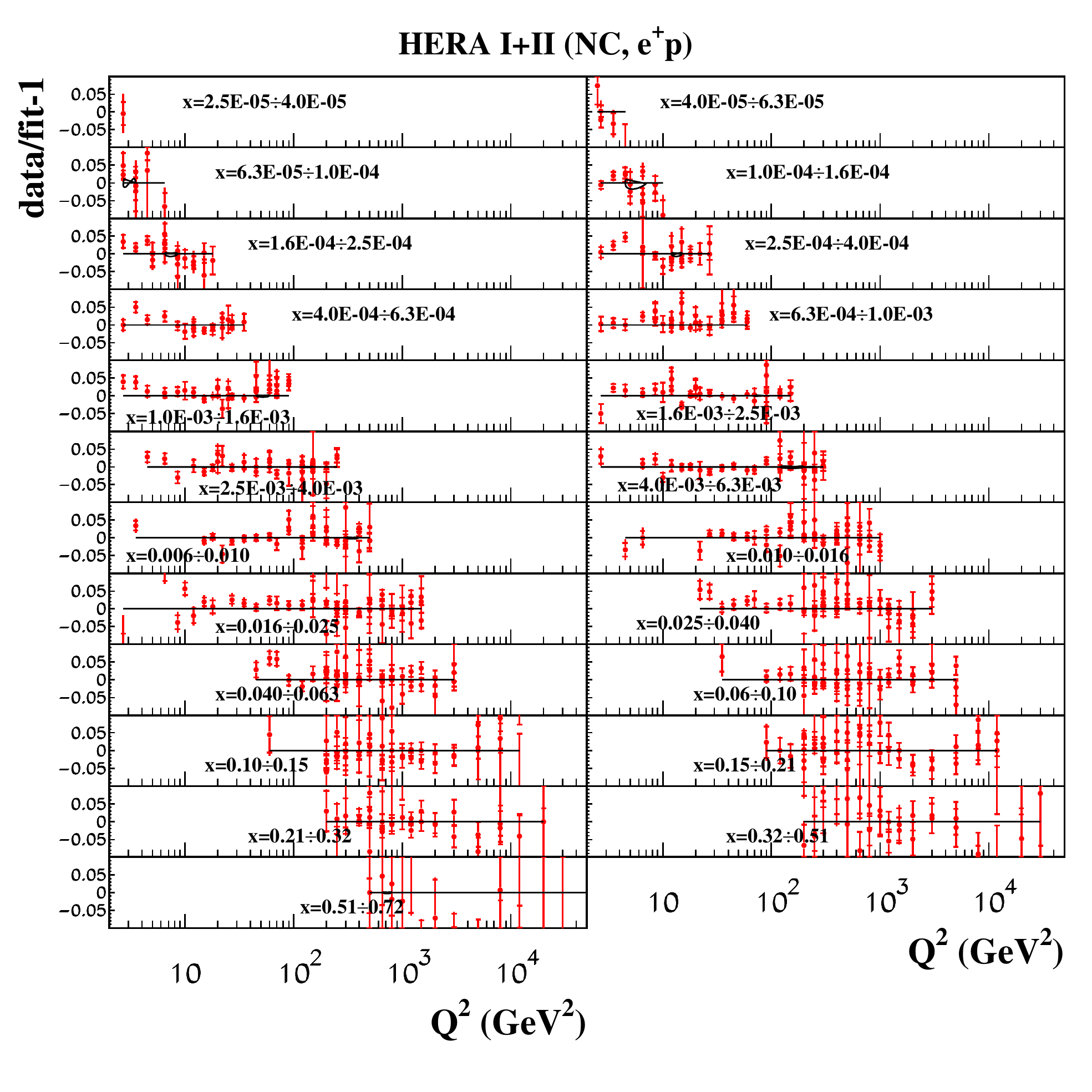}}
  \caption{\small
    \label{fig:hera-ncep}
    The pulls versus the momentum transfer $Q^2$ 
    for the final HERA NC
    $e^+p$ inclusive DIS data~\cite{Abramowicz:2015mha}
    in bins of Bjorken $x$ with respect to the present NNLO fit.
}
\end{figure}

\begin{figure}[t!]
\centerline{
  \includegraphics[width=16.0cm]{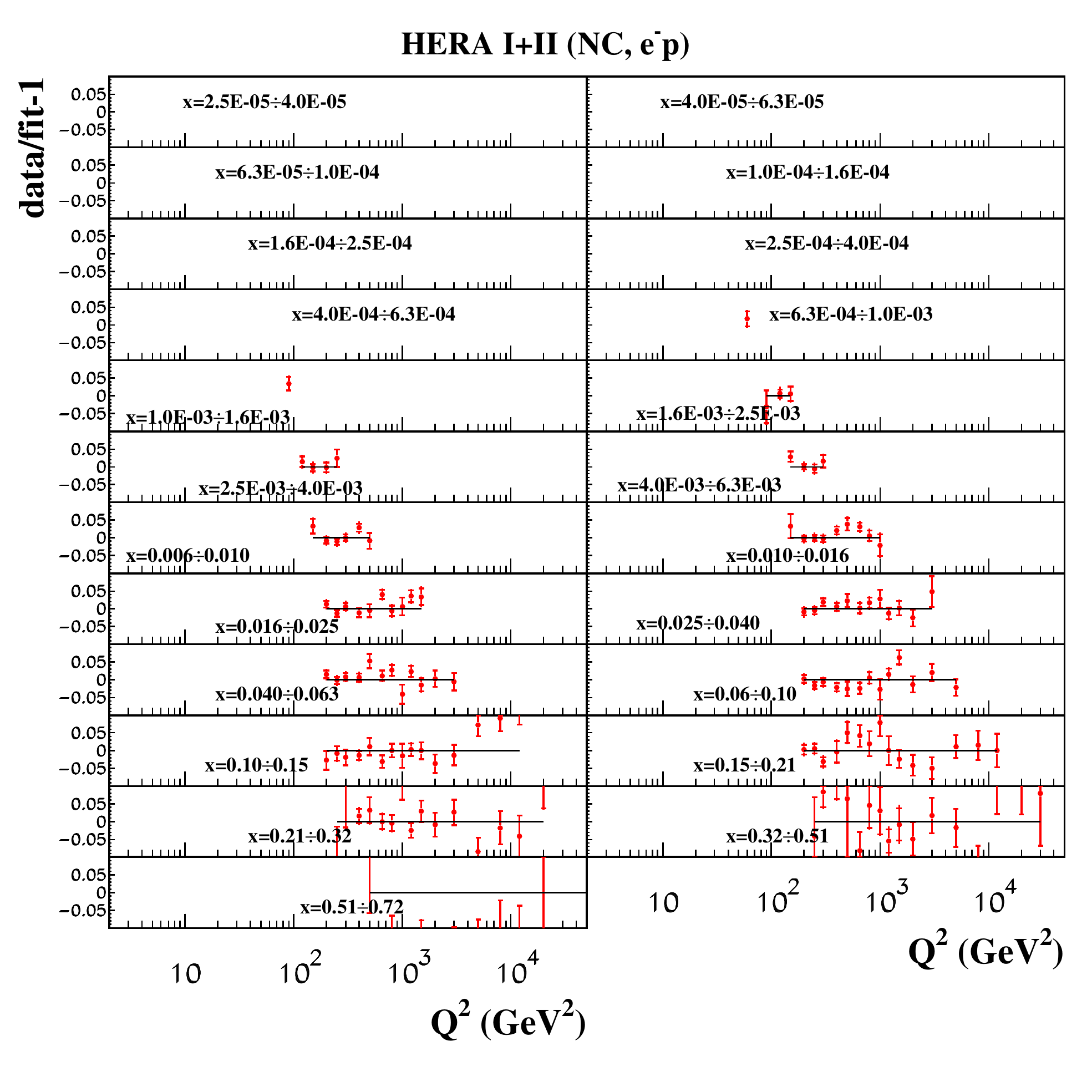}}
  \caption{\small
    \label{fig:hera-ncem}
    The same as in Fig.~\ref{fig:hera-ncep} for the NC 
    $e^-p$ inclusive DIS data~\cite{Abramowicz:2015mha}.
}
\end{figure}

\begin{figure}[t!]
\centerline{
  \includegraphics[width=16.0cm]{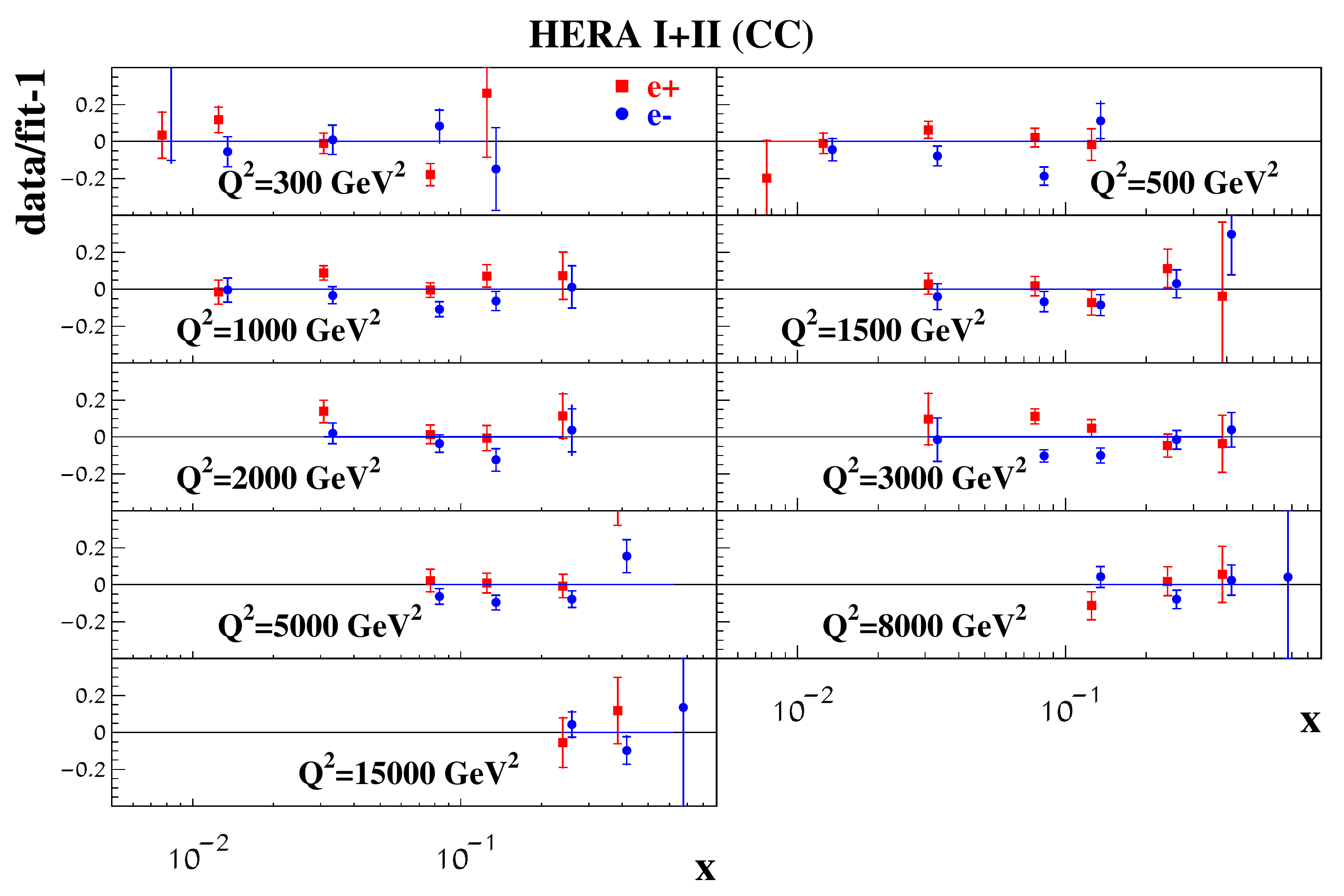}}
  \caption{\small
    \label{fig:hera-cc}
    The same as in Fig.~\ref{fig:hera-ncep} for the CC 
    $e^+p$ (squares) and $e^-p$ (circles) inclusive DIS 
    data~\cite{Abramowicz:2015mha} versus Bjorken $x$ 
    in bins of the momentum transfer $Q^2$. 
}
\end{figure}

The data sets newly included into the present analysis are smoothly accommodated in general, 
while keeping the quality of the fixed-target BCDMS, NMC and SLAC data included in the 
earlier ABM12 fit.
In particular, this applies to the HERA inclusive DIS data obtained 
from the combination of the statistics of run I and II~\cite{Abramowicz:2015mha}.
No trend can be observed in the pulls of this sample plotted in Figs.~\ref{fig:hera-ncep}--\ref{fig:hera-cc}, 
although the fluctuations 
in the central values of the data extend somewhat beyond the published uncertainties. 
As a result, these fluctuations prevent a statistically ideal description of the inclusive HERA data 
yielding values of $\chi^2/NDP$ slightly bigger than one. However, 
the fit cannot be improved in any essential way by further relaxing the fitted PDF shape. 

We have also checked the combined HERA inclusive data with varying cuts on $Q^2$.
Due to bigger errors in the data at large $Q^2$ the value of $\chi^2/NDP$ is smaller 
for the variants of the fit with more stringent cuts on $Q^2$.  
We find $1350/1092 = 1.24$ and $1225/1007 = 1.22$ 
for the cuts of $Q^2 > 5$~GeV$^2$ and $Q^2 > 10$~GeV$^2$, respectively.  
The same conclusion was drawn in a previous QCD analysis~\cite{Abramowicz:2015mha}, 
however, the values of $\chi^2$ reported in Ref.~\cite{Abramowicz:2015mha} 
are somewhat smaller than ours due to the limited number of data sets employed in that 
analysis.

\subsubsection{Drell-Yan data}

Due to the large amount of the DY data from Tevatron and the LHC 
a precision determination of the light-quark PDFs in a wider kinematic range in $x$ 
than ever before becomes possible, see also \cite{Alekhin:2015cza}.
The quality of the ABMP16 fit for the Drell-Yan data description is summarized in Tab.~\ref{tab:dydata}.
Data sets of lower accuracy, which have become obsolete and data sets
superseded are not listed there. 
Instead, we refer to the review~\cite{Accardi:2016ndt} for further comparisons 
concerning the status of Drell-Yan data in PDF fits. 

\begin{figure}[b!]
\centerline{
  \includegraphics[width=8.75cm]{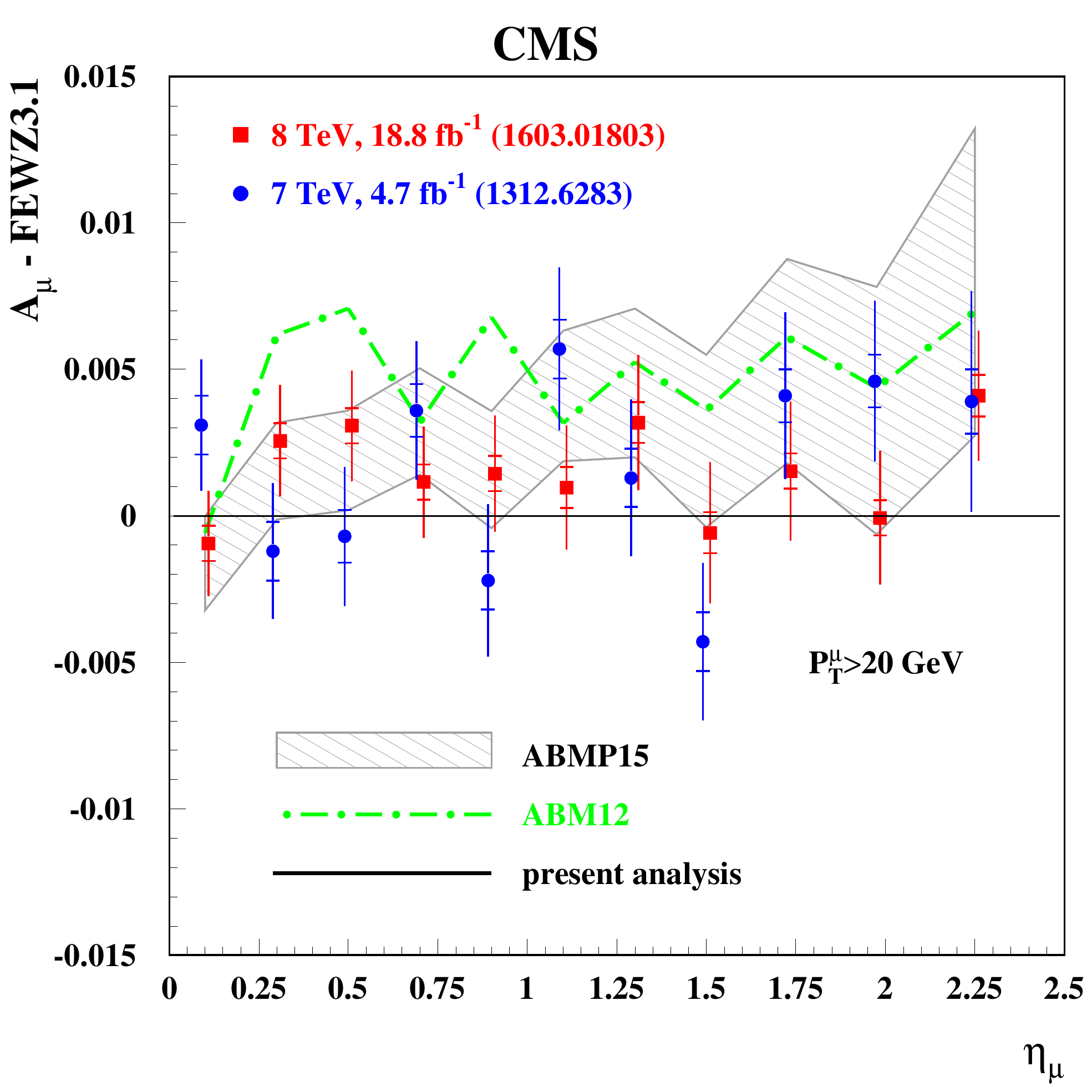}}
  \caption{\small
    \label{fig:cms-asym}
    The pulls for the CMS data on the muon charge asymmetry $A_{\mu}$ in inclusive
    $pp \to W^\pm+X \to \mu^\pm \nu + X$ production at 
    $\sqrt s = 7$~TeV~\cite{Chatrchyan:2013mza} (circles) and
    $8$~TeV~\cite{Khachatryan:2016pev} (squares) 
    with the muon transverse momentum $P_T^\mu>20$~GeV 
    and as a function of the muon pseudo-rapidity $\eta_\mu$ with respect to  our fit.
    The  ABM12~\cite{Alekhin:2013nda} central predictions for the CMS data at 
    $\sqrt s = 8$~TeV~\cite{Khachatryan:2016pev} obtained with {\tt FEWZ} 
    (version 3.1)~\cite{Li:2012wna,Gavin:2012sy} (dotted dashes) and the 
    uncertainty band for the ABMP15 ones~\cite{Alekhin:2015cza} (hatch) with respect to  our 
    fit are given for comparison. 
}
\end{figure}

\begin{figure}[th!]
\centerline{
  \includegraphics[width=16.0cm]{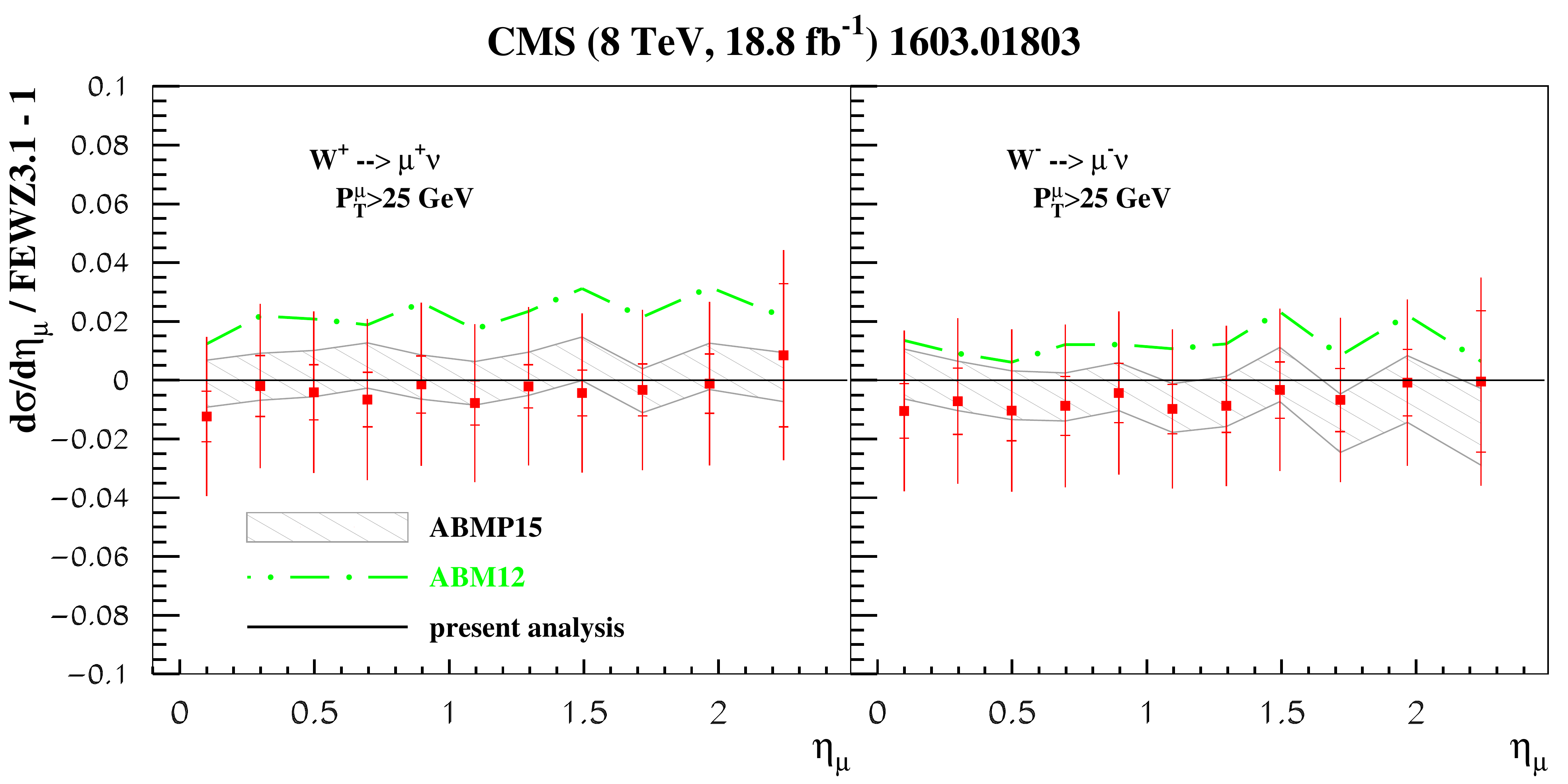}}
  \caption{\small
    \label{fig:cms-w}
    The same as Fig.~\ref{fig:cms-asym} for the
    CMS data on the cross section of inclusive $W$-boson production 
    at $8$~TeV~\cite{Khachatryan:2016pev}. 
}
\end{figure}

In general the data sets in Tab.~\ref{tab:dydata} with a total $NDP = 172$ 
can be smoothly accommodated, although the values of $\chi^2/NDP$ obtained 
for individual data sets are bigger than one in some places. 
This is the case, for instance, for the  
CMS data on the muon charge asymmetry collected at the collision energy of $\sqrt{s}=7$~TeV  
shown in Fig.~\ref{fig:cms-asym}, 
which yields a value of $\chi^2/NDP\sim 2$.   
Similar CMS data for $\sqrt{s}=8$~TeV, however, are much smoother, cf. the
pulls in Figs.~\ref{fig:cms-asym} and~\ref{fig:cms-w}, and a good value of $\chi^2/NDP$ is achieved in this case. 
Therefore, the observed fluctuations in the CMS data for $\sqrt{s}=7$~TeV 
should rather be attributed to experimental systematic effects than to any shortcomings in the fitted PDFs.

\begin{figure}[th!]
\centerline{
  \includegraphics[width=8.75cm]{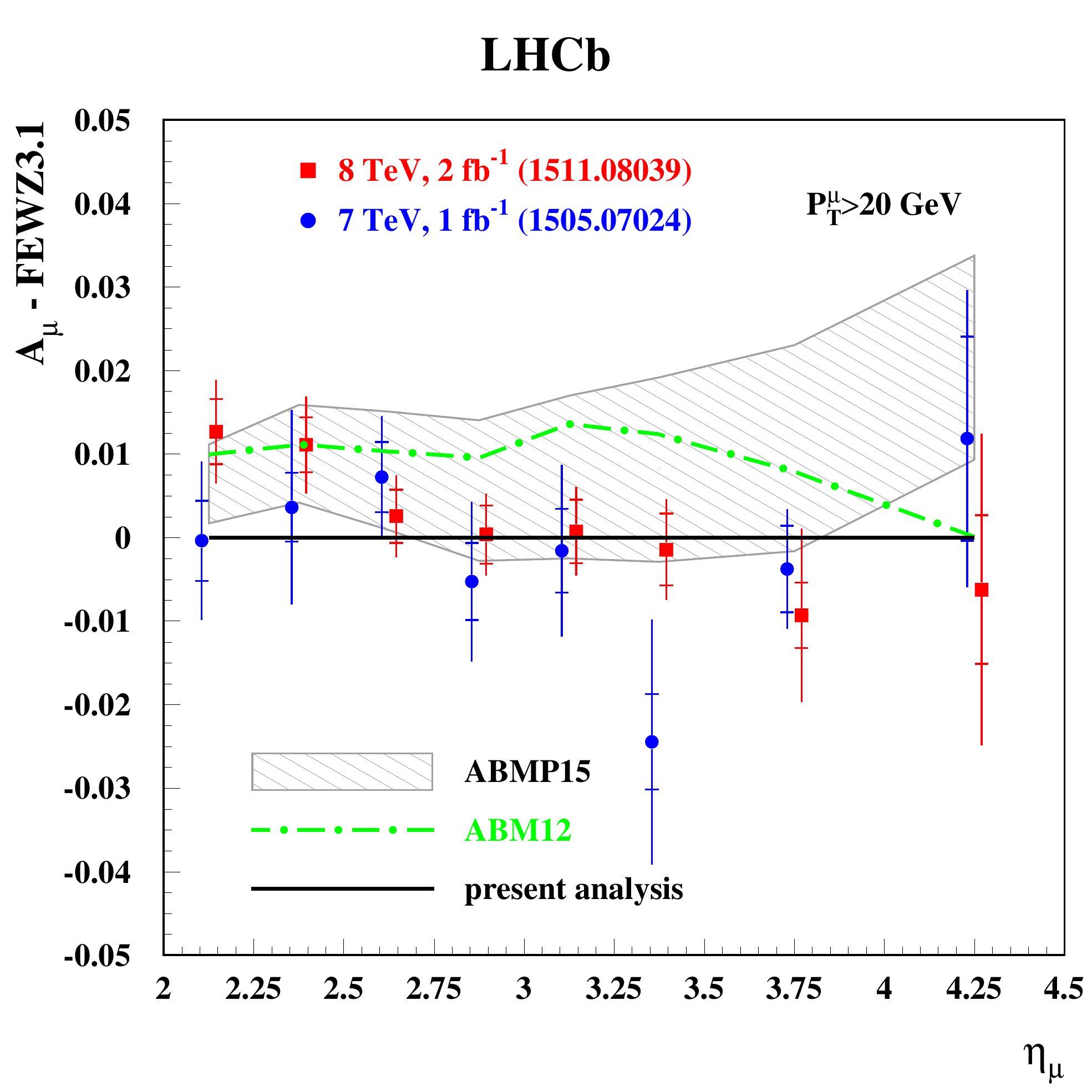}}
  \caption{\small
    \label{fig:lhcb-asy}
    The same as in Fig.~\ref{fig:cms-asym} for 
    the LHCb data on the muon charge asymmetry $A_\mu$ in inclusive
    $pp \to W^\pm+X \to \mu^\pm \nu + X$ production at 
    $\sqrt s = 7$~TeV~\cite{Aaij:2015gna} (circles) and
    $8$~TeV~\cite{Aaij:2015zlq} (squares).
    The data at
    $\eta_\mu = 3.275$ and $\sqrt s = 7$~TeV are not used in the fit.
}
\end{figure}

The pulls for the LHCb data on the muon charge asymmetry collected at $\sqrt{s}=7$~TeV~\cite{Aaij:2015gna}
are shown in Fig.~\ref{fig:lhcb-asy}.
They display an irregularity at pseudo-rapidity $\eta_\mu = 3.275$, 
which is not confirmed by the LHCb data at $\sqrt{s}=8$~TeV~\cite{Aaij:2015zlq}. 
Moreover, this spike at $\eta_\mu = 3.275$ coincides with fluctuations in the correction for final-state radiation 
which has been applied to the LHCb data, cf. Fig.~5 in Ref.~\cite{Alekhin:2015cza}.
The two data points for $W^+$- and $W^-$-boson production corresponding 
to this spike contribute about 13 units to the value of $\chi^2$ and, 
in line with our earlier analysis~\cite{Alekhin:2015cza}, we discard these two data points from the fitted set. 
This has only marginal impact on the fit results.

\begin{figure}[th!]
\centerline{
  \includegraphics[width=16.0cm]{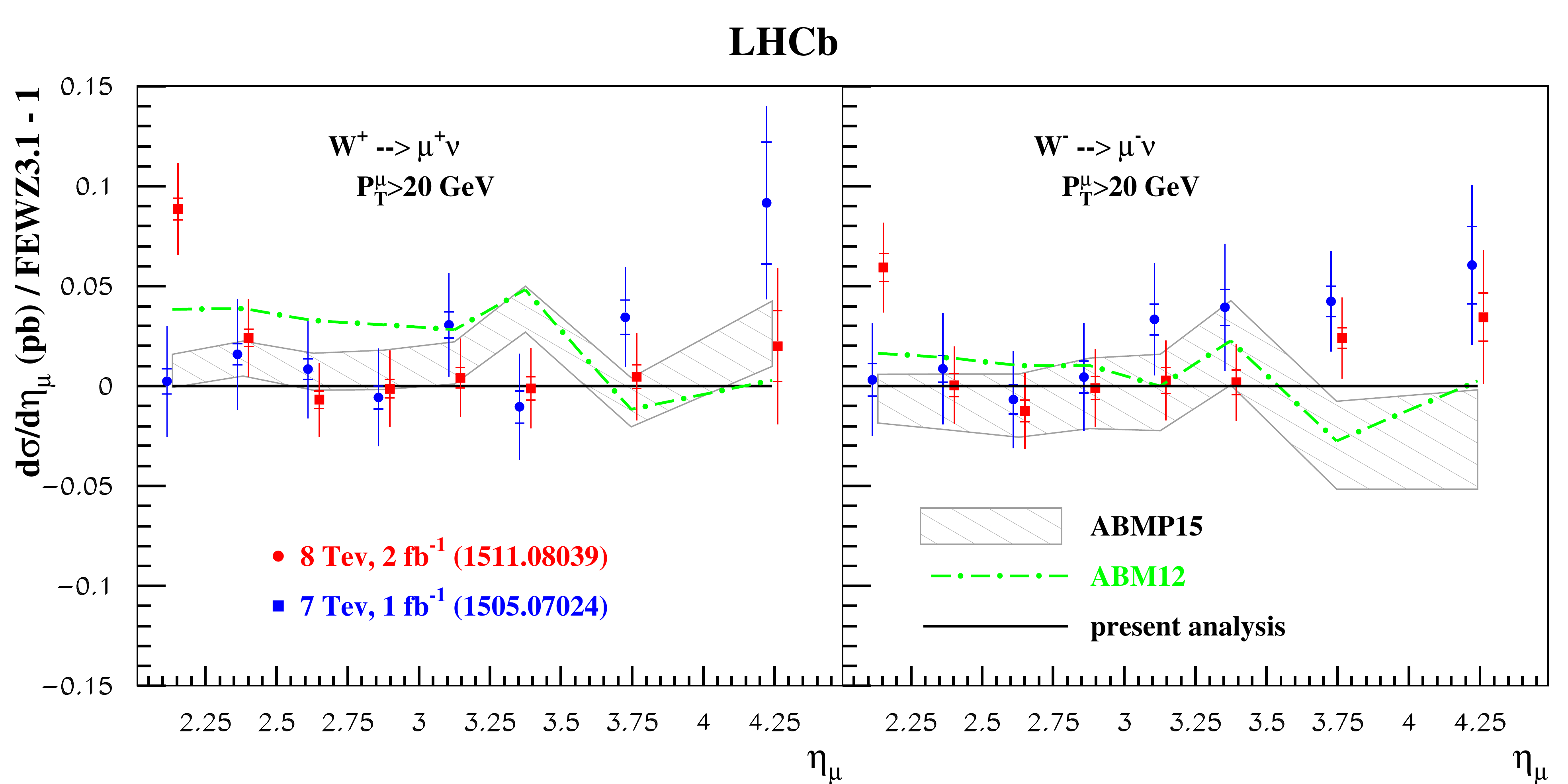}}
  \caption{\small
    \label{fig:lhcb-pulls}
    The same as in Fig.~\ref{fig:cms-asym} for 
    LHCb data on the cross section of inclusive $W$-boson production 
    in the $pp$ collision (left: $pp \to W^++X \to \mu^+ \nu + X$, 
    right: $pp \to W^-+X \to \mu^- \nu + X$)
    at $\sqrt s = 7$~TeV~\cite{Aaij:2015gna} (circles) and $\sqrt s = 8$~TeV~\cite{Aaij:2015zlq} (squares) with the muon transverse momentum 
    $P_T^\mu>20$~GeV 
    and as a function of the muon pseudo-rapidity $\eta_\mu$ with respect to  our fit.
    The points with the lowest $\eta_\mu = 2.125$ at $\sqrt s = 8$~TeV
    are not used in the fit.
}
\end{figure}

The pulls for the LHCb data on the $W$-production at $\sqrt{s}=8$~TeV~\cite{Aaij:2015zlq} 
are displayed in Fig.~\ref{fig:lhcb-pulls}.
They exhibit an excess at $\eta_\mu = 2.125$ both in $\mu^+$ and $\mu^-$ channels, 
while the muon charge asymmetry remains smooth. 
Since these two points also give a quite sizeable contribution to the value of $\chi^2$, 
about 14 units, we discard them from the fitted set. 
Similar to the spike in LHCb data at $\sqrt{s}=7$~TeV, 
we find again only a marginal impact of this filtering procedure on the results.


\begin{figure}[th!]
\centerline{
  \includegraphics[width=16.0cm]{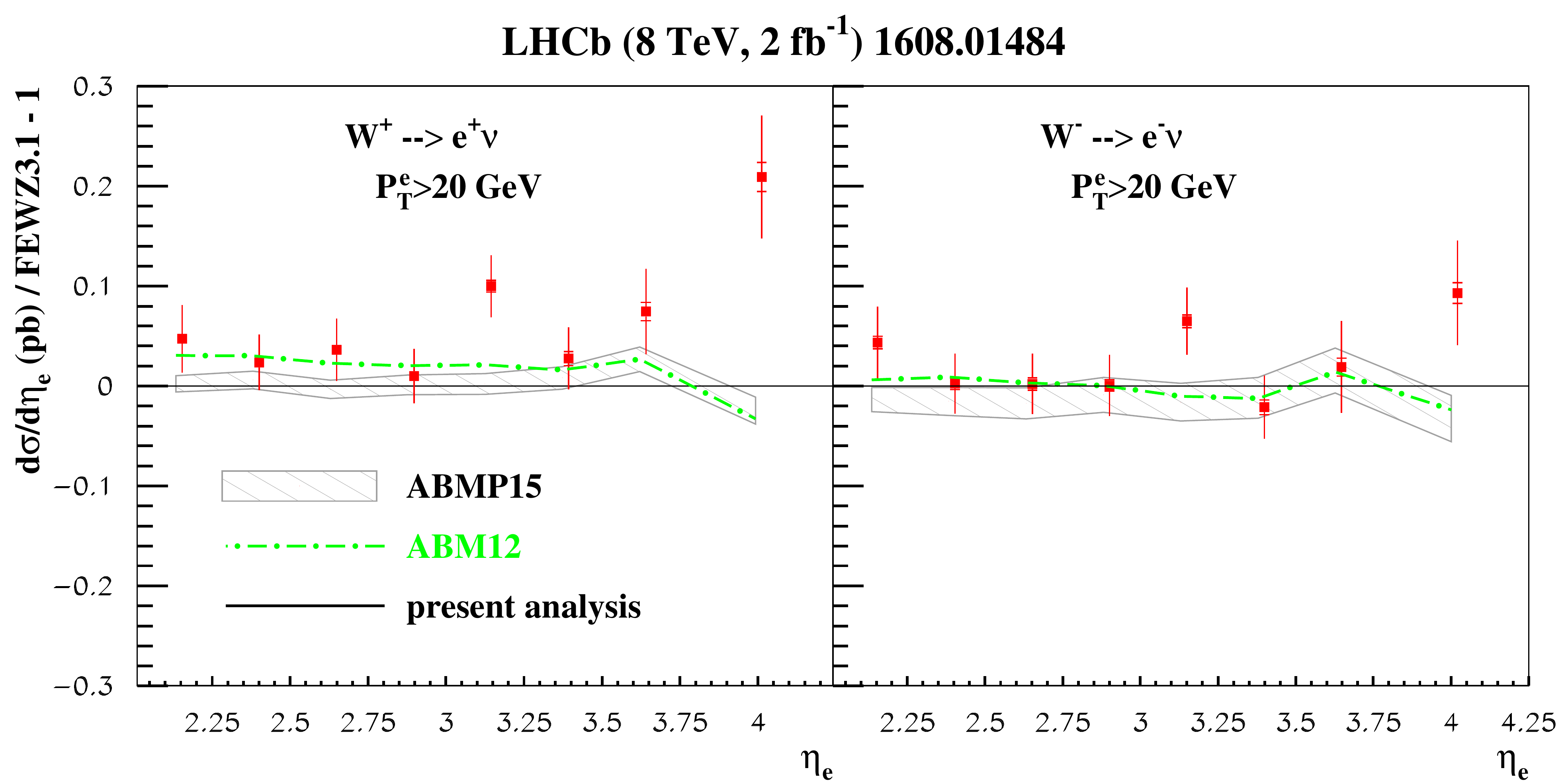}}
  \caption{\small
    \label{fig:lhcbwe}
    The same as Fig.~\ref{fig:lhcb-pulls} for the
LHCb data on the cross section of inclusive 
$pp \to W^++X \to e^+ \nu + X$ (left) and 
$pp \to W^- + X \to e^- \nu + X$ (right)
production at $8$~TeV~\cite{Aaij:2016qqz}. 
}
\end{figure}

The LHCb data on the electron charge asymmetry
collected at $\sqrt{s}=8$~TeV~\cite{Aaij:2016qqz}
 are in broad agreement with the NNLO QCD predictions 
based on our ABMP16 PDFs as demonstrated in Fig.~\ref{fig:lhcbwe}.
However, significant fluctuations occur in some places, in particular 
for the pseudo-rapidity bin at $\eta_e=4$, which is the biggest one available in this sample. 
The uncertainties in the data are dominated by the systematic ones, which are strongly correlated. 
This prevents us from achieving a reasonable value of $\chi^2$ in the fit. 
Moreover, the electron data also demonstrate a different trend as compared to the muon ones.
As a consequence, we do not include the LHCb electron set from $\sqrt{s}=8$~TeV
into the fit until these issues are resolved. 

\begin{figure}[th!]
\centerline{
  \includegraphics[width=8.45cm]{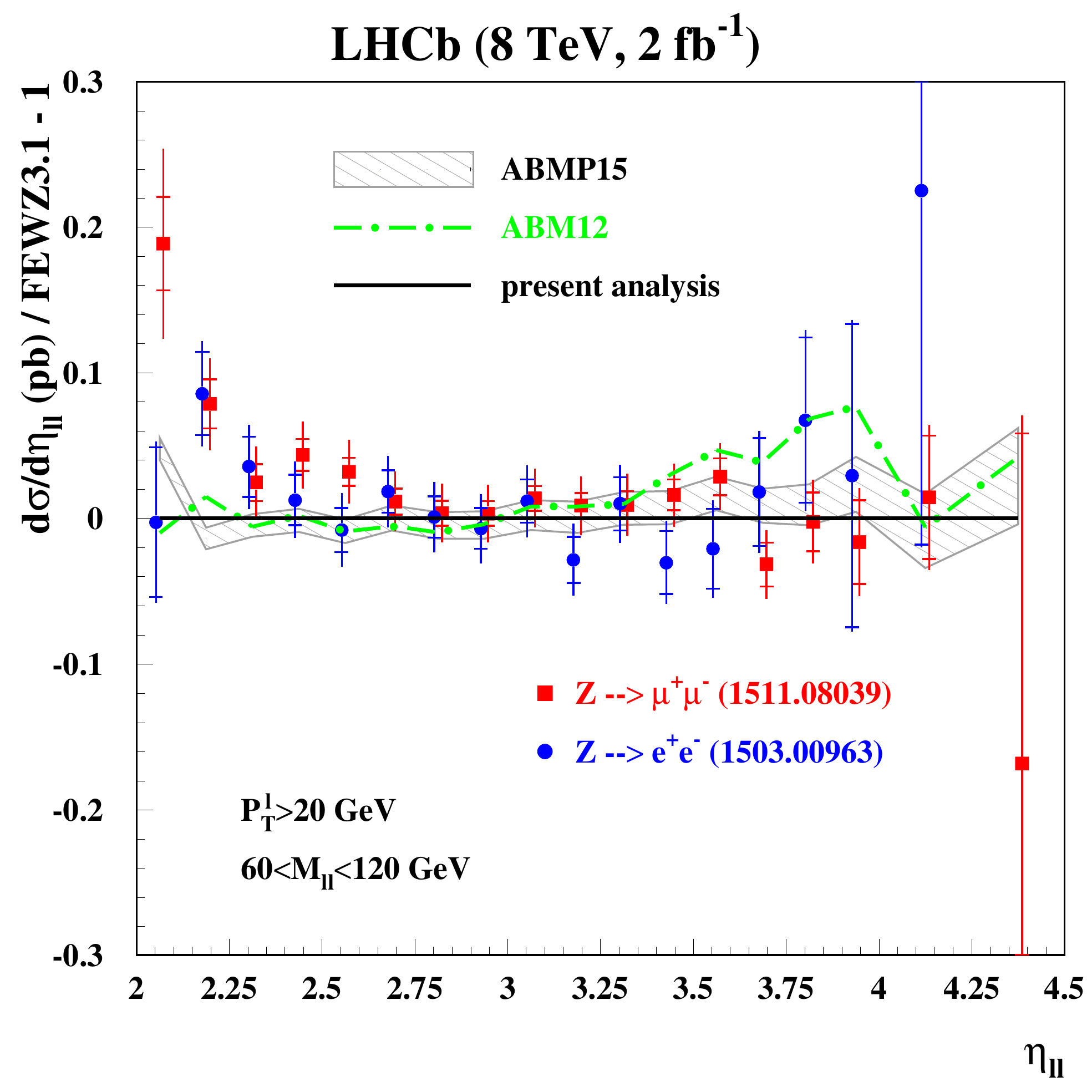}
  \includegraphics[width=8.45cm]{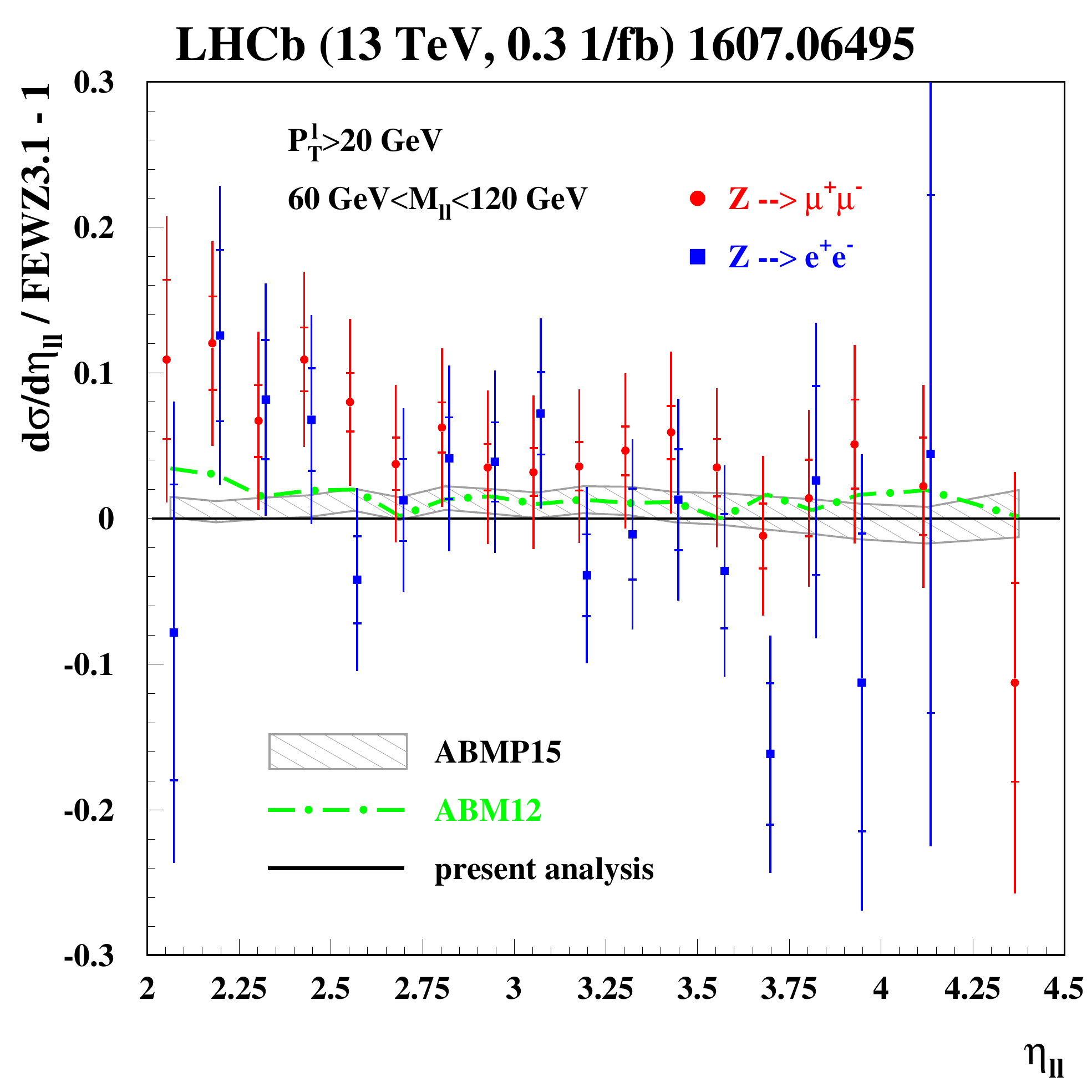}}
\vspace*{-4mm}
  \caption{\small
    \label{fig:lhcb-z}
    Left panel: 
    The same as in Fig.~\ref{fig:lhcb-pulls} for 
    the LHCb data on the inclusive 
    $pp \to Z+X \to l^+ l^- + X$ production at 
    $\sqrt s = 8$~TeV for the muon~\cite{Aaij:2015zlq} (circles) 
    and electron~\cite{Aaij:2015vua} 
    (squares) decay modes with the lepton transverse momentum 
    $P_T^l>20$~GeV, the lepton pair mass  
    60~GeV$<M_{ll}<$120~GeV and as a function of the 
    lepton pair pseudo-rapidity $\eta_{ll}$.
    Right panel: 
    The same for the LHCb data on the inclusive 
    $pp \to Z+X \to l+ l^- + X$ production at 
    $\sqrt s = 13$~TeV~\cite{Aaij:2016mgv}.
}
\vspace*{5mm}
\centerline{
  \includegraphics[width=16.0cm]{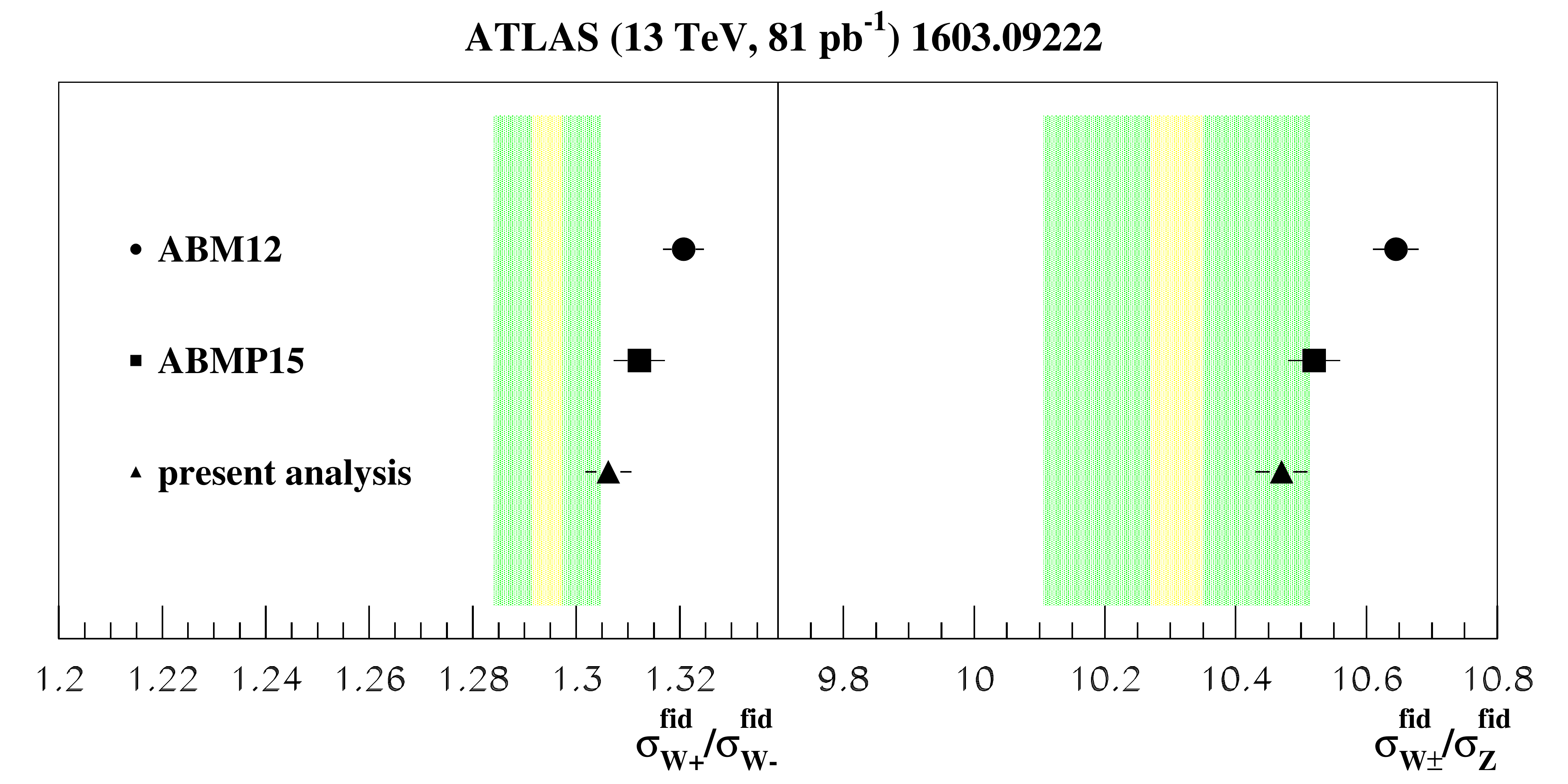}}
\vspace*{-3mm}
  \caption{\small
    \label{fig:WandZ-ratio}
    Cross section ratios for the production of $W^\pm$- and $Z$-bosons in
    the fiducial volume, 
    $\sigma_{W^+}/\sigma_{W^-}$ (left) and$\sigma_{W^\pm}/\sigma_{Z}$ (right), 
    in comparison to ATLAS data~\cite{Aad:2016naf}
    at $\sqrt s = 13$~TeV together with their $1\sigma$ PDF uncertainties 
    using the results of our NNLO fit (triangles)
    as well as ABM12~\cite{Alekhin:2013nda} (circles)
    and ABMP15~\cite{Alekhin:2015cza} (squares).
    The inner (yellow) band denotes the statistical uncertainty 
    of the ATLAS data~\cite{Aad:2016naf} and the outer (green) 
    one the combined uncertainty due to statistics and systematics.
    The ABM12 predictions are larger than the ones presented in  
    Ref.~\cite{Aad:2016naf} due to different programs used, 
    {\tt FEWZ} (version 3.1)~\cite{Li:2012wna,Gavin:2012sy} and 
    {\tt DYNNLO} (version 1.5)~\cite{Catani:2007vq,Catani:2009sm}, respectively. 
  }
\end{figure}

In contrast, the $\sqrt{s}=8$~TeV LHCb data on the $Z$-boson production 
and decay in the electron mode~\cite{Aaij:2015vua} are 
generally in a good agreement with the fit and also with the ones for the muon decay mode~\cite{Aaij:2015zlq} as shown in Fig.~\ref{fig:lhcb-z}.  
The muon channel data are somewhat enhanced at small rapidity, 
similar to the LHCb $W$-boson sample at $\sqrt{s}=8$~TeV. 
However, this enhancement is statistically not significant due to the limited 
accuracy of the $Z$-boson data~\cite{Aaij:2015zlq}. The pulls 
for the first LHCb $Z$-boson data obtained at the collision energy of $\sqrt{s}=13$~TeV
are given in Fig.~\ref{fig:lhcb-z}.
These data are also broadly in agreement with the present fit and with our earlier predictions
based on the ABM12 and ABMP15 PDFs. 
However, the $Z$-boson data collected for the electron decay mode are subject to more significant fluctuations 
and for this data set we achieve only a value of $\chi^2/NDP \simeq 2$ taking the ABMP16 PDFs. 
This confirms a tendency, that the description of the electron LHCb $W$- and $Z$-boson data 
is inferior to the muon ones.
The uncertainties in the existing LHCb data at $\sqrt{s}=13$~TeV are still 
quite big as compared to the earlier data sets at lower collision energies 
and to the PDF uncertainties in the current theoretical predictions, 
cf.~Fig.~\ref{fig:lhcb-z}. 
This puts a limitation on a potential impact of these data on the 
PDF extraction and, therefore, we do not include them into the fit.

Finally, we briefly discuss cross section ratios for the production of $W^\pm$- and $Z$-bosons 
integrated over the fiducial volume.
The recent ATLAS data on those ratios at the collision energy at $\sqrt s = 13$~TeV~\cite{Aad:2016naf} 
is shown in Fig.~\ref{fig:WandZ-ratio}. 
Our earlier ABM12 predictions somewhat overshoot the data,
whereas the agreement with the more recent ABMP15 PDFs is better. 
For the present analysis a good agreement within the uncertainties is achieved as quantified
in Fig.~\ref{fig:WandZ-ratio}.

\subsubsection{Data on heavy-quark production}

\begin{figure}[th!]
\centerline{
  \includegraphics[width=13.0cm]{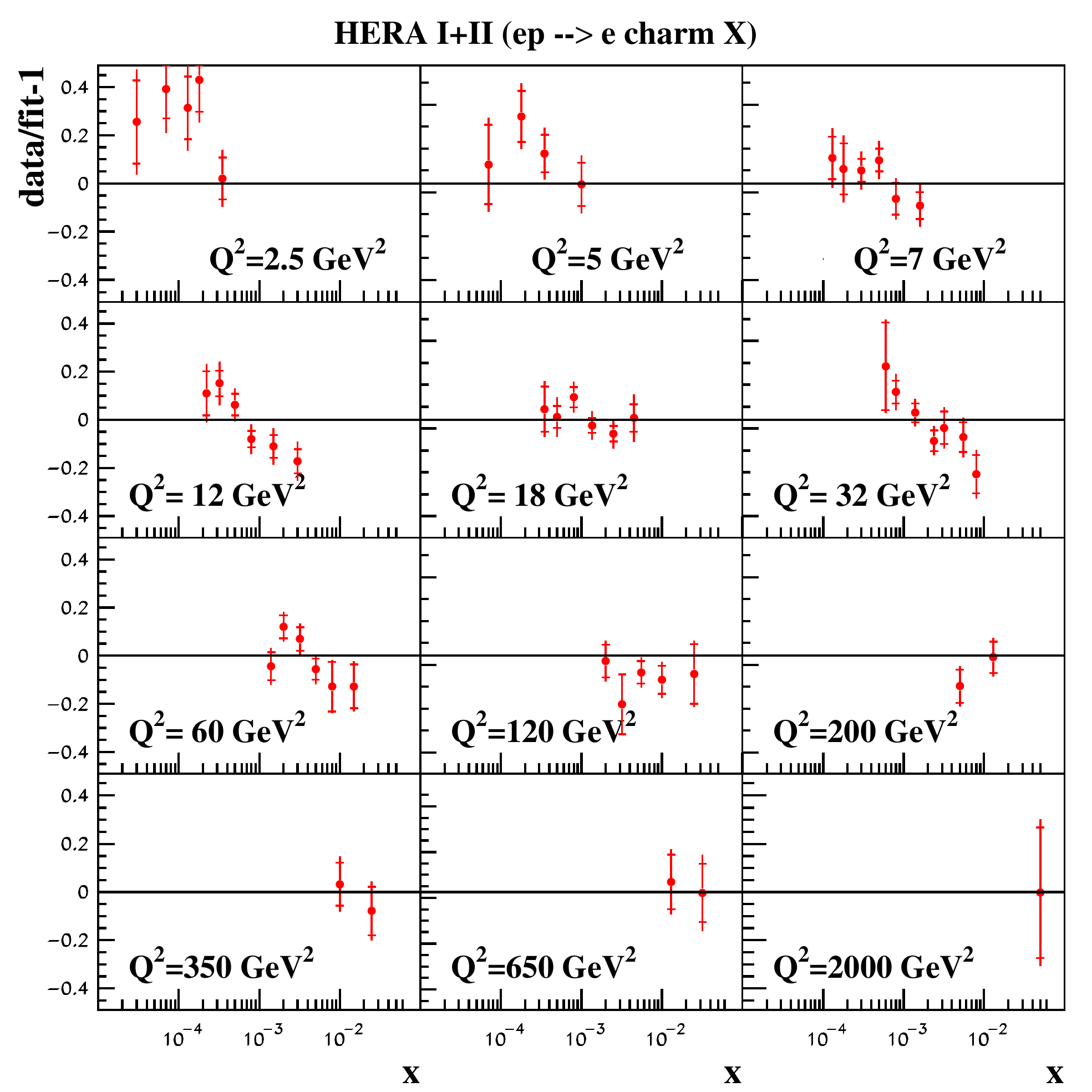}}
  \caption{\small
    \label{fig:hera-ccbar}
    The same as in Fig.~\ref{fig:hera-ncep} for the NC 
    DIS inclusive charm-quark production 
    data~\cite{Abramowicz:1900rp} versus Bjorken $x$ 
    in bins of the momentum transfer $Q^2$. 
  }
\end{figure}

\begin{figure}[th!]
\centerline{
  \includegraphics[width=13.0cm]{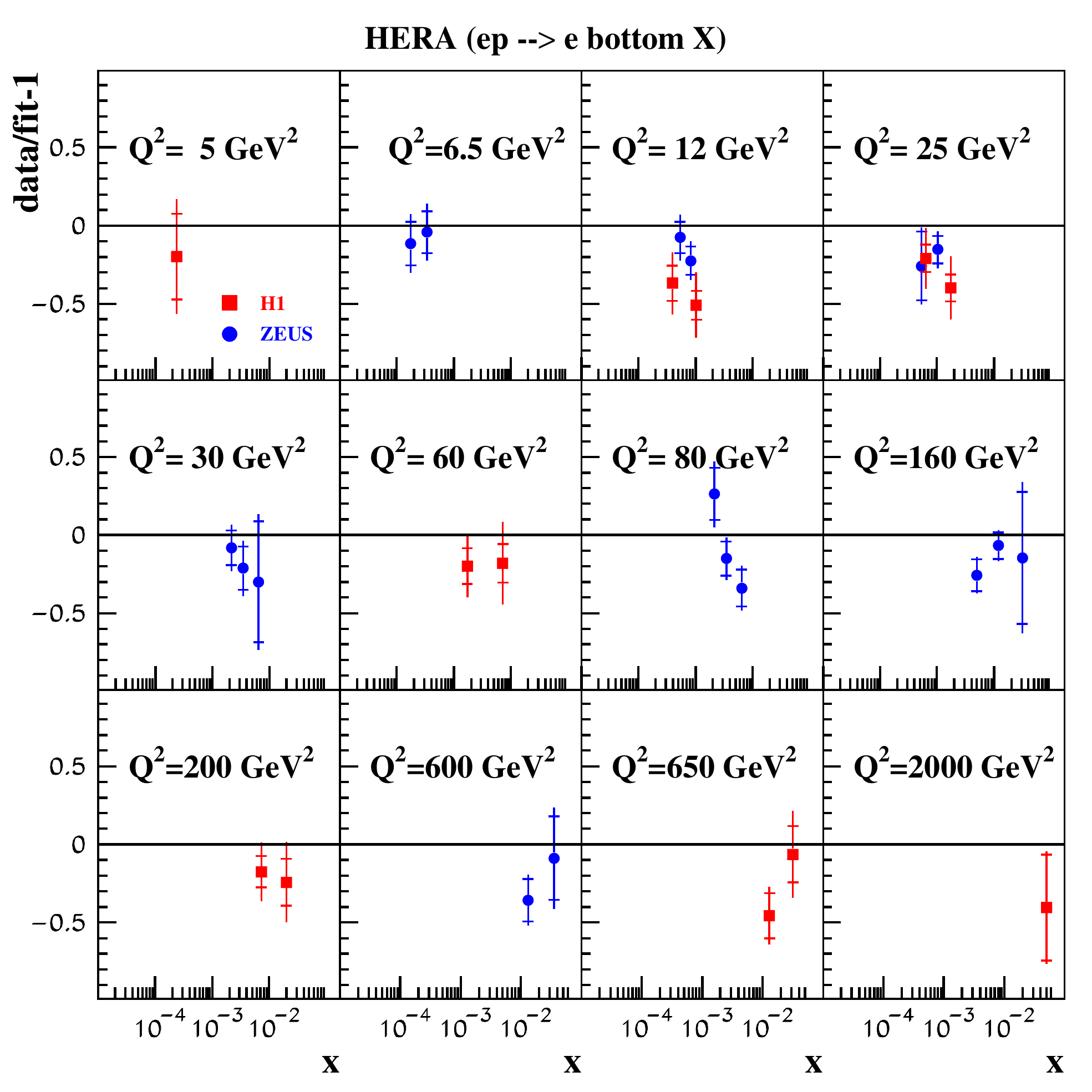}}
  \caption{\small
    \label{fig:hera-bbbar}
    The same as Fig.~\ref{fig:hera-ccbar} for the NC DIS inclusive bottom-quark
    production data from the H1~\cite{Aaron:2009af} (squares)
    and ZEUS~\cite{Abramowicz:2014zub} (circles) collaborations at HERA.
  }
\end{figure}

\begin{figure}[th!]
\centerline{
  \includegraphics[width=15.0cm]{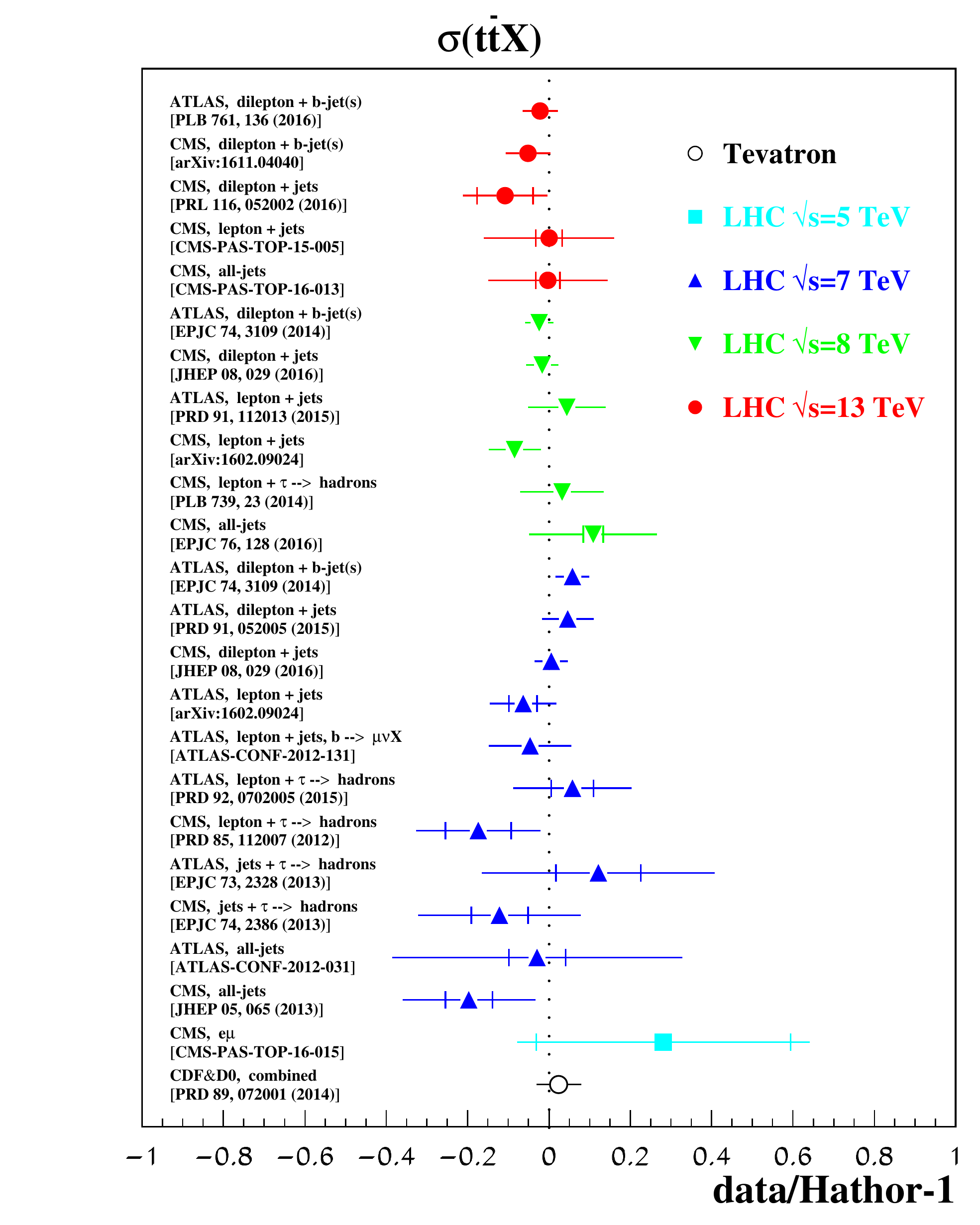}}
  \caption{\small
    \label{fig:ttbar-pulls}
    The pulls for data on top-quark pair production from the Tevatron (CDF and D{\O} 
    collaborations)
    at $\sqrt s = 1.96$~TeV 
    and the LHC (ATLAS and CMS collaboration) at $\sqrt s = 5, 7, 8$ and $13$~TeV 
    with respect to our NNLO fit.
    The NNLO QCD predictions have been obtained with 
    {\tt Hathor}~\cite{Aliev:2010zk}. 
  }
\end{figure}

\begin{figure}[th!]
\centerline{
  \includegraphics[width=13.0cm]{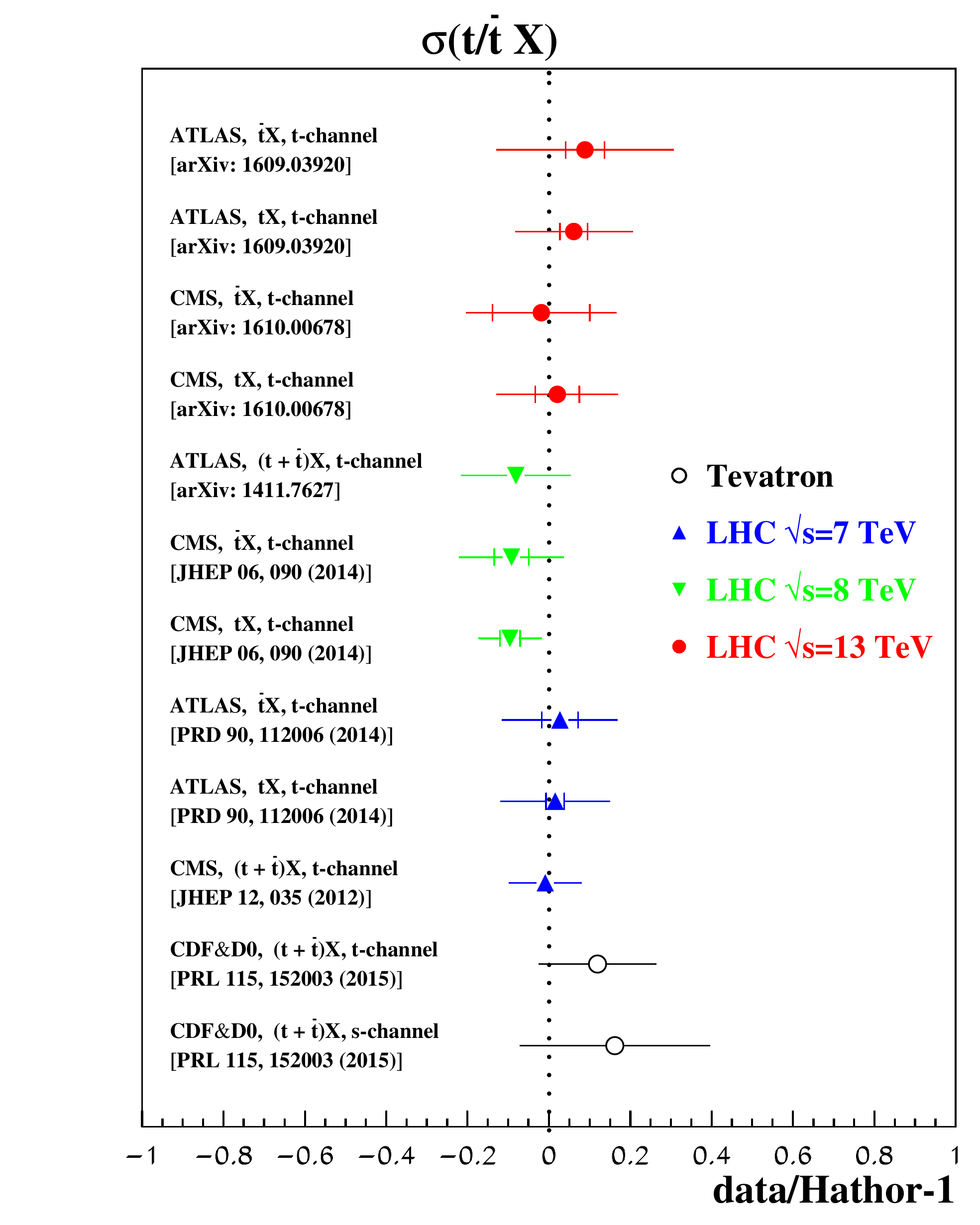}}
  \caption{\small
    \label{fig:single-top-pulls}
    The same as Fig.~\ref{fig:ttbar-pulls} for data on single-top production 
    in the $s$- and $t$-channel for final states with $t$- and ${\bar t}$-quarks.
    The approximate NNLO QCD predictions have been computed with {\tt Hathor}~\cite{Kant:2014oha} 
    as described in Sec.~\ref{sec:theory}.
  }
\end{figure}

The theoretical framework of Sec.~\ref{sec:ccbar-dis} provides 
an excellent description of the data on $c$- and bottom-quark production 
in the NC electron-proton DIS collected by the H1 and ZEUS experiments at HERA
in range of $Q^2$ available, 
cf. Figs.~\ref{fig:hera-ccbar},~\ref{fig:hera-bbbar}. 
The same applies to the charm-quark production in CC neutrino-nucleon DIS
measured in the fixed-target experiments. 
In particular, the NOMAD~\cite{Samoylov:2013xoa} and 
CHORUS~\cite{KayisTopaksu:2011mx} data, which were earlier smoothly included
into an updated version~\cite{Alekhin:2014sya} of the ABM12 fit
are also well described by the present PDFs, cf. Tab.~\ref{tab:datahq}.
Finally, the newly included Tevatron and LHC data on $t\bar t$ and single-$t$ production 
are in good agreement with the fit, cf. Figs.~\ref{fig:ttbar-pulls},~\ref{fig:single-top-pulls}.

\subsection{PDF improvement}
\label{sec:pdfs}

In the present analysis we parameterize PDFs in the scheme with 
$n_f=3$ light flavors at a starting scale $\mu_0=3~{\rm GeV}$ of the QCD
evolution by the following form  
\begin{eqnarray}
\label{eq:val}
  x q_{v}(x,\mu_0^2)
  &=&
  \displaystyle
  \frac{ 2 \delta_{qu}+\delta_{qd}} {N^{v}_q}x^{a_{q}}(1-x)^{b_q}x^{P_{qv}(x)}
  \, 
\end{eqnarray}
for the valence quark distributions $q_{v}$, 
where $q = u,d$ and $\delta_{qq^\prime}$ denotes the Kronecker symbol,
\begin{equation}
  \label{eq:sea}
  xq_s(x,\mu_0^2) \,=\, x \bar{q}_s(x,\mu_0^2) =A_{qs}(1-x)^{b_{qs}}x^{a_{qs}P_{qs}(x)}\, 
\end{equation}
for the sea quark distributions, where $q=u,d,s$, and 
\begin{eqnarray}
\label{eq:glu}
  xg(x,\mu_0^2)
  &=&
  A_{g}x^{a_{g}}(1-x)^{b_{g}} x^{a_{g}\, P_{g}(x) }
  \, 
  \qquad
\end{eqnarray}
for gluons. While the small- and large-$x$ PDF asymptotic is defined by 
the exponents $a$ and $b$, respectively, 
the functions $P_p(x)$ take a general form
\begin{equation}
\label{eq:poly}
P_{p}(x) \,=\, (1+\gamma_{-1,p}\ln x)
\left( 1 + \gamma_{1,p}x + \gamma_{2,p}x^2 + \gamma_{3,p}x^3 \right)\, ,
\end{equation}
where $p=qv,qs,g$.
This allows for flexibility of the PDFs in the entire range of $x$.  The parameters $N_q^v$ and 
$A_g$ are determined from the sum rules for fermion number and momentum conservation, 
respectively, and $A_{qs}, a_{qs}, b_{qs}, \gamma_{qs}$ are fitted to the data.
The functional form in Eqs.~(\ref{eq:val})--(\ref{eq:poly}) provides sufficient flexibility 
over the entire $x$-range with respect to the combined data within the ABMP16 analysis.
We have checked that the quality of the fit does not improve when allowing for
additional terms in Eq.~(\ref{eq:poly}). 
The parameter values obtained with their $1\sigma$ uncertainties corresponding to 
the statistical and systematic errors in the data 
are given in Tab.~\ref{tab:fitvalues}. The respective correlations are listed
in Appendix~\ref{sec:appA}.

\input{table-pdf-parameters}

\begin{figure}
\begin{center}
\vspace*{-10mm}
\includegraphics[width=0.95\textwidth,height=0.9\textheight]{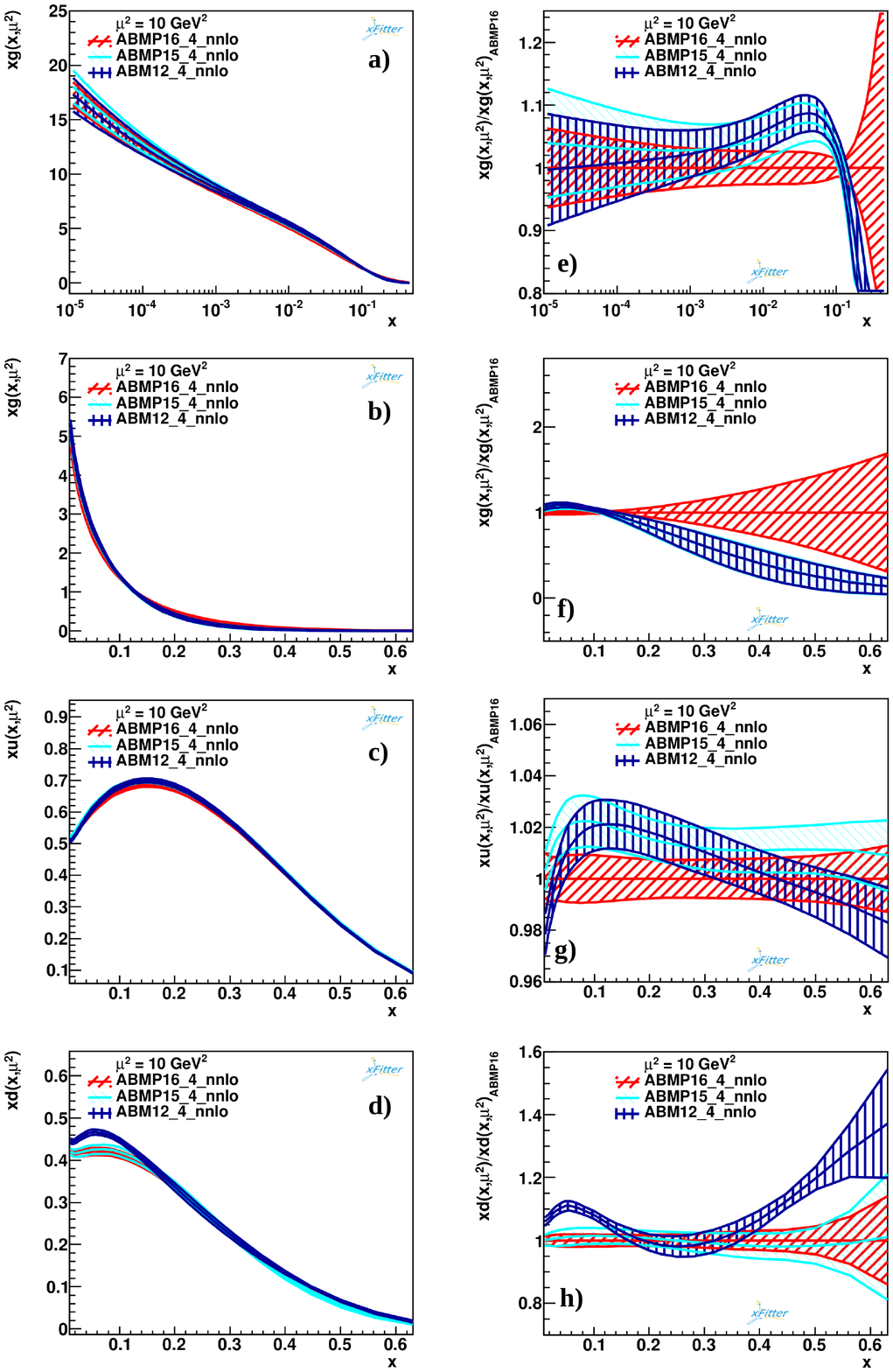}
\caption{\small 
  \label{fig:abm16_vs_abm12_01}
The distributions of $n_f=4$ flavor gluons $xg(x,\mu^2)$ 
({\bf a}: logarithmic $x$-scale, {\bf b}: linear $x$-scale), up-quarks 
$xu(x,\mu^2)=x[u_v(x,\mu^2)+u_s(x,\mu^2)]$ ({\bf c}) and down-quarks 
$xd(x,\mu^2)=x[d_v(x,\mu^2)+d_s(x,\mu^2)]$ ({\bf d})
with their $1\sigma$ uncertainties at the factorization scale 
$\mu^2=10~{\rm GeV}^2$ versus $x$ for NNLO ABMP16 (right-tilted hatch),
ABMP15~\cite{Alekhin:2015cza} (left-tilted hatch)
and ABM12~\cite{Alekhin:2013nda} (vertical hatch) PDFs
in the $n_f=4$ flavor scheme. The same
distributions normalized to the central ABMP16 values 
are given for comparison ({\bf e-h}).
}
\end{center}
\end{figure}

\begin{figure}
\begin{center}
\vspace*{-10mm}
  \includegraphics[width=0.95\textwidth,height=0.9\textheight]{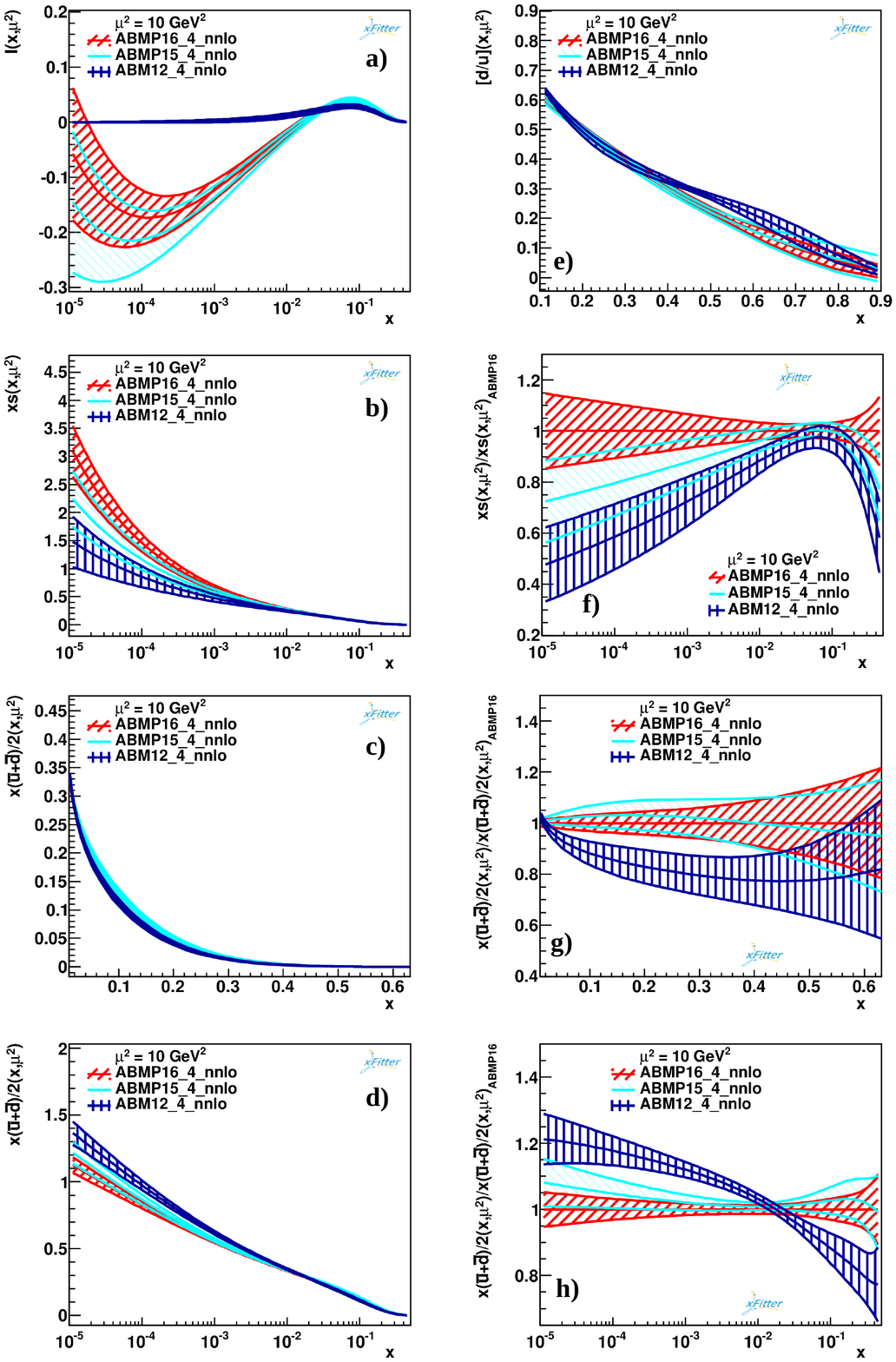}
  \caption{\small 
    \label{fig:abm16_vs_abm12_02}
    The same as in Fig.~\ref{fig:abm16_vs_abm12_01} for the sea-quark iso-spin asymmetry 
$I(x,\mu^2)=x[\bar d_s(x,\mu^2) - \bar u_s(x,\mu^2)]$ ({\bf a}), 
symmetrized strange sea $s(x,\mu^2)=[ s_s(x,\mu^2) +  \bar s_s(x,\mu^2)]/2$
 ({\bf b}), the non-strange sea 
$[x(\bar u + \bar d)/2](x,\mu^2)=
x[\bar u_s(x,\mu^2) + \bar d_s(x,\mu^2)]/2$ 
({\bf c}: linear $x$-scale, {\bf d}: logarithmic $x$-scale),
and the ratio
$[d/u](x,\mu^2)=[ d_v(x,\mu^2) +  d_s(x,\mu^2)]
/[ u_v(x,\mu^2) + u_s(x,\mu^2)]$ ({\bf e}).
The distributions of strange sea $s(x,\mu^2)$
({\bf f}) and the non-strange sea $[x(\bar d +\bar u)/2](x,\mu^2)$ 
({\bf g}: linear $x$-scale, {\bf h}: logarithmic $x$-scale) 
normalized to the central ABMP16 values are given for comparison. 
 }
\end{center}
\end{figure}

A representative comparison of the PDFs obtained in the present analysis 
with our earlier ABM12 and ABMP15 parametrizations 
demonstrates the recent improvements and is given in Figs.~\ref{fig:abm16_vs_abm12_01} and 
\ref{fig:abm16_vs_abm12_02}.\footnote{The plots are generated in the
  {\texttt{xFitter}} framework~\cite{Alekhin:2014irh} using our PDFs grids in
  the format of the {\tt LHAPDF} library (version 6)~\cite{Buckley:2014ana}, cf. 
  Sec.~\ref{sec:lhapdf}.}
In particular, the DY data added to the fit 
extend the range of $x$-values probed and, being complementary to the DIS sample used, 
help to disentangle quark distributions at small and large $x$. 
This mainly improves the accuracy of the down-quark distribution, which is 
commonly determined from a combination of the data on DIS off proton 
and deuteron targets. As detailed in Sec.~\ref{sec:data}, 
the down- and up-quark distributions in the present analysis 
are separated by using a combination of data on electron-proton DIS 
together with those on $W^{\pm}$- and $Z$-production in (anti)proton-proton collisions.
In consequence the deuteron data from DIS fixed-target experiments 
are not used anymore, which eliminates any errors related to the modeling of
nuclear effects in deuterium in the ABMP16 analysis.

The $d/u$ ratio obtained in this way is comparable 
with our earlier ABM12 determination within uncertainties.
At $x \lesssim 0.4$ its accuracy is improved due to the impact of the
DY LHC data in the central-rapidity region while it is
comparable to the one obtained in the ABM12 analysis for $x\gtrsim0.4$.  
The central values of the $d/u$ ratio obtained in both cases basically agree
within the errors. 
This proves that the nuclear corrections on the basis of the Kulagin-Petti model~\cite{Kulagin:2004ie,Kulagin:2007ju}
employed in the ABM12 analysis provide a consistent treatment~\cite{Alekhin:2016giu}.
The enhanced statistical potential of the DY data also appears
in disentangling the light-quark PDFs at small $x$.
The parameterization Eq.~(\ref{eq:sea}) allows for non-zero values of the sea
iso-spin asymmetry $I(x)=x[\bar d(x) - \bar u(x)]$ at small $x$, 
in contrast to the ABM12 one, which was based on the Regge-like asymptotic of
$I(x)\sim x^{0.7}$. 
By releasing the Regge-like constraint on $I(x)$ we find that 
negative values at $x\sim 10^{-4}$ are preferred by the data, while 
a turnover in its shape is observed at $x \lesssim 10^{-4}$. 
Therefore $I(x)$ may still be comparable with zero at smaller values $x \sim 10^{-6}$.

The strange sea distribution is traditionally determined by data on charm
production in the neutrino-induced DIS. 
In the present analysis we extend this sample with the measurements performed
by the NOMAD and CHORUS experiments, cf. Tab.~\ref{dis_dy_table}. This allows to improve 
the accuracy of the strange sea in a wide range of $x$, particularly due to the NOMAD 
data, which probe the range of $x\approx 0.02\div 0.75$. 
The newly added data on the $W^{\pm}$- and $Z$-boson production also 
help to disentangle the strange sea because of their particular contribution 
to the NC and CC processes. 
This improvement concerns mainly the small-$x$ region, which is 
poorly constrained by the existing fixed-target DIS data. 
As a result we obtain a continuous improvement in the 
accuracy of the strange sea at small-$x$ from the ABM12 to the ABMP15 PDFs and further to the present analysis 
due to the gradual increase in the statistical significance of the DY data sample used in those fits.
Furthermore, the central value of the small-$x$ strange sea distribution is
larger compared to the ABM12 one and comes into agreement with the ATLAS results 
based on a QCD analysis of their own data on the $W^{\pm}$- and $Z$-production 
in combination with the HERA inclusive DIS ones~\cite{Aad:2012sb,Aaboud:2016btc}. 
However, at $x \gtrsim 0.01$ we still observe a suppression of the strange sea as compared to 
the non-strange one by factor of $\sim 0.5$. 
This finding is in contrast to the results~\cite{Aad:2012sb,Aaboud:2016btc} 
which claim $\mbox{SU}(3)$ universality of the light-quark PDFs over a wide range of $x$. 
The difference of our analysis with Refs.~\cite{Aad:2012sb,Aaboud:2016btc} 
is evidently correlated with the difference in the obtained iso-spin asymmetries $I(x)$. 
For the ATLAS determination $I(x)$ is negative at $x\sim 0.1$, while in our case 
$I(x)$ is positive as suggested by the fixed-target DY data of the Fermilab E-866
experiment~\cite{Towell:2001nh} included into the fit. 
Therefore, a consolidation of our results with those of ATLAS would require
a critical appraisal of the E-866 results.  
This issue can be also reconciled in the future by measuring the associated $W+c$ production in 
proton-proton collisions at the LHC. 
The existing ATLAS~\cite{Aad:2014xca} and CMS~\cite{Chatrchyan:2013uja} data
somewhat overshoot the theory predictions at NLO in QCD based on our PDFs~\cite{Alekhin:2014sya}. 
However, the discrepancy is well within the experimental uncertainties and, moreover, 
might vanish once QCD corrections at NNLO accuracy become available.

\begin{figure}[b!]
\centerline{
  \includegraphics[width=16.0cm]{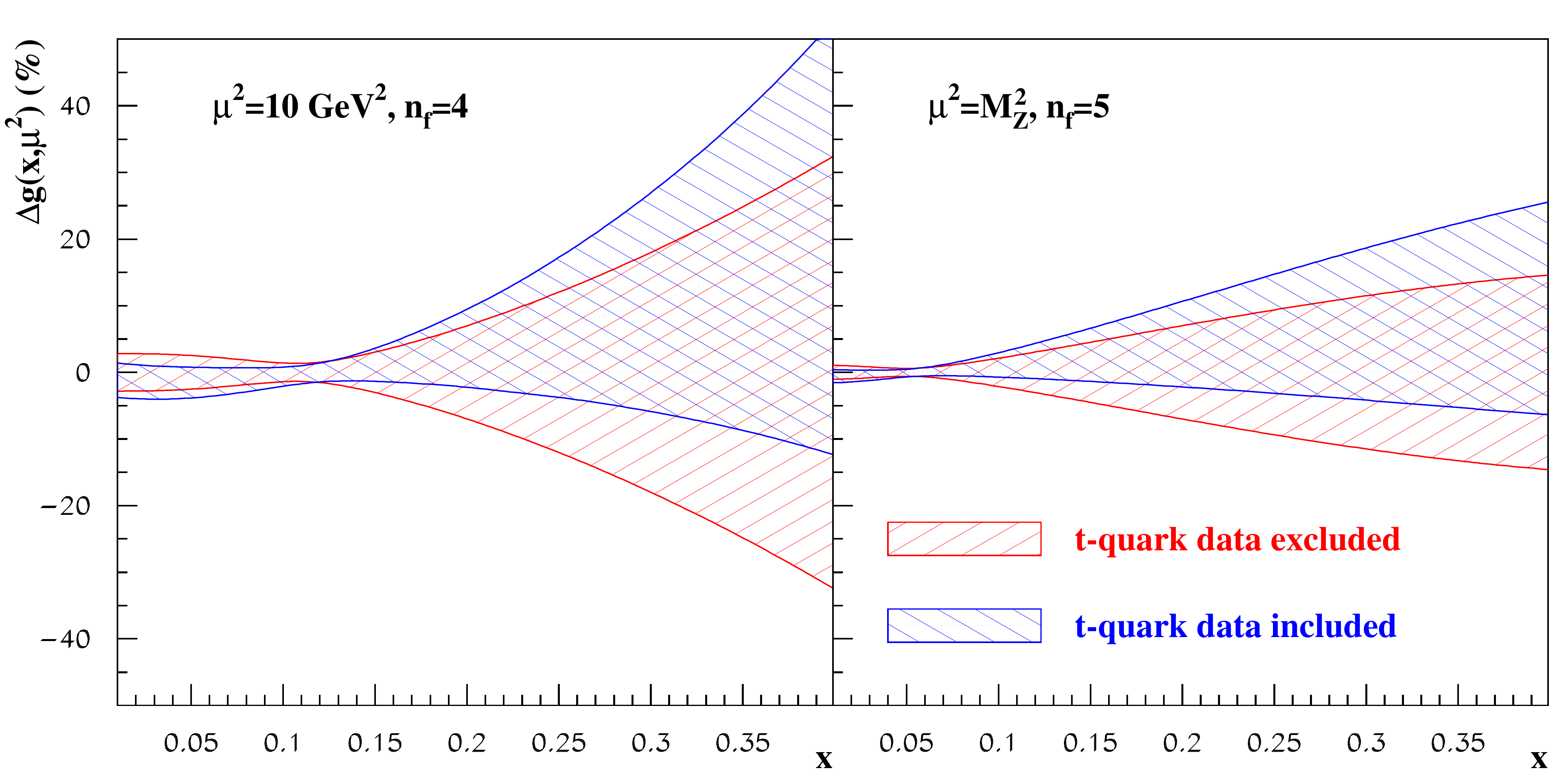}}
  \caption{\small
    \label{fig:glut}
    The $1\sigma$ relative uncertainty for the $n_f=4$ flavor gluon distribution 
    at the factorization scale $\mu^2~=~10~{\rm GeV}^2$ (left panel) and 
    for the $n_f=5$ flavor gluon distribution 
    at the factorization scale $\mu^2=M_Z^2$ (right panel).
    The nominal fit of the present analysis (left-tilted hatch) is compared to 
    a variant where the $t$-quark data of Tabs.~\ref{tab:data-tt} and \ref{tab:data-inp}
    has been excluded (right-tilted hatch).
  }
\end{figure}

\begin{figure}[t!]
\centerline{
  \includegraphics[width=16.0cm]{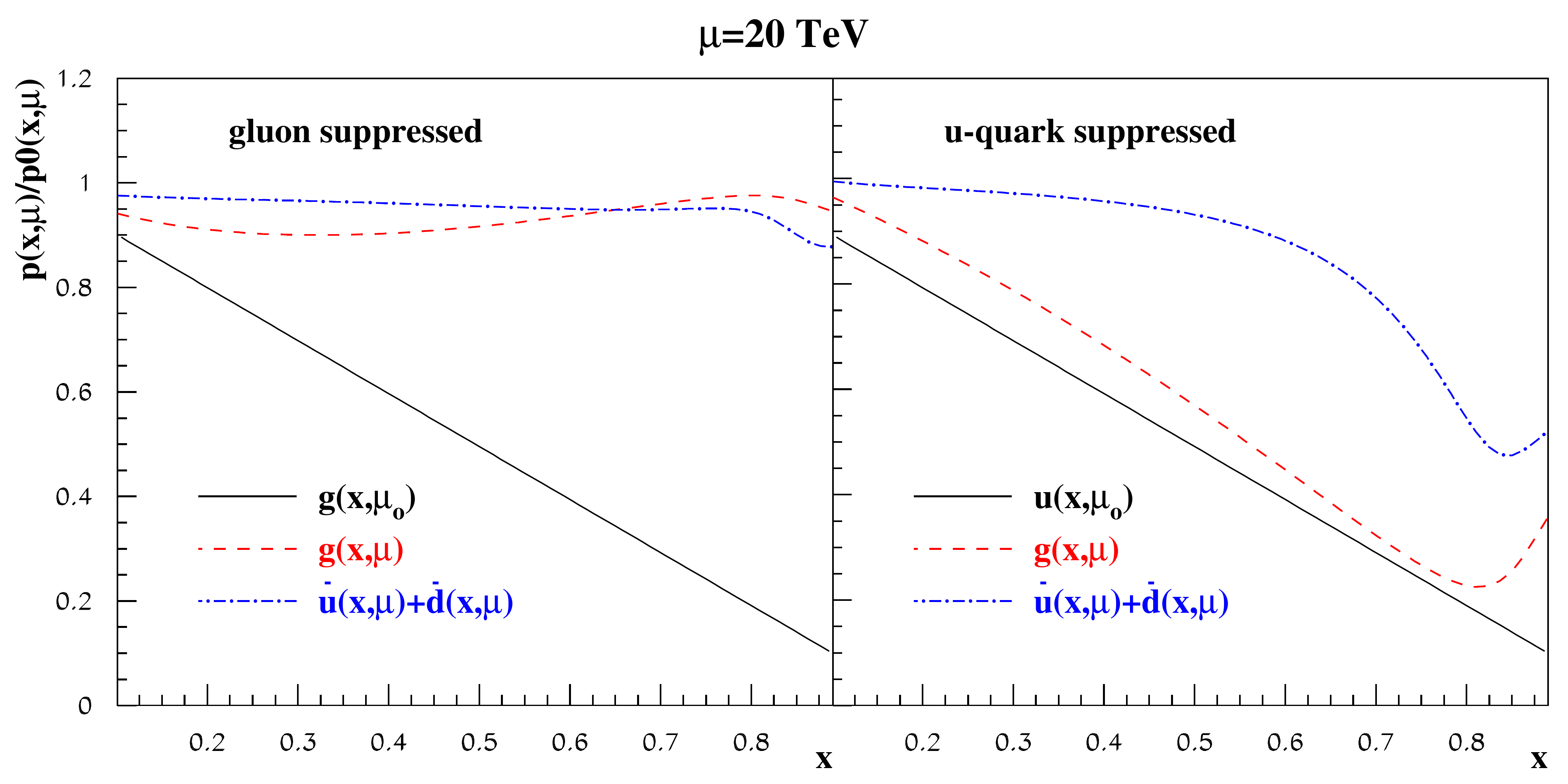}}
  \caption{\small
    \label{fig:psup}
    A response of the gluon (dashes) and non-strange sea 
    (dashed dots) distributions 
    evolved from the factorization scale $\mu_0=3~{\rm GeV}$
    to $\mu=20~{\rm  TeV}$
    to the suppression of the initial gluon distribution $g(x,\mu_0)$
    (left panel) and up-quark distribution $u(x,\mu_0)$ (right panel) by a factor 
    of $(1-x)$ (solid lines) given as a ratio of PDFs $p(x, \mu)$ and $p0(x,\mu)$ 
    with and without suppression, respectively.  
  }
\end{figure}
\begin{figure}
\begin{center}
\vspace*{-10mm}
\includegraphics[width=0.95\textwidth]{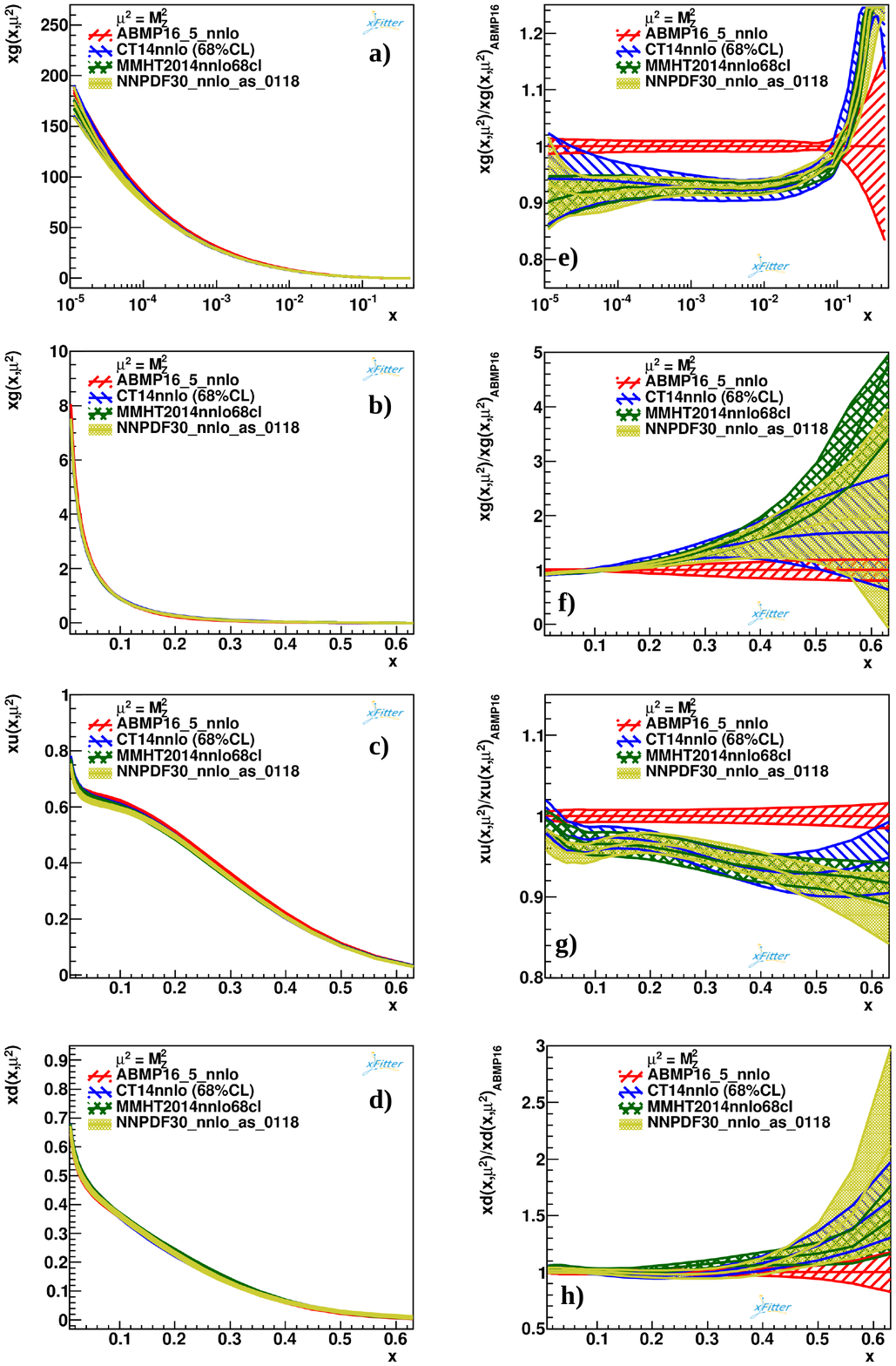}
\caption{\small 
  \label{fig:abm16_vs_others1_01}
   The same as in Fig.~\ref{fig:abm16_vs_abm12_01} for 
the absolute values of the $n_f=5$ flavor ABMP16 (right-tilted hatch), 
CT14~\cite{Dulat:2015mca} (left-tilted hatch), 
   MMHT14~\cite{Harland-Lang:2014zoa} (left-right tilted hatch)
 and NNPDF3.0~\cite{Ball:2014uwa} (shaded area) PDFs at the 
factorization scale $\mu^2=M_Z^2$.
}
\end{center}
\end{figure}
\begin{figure}
\begin{center}
\vspace*{-10mm}
\includegraphics[width=0.95\textwidth]{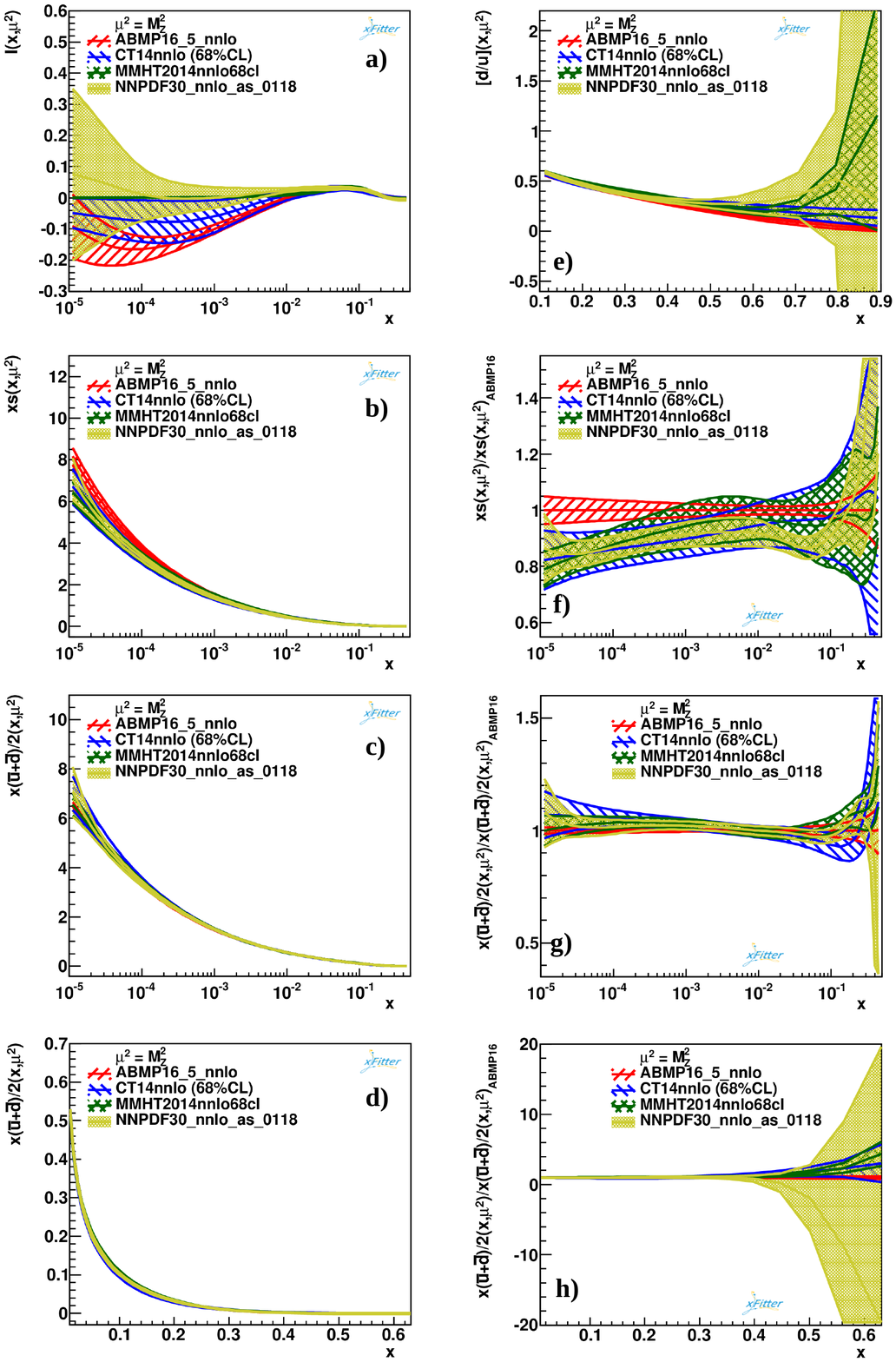}
\caption{\small 
  \label{fig:abm16_vs_others1_02}
   The same as in Fig.~\ref{fig:abm16_vs_abm12_02} for 
the absolute values of the $n_f=5$ flavor ABMP16 (right-tilted hatch), 
CT14~\cite{Dulat:2015mca} (left-tilted hatch), 
   MMHT14~\cite{Harland-Lang:2014zoa} (left-right tilted hatch)
 and NNPDF3.0~\cite{Ball:2014uwa} (shaded area) PDFs at the 
factorization scale $\mu^2=M_Z^2$.
}
\end{center}
\end{figure}
\begin{figure}
\begin{center}
\vspace*{-10mm}
\includegraphics[width=0.95\textwidth]{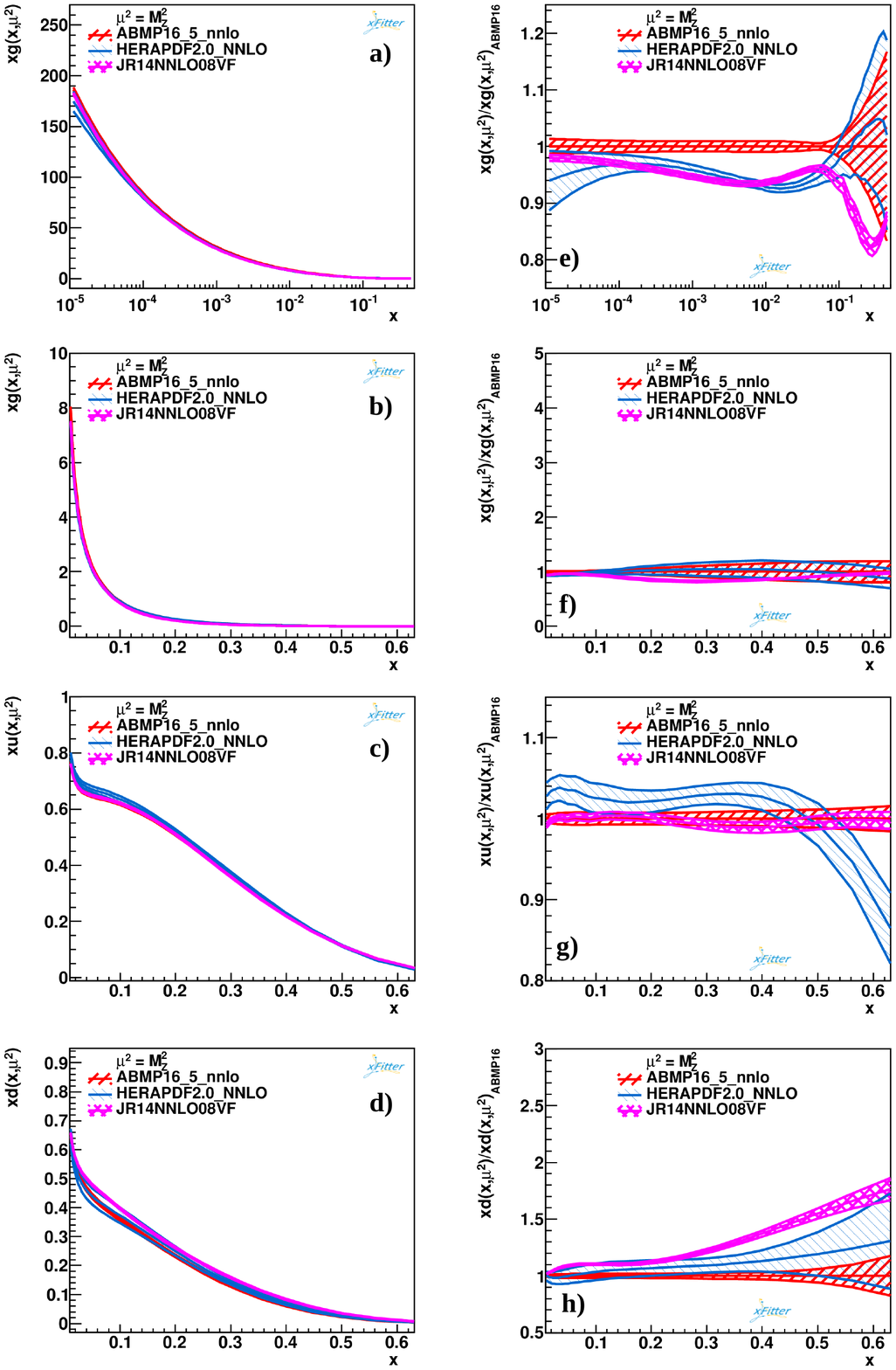}
\caption{\small 
  \label{fig:abm16_vs_others2_01}
   The same as in Fig.~\ref{fig:abm16_vs_abm12_01} for 
the absolute values of the $n_f=5$ flavor ABMP16 (right-tilted hatch), 
HERAPDF2.0~\cite{Abramowicz:2015mha} (left-tilted hatch) and
  JR14~\cite{Jimenez-Delgado:2014twa} (left-right tilted hatch) PDFs
 at the factorization scale $\mu^2=M_Z^2$.
}
\end{center}
\end{figure}

\begin{figure}
\begin{center}
\vspace*{-10mm}
\includegraphics[width=0.95\textwidth]{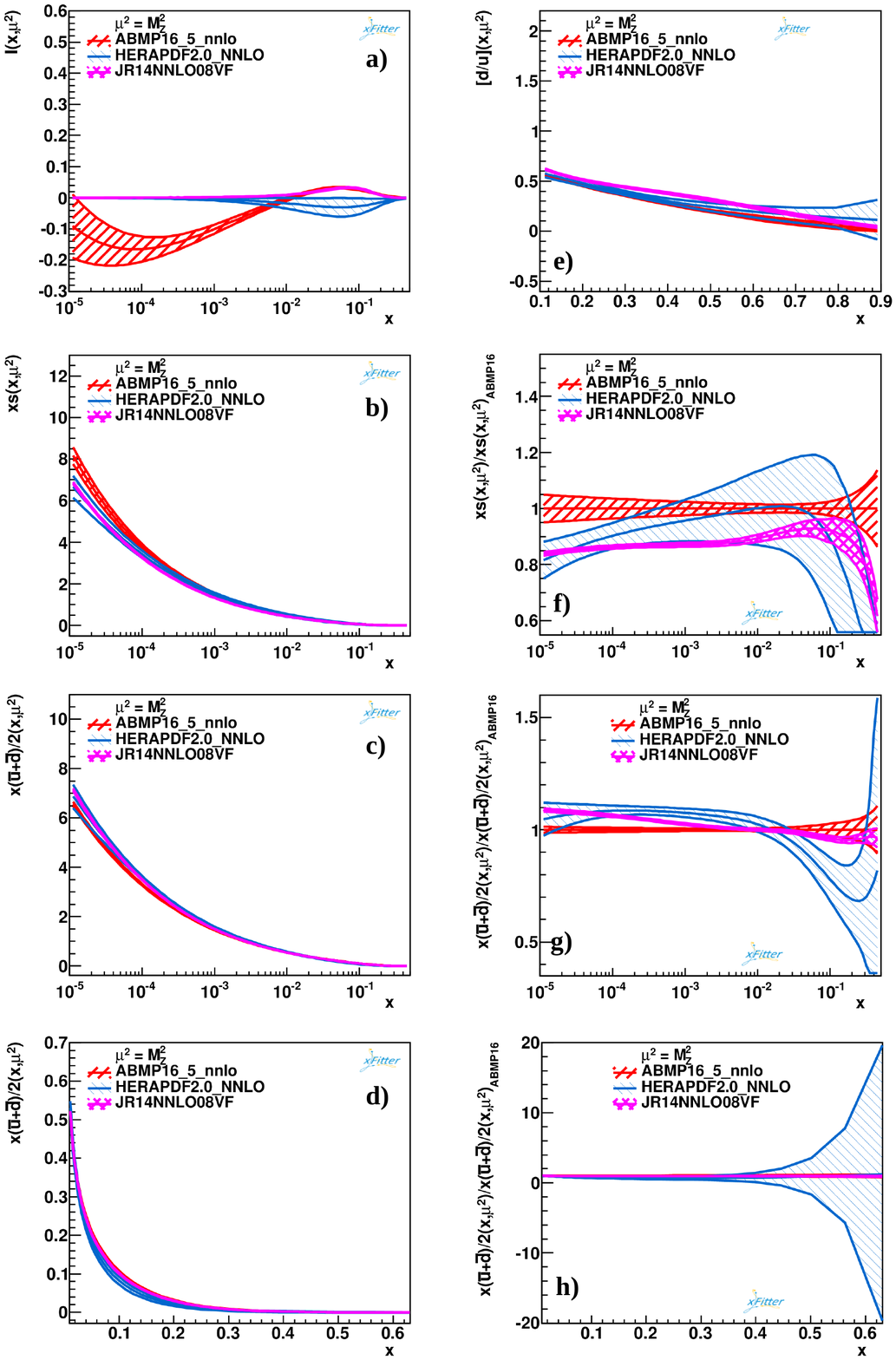}
\caption{\small 
  \label{fig:abm16_vs_others2_02}
   The same as in Fig.~\ref{fig:abm16_vs_abm12_02} for 
the absolute values of the $n_f=5$ flavor ABMP16 (right-tilted hatch), 
HERAPDF2.0~\cite{Abramowicz:2015mha} (left-tilted hatch) and
  JR14~\cite{Jimenez-Delgado:2014twa} (left-right tilted hatch) PDFs at the 
factorization scale $\mu^2=M_Z^2$.
}
\end{center}
\end{figure}

The data of Tabs.~\ref{tab:data-tt} and \ref{tab:data-inp} 
on hadronic $t$-quark production employed in the present analysis 
provide additional constraints on the PDFs, in particular on the gluon
distribution at large $x$.  
Due to this new input the latter increases at $x \gtrsim 0.1$ by 10--20\%, 
depending on the factorization scale, which is well within its uncertainty, cf. Fig.~\ref{fig:glut}. 
Here, the data on $t \bar{t}$-production play a major role since the
production process is mostly driven by initial gluons. 

It is worth noting that the uncertainty in the 
gluon distribution decreases as the factorization scale increases. 
This happens because the large-scale gluon PDF receives also sizable contributions from 
the quark PDFs during the QCD evolution in the singlet sector due to the splitting function $P_{gq}$.
In this way the large-$x$/large-scale gluon PDF benefits from an 
accurate determination of large-$x$ quark PDFs.
The latter, in turn, are well constrained by the DIS and DY data used in the present analysis. 
This peculiar feature is illustrated in Fig.~\ref{fig:psup}. 
We compare the response of the gluon distribution when evolved to a large scale of $\mu=20~{\rm TeV}$ 
to a model suppression of the gluon and up-quark PDFs by factor of $(1-x)$ at the initial
scale of the evolution.
Indeed, as Fig.~\ref{fig:psup} shows, the sensitivity of the large-scale gluon PDF 
to its modification at the initial scale is marginal.
If, however, the up-quark distribution is modified by the factor $(1-x)$ at the initial
scale, this suppression is basically reproduced by the gluon PDF at $\mu=20~{\rm TeV}$.
The large-scale sea-quark distributions are also sensitive 
to modifications of the quark PDFs at small scales, although to a lesser extent. 
In this case the effect originates from a two-step process in the singlet
sector during evolution due to the splitting function $P_{qg}$ acting on 
the modified gluon PDF.
Such an interplay is especially pronounced at big scales 
which are currently explored in searches for potential new effects beyond the Standard Model. 

A comparison of the 
PDFs determined in the present analysis with other available PDF sets 
at the scale of the $Z$-boson mass squared, $M_Z^2$, is presented 
in Figs.~\ref{fig:abm16_vs_others1_01}--\ref{fig:abm16_vs_others2_02}. 
The discrepancies observed appear mainly due to differences in the 
data sets and in the theoretical framework used, in 
particular for the description of the heavy-quark NC DIS production, cf. 
Ref.~\cite{Accardi:2016ndt}. 
Firstly, the up- and down-quark distributions in the present
analysis exhibit an improved precision as a result of employing the latest DY and DIS data samples. 
Consequently, these distributions have typically smaller uncertainties compared to other PDFs, which 
do not yet use all currently available DY and DIS data. In part due to 
this improvement, as we discuss above,
the ABMP16 gluon distribution, now also constrained by the $t$-quark data 
from LHC, 
has overall smaller uncertainties as compared with other PDFs,
although their central 
values are typically bigger(smaller) than ours at $x\gtrsim 0.1( \lesssim 0.1)$.
The sea-quark iso-spin asymmetry 
$I(x)$ is in a good agreement with the CT14 and NNPDF3.0 PDFs, the latter 
having largest uncertainties at low $x$. 
In contrast, for the MMHT14 and JR14 PDFs the same distribution is vanishing, i.e., 
$I(x) \simeq 0$, at low $x$ due to possible restrictions in the chosen parametrization. Also the HERAPDF2.0 result for $x(\bar d - \bar u)$ 
indicates a different shape at $x \gtrsim 0.01$ due to the fact that only HERA
data are used in fitting those PDFs. 
However, the difference is covered by the large uncertainty in this distribution. 
Finally, the improved $s$-quark distribution due to the latest precise NOMAD~\cite{Samoylov:2013xoa} and 
CHORUS~\cite{KayisTopaksu:2011mx} data is generally in good agreement with all PDFs 
with the exception of HERAPDF2.0 which, as mentioned before, is determined
solely from HERA data which constrain strange sea quarks only weakly.

\subsection{Higher twist}
\label{sec:ht}

\begin{figure}[t!]
\centerline{
  \includegraphics[width=16.0cm]{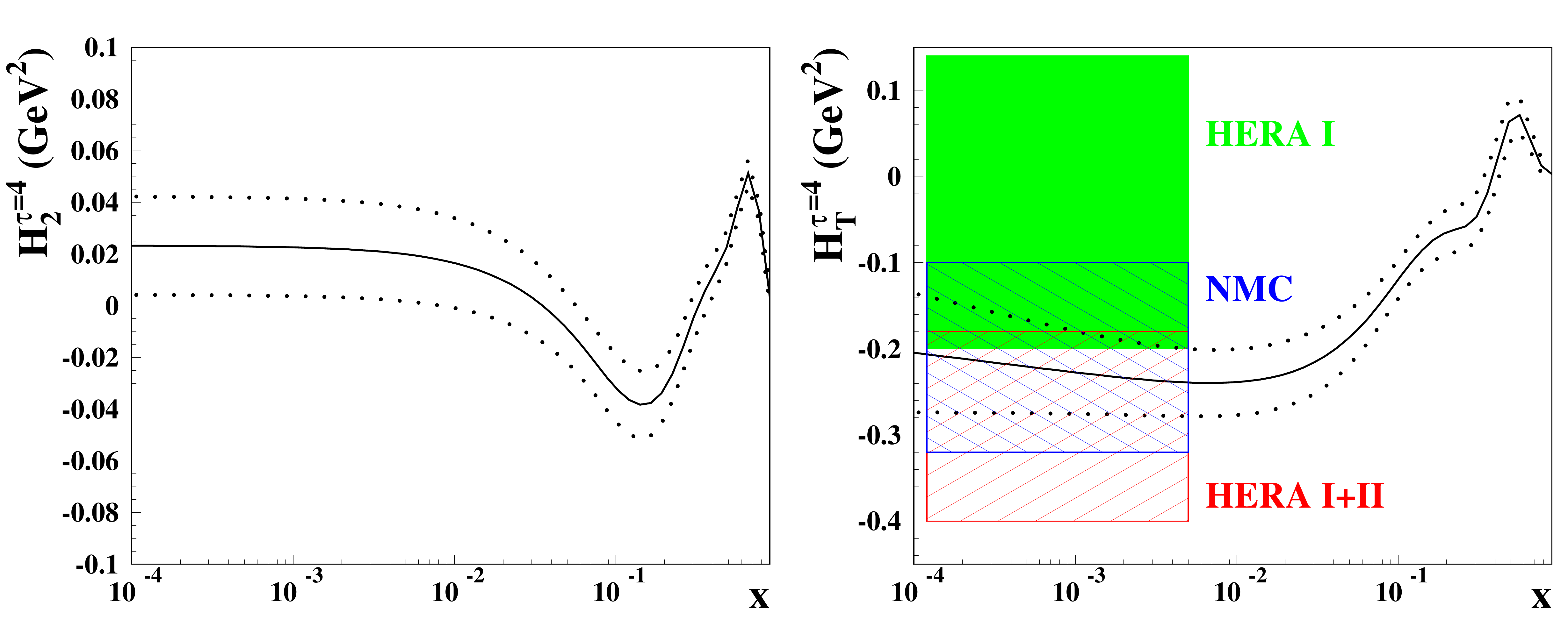}}
  \caption{\small
    \label{fig:ht}
    The coefficients $H_2^{\tau=4}$ (left) and $H_T^{\tau=4}$
    (right) in the higher twist terms of the inclusive DIS structure 
    functions, cf. Eq.~(\ref{eq:htwist}), as a function of $x$. 
    Shown are the central values (solid) and the 1$\sigma$ bands 
    (dots). The boxes at $x \lesssim 5 \cdot 10^{-3}$ display 
    the low-$x$ asymptotic of $H_T^{\tau=4}$
    preferred by the individual data sets indicated  
    (right-tilted hatch: combined HERA run I+II~\cite{Abramowicz:2015mha},
    left-tilted hatch: NMC~\cite{Arneodo:1996qe}, shaded: combined HERA run I~\cite{Aaron:2009aa}).
  }
%
\centerline{
  \includegraphics[width=8.75cm]{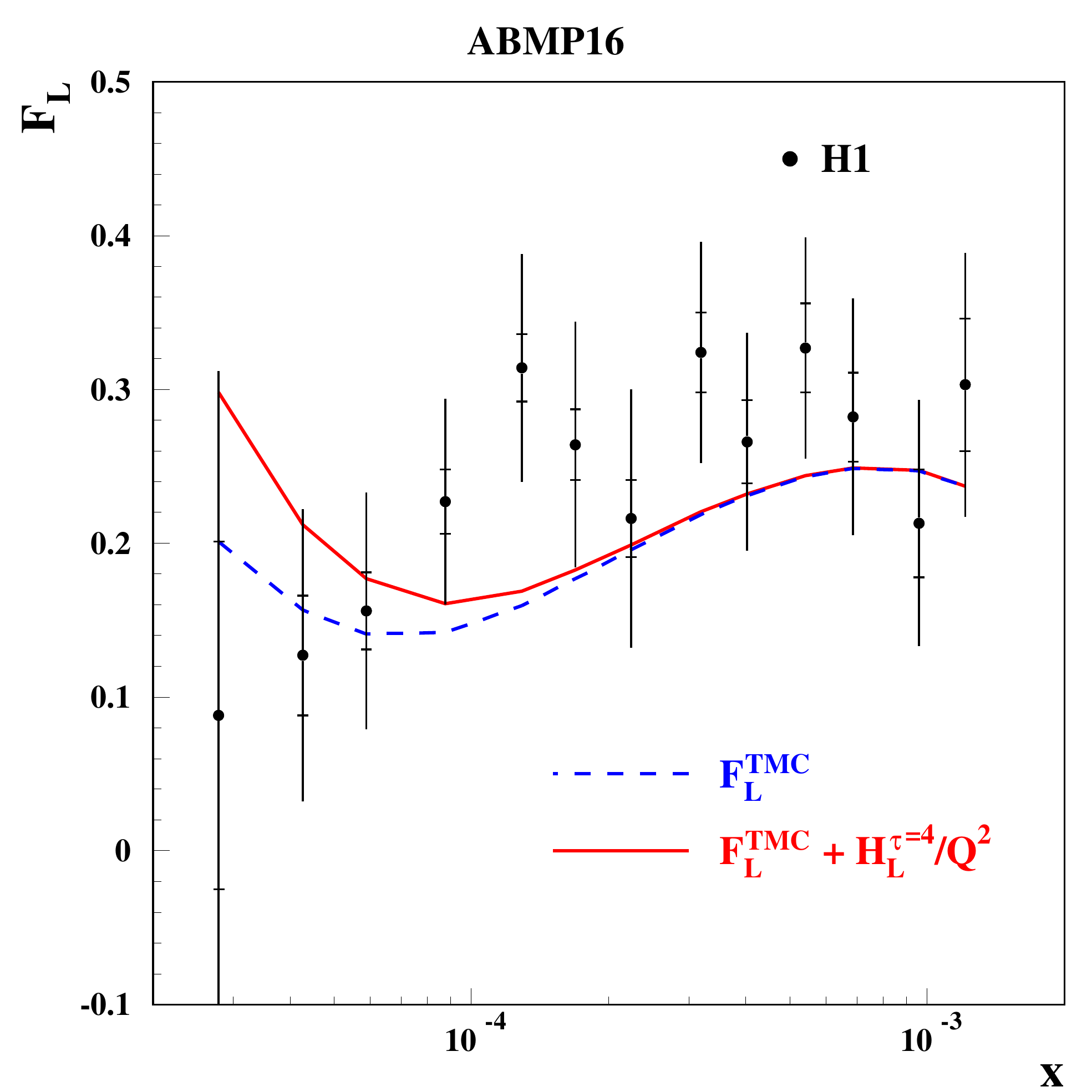}}
  \caption{\small
    \label{fig:FL}
    The data on the longitudinal structure functions $F_L$ as a function of $x$  
    by the H1 collaboration at HERA~\cite{Collaboration:2010ry} 
    compared to the predictions of our analysis at NNLO in QCD 
    based on either including higher twist terms (solid) or setting 
    them to zero (dashes), cf. Eq.~(\ref{eq:htwist}).
  }
\end{figure}

We parameterize the coefficients $H_{i}^{\tau=4}$ for $i=2,T$ 
of the higher twist (HT) terms in the DIS nucleon structure functions $F_{i}$  
in Eq.~(\ref{eq:htwist}) as follows
\begin{equation}
\label{eq:htspline}
H_{i}^{\tau=4}(x) \,=\, x^{\alpha_{i}} S_{i}(x)
\, ,
\qquad\qquad
i \,=\, 2,T
\, ,
\end{equation}
where $S_{i}(x)$ are the cubic splines defined at the knots
$\{[x_k,H_{i}^{\tau=4}(x_k)]:k=1,7\}$ and $\{x_k\}=(0,0.1,0.3,0.5,0.7,0.9,1)$.
The constraint $H_{i}^{\tau=4}(1)=0$ is imposed due to poor coverage 
of the region of $x\rightarrow 1$ by the existing data. 
The remaining knot values $H_{i}^{\tau=4}(x_k)$ for $k=1,\dots , 6$ are taken as fit parameters 
and the result of the fit with Eq.~(\ref{eq:htspline}) 
is presented in Tab.~\ref{tab:hts} and Fig.~\ref{fig:ht}.  

\input{table-hts}

A particular benefit of the present analysis is the possibility to 
study the small-$x$ shape of the HT terms due to the improved statistical 
significance of the inclusive HERA run I+II data. 
With this input we find a sizeable deviation of $H_{T}^{\tau=4}$ from 0 at small $x$. 
The small-$x$ shape of $H_{i}^{\tau=4}$ is controlled by the combination of the NMC and HERA data. 
Both data sets prefer a negative value of $H_{T}^{\tau=4}$ at small $x$, as we find in the variants of 
of our analysis when either the NMC or the HERA data set is dropped. 
A similar check for the HERA run I data~\cite{Aaron:2009aa} demonstrates 
that the small-$x$ value of $H_{T}^{\tau=4}$ is significantly bigger than the one
preferred by the HERA run I+II data, although both are comparable within the uncertainties.
Nonetheless, for this reason the ABMP16 HT terms are different from 
the earlier ABM12 determination of the HT terms based on the HERA run I data, which
where consistent with zero at small $x$~\cite{Alekhin:2013nda}.
For larger values $x\gtrsim 0.1$, both results, ABMP16 and ABM12 agree with each other.

The value of the small-$x$ exponent $\alpha_T = 0.05 \pm 0.07$ found in the fit 
defines a shallow falloff of $H_{T}^{\tau=4}$ at $x\rightarrow 0$, see Fig.~\ref{fig:ht}.
However its statistically significant deviation from zero persists down to $x\sim 10^{-5}$. 
At the same time the fitted value of $H_{2}^{\tau=4}$ is comparable to zero at small $x$.
Therefore, the constraint $\alpha_2=0$ is imposed. 
This results in a sizeable contribution of the HT terms to the structure function $F_L$
at small $x$ and $Q^2$, cf. Fig.~\ref{fig:FL}. 
The excess in $F_L$ due to the HT found in~\cite{Abt:2016vjh} is more essential 
than the one observed in our case. 
This might be explained by the particular parameterization of the HT terms assumed in Ref.~\cite{Abt:2016vjh}
and discussed in Sec.~\ref{sec:hts}, which implies a strong effect of the QCD evolution on the small-$x$ HT terms. 
In our analysis such an effect does not appear.

\begin{figure}[t!]
\centerline{
  \includegraphics[width=9.0cm]{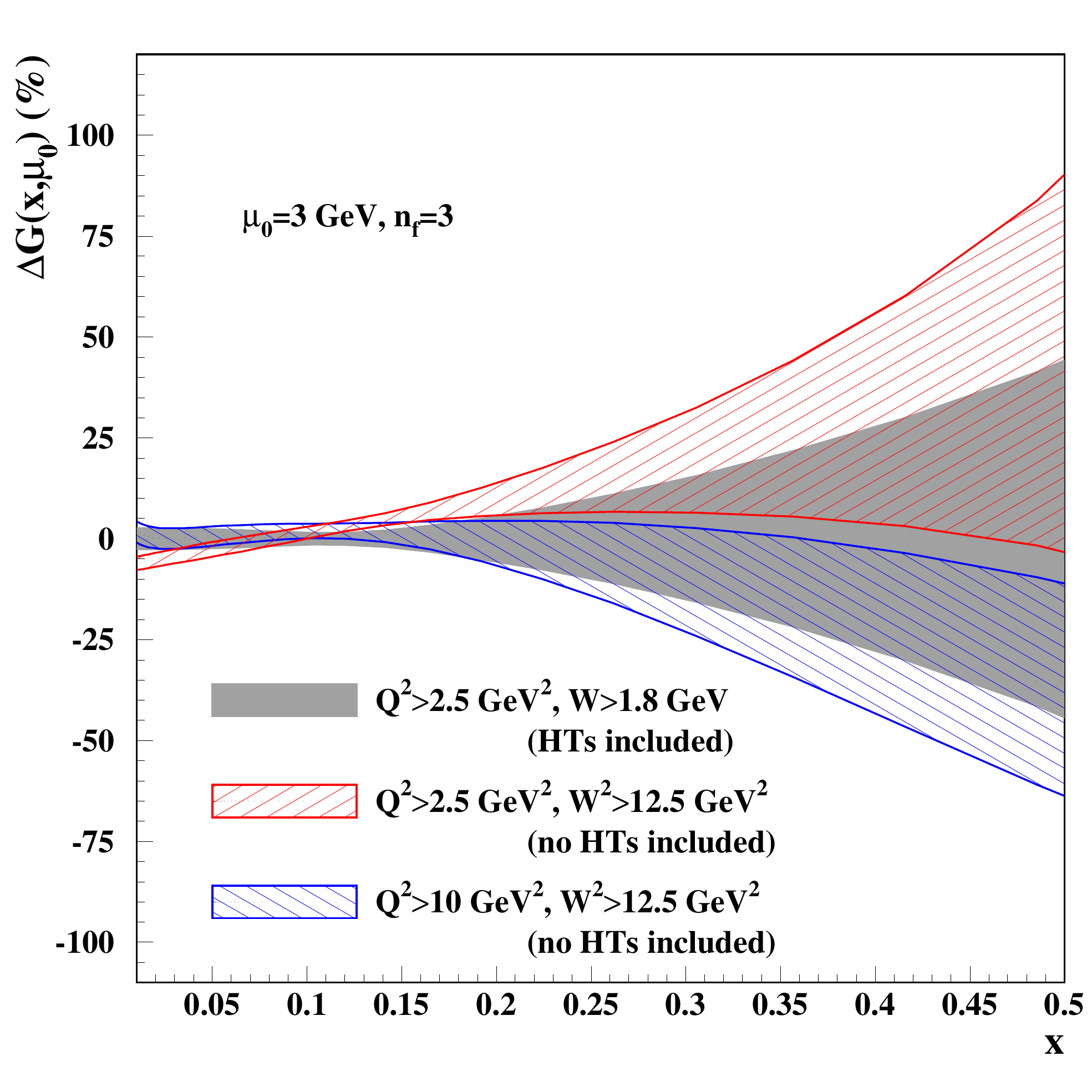}}
  \caption{\small
    \label{fig:gluhigh}
    The $1\sigma$ relative uncertainty in the $n_f=3$ flavor gluon distribution 
    at the starting scale of the QCD evolution $\mu_0=3~{\rm GeV}$ 
    obtained in the present analysis with the HT terms taken into account and the 
    cuts of $Q^2 > 2.5~{\rm GeV}^2$ and $W > 1.8~{\rm GeV}$ imposed on the inclusive 
    DIS data  (shaded area). 
    In comparison two variants of the analysis with no HT taken into account and 
    the cuts of $Q^2 > 2.5~{\rm GeV}^2$ and $W^2 > 12.5~{\rm GeV}^2$ (right-tilted hatch) 
    and $Q^2 > 10~{\rm GeV}^2$ and $W^2 > 12.5~{\rm GeV}^2$ (left-tilted hatch) 
    are shown. 
  }
\end{figure}

The interplay between the leading twist and the HT terms is essential for the
determination of the PDFs. 
To illustrate this, we consider two variants of our analysis when no HT terms
are taken into account and more stringent kinematic cuts are imposed on the inclusive
DIS data as follows  
\begin{equation}
  \label{eq:cut1}
  Q^2 > 2.5~{\rm GeV}^2\, , \qquad W^2 > 12.5~{\rm GeV}^2 
  \, , 
\end{equation}
and
\begin{equation}
  \label{eq:cut2}
  Q^2 > 10~{\rm GeV}^2\, , \qquad W^2 > 12.5~{\rm GeV}^2 
  \, ,
\end{equation}
where $Q^2$ is the momentum transferred squared and $W$ denotes 
the invariant mass of the hadronic system, cf. Eq.~(\ref{eq:wdef}). 
The gluon distributions obtained in these three fits are compared 
in Fig.~\ref{fig:gluhigh}. 
In case of the cut in Eq.~(\ref{eq:cut1}) the gluon PDF increases 
by $1\sigma$ compared to our nominal one. 
Moreover, the central value differs by up to 50\% at $x<0.5$. 
For the fit with the cut in Eq.~(\ref{eq:cut2}) the difference is twice smaller. 
This means, that the HT contribution cannot be eliminated by the commonly used cut of
$W^2 > 12.5~{\rm GeV}^2$ and that the HT terms are still substantial in the region of
$Q^2 = 2.5 \div 10~{\rm GeV}^2$, see also Refs.~\cite{Alekhin:2011ey,Alekhin:2013nda}.

\subsection{The strong coupling constant}
\label{sec:alps}

In the present analysis the value of the strong coupling constant $\alpha_s$ is
fitted simultaneously with the leading-twist PDFs and the twist-four terms.
The DIS and $t\bar{t}$-production data sets play the most important role in this determination.
To study the particular significance of these two samples
we consider a variant of the present analysis without the $t$-quark data included. 
In this case the value of 
\begin{eqnarray}
  \label{eq:asdis}
  \alpha_s^{(n_f=5)}(M_Z)&=& 0.1145 \pm 0.0009 
  \, 
\end{eqnarray}
at NNLO in QCD for the scheme with $n_f=5$ light flavors is obtained. 
This value is basically defined by the data from the fixed-target 
SLAC, BCDMS, and NMC experiments and from the HERA collider. 
To separate the contribution of each of these four data sets to the average 
Eq.~(\ref{eq:asdis}) we also perform variants of the fit 
when only one of them is included and all other DIS data sets are discarded.
However, the HT terms discussed above in Sec.~\ref{sec:ht} 
can be determined consistently only if the SLAC data included. 
Therefore, in the variants of the fit based on either the BCDMS, NMC, or the HERA data 
set the HT coefficients are fixed to those values which have been obtained in our nominal analysis.
Only in the variant based on the SLAC data the HT terms are fitted. 

\begin{figure}[th!]
\centerline{
  \includegraphics[width=8.75cm]{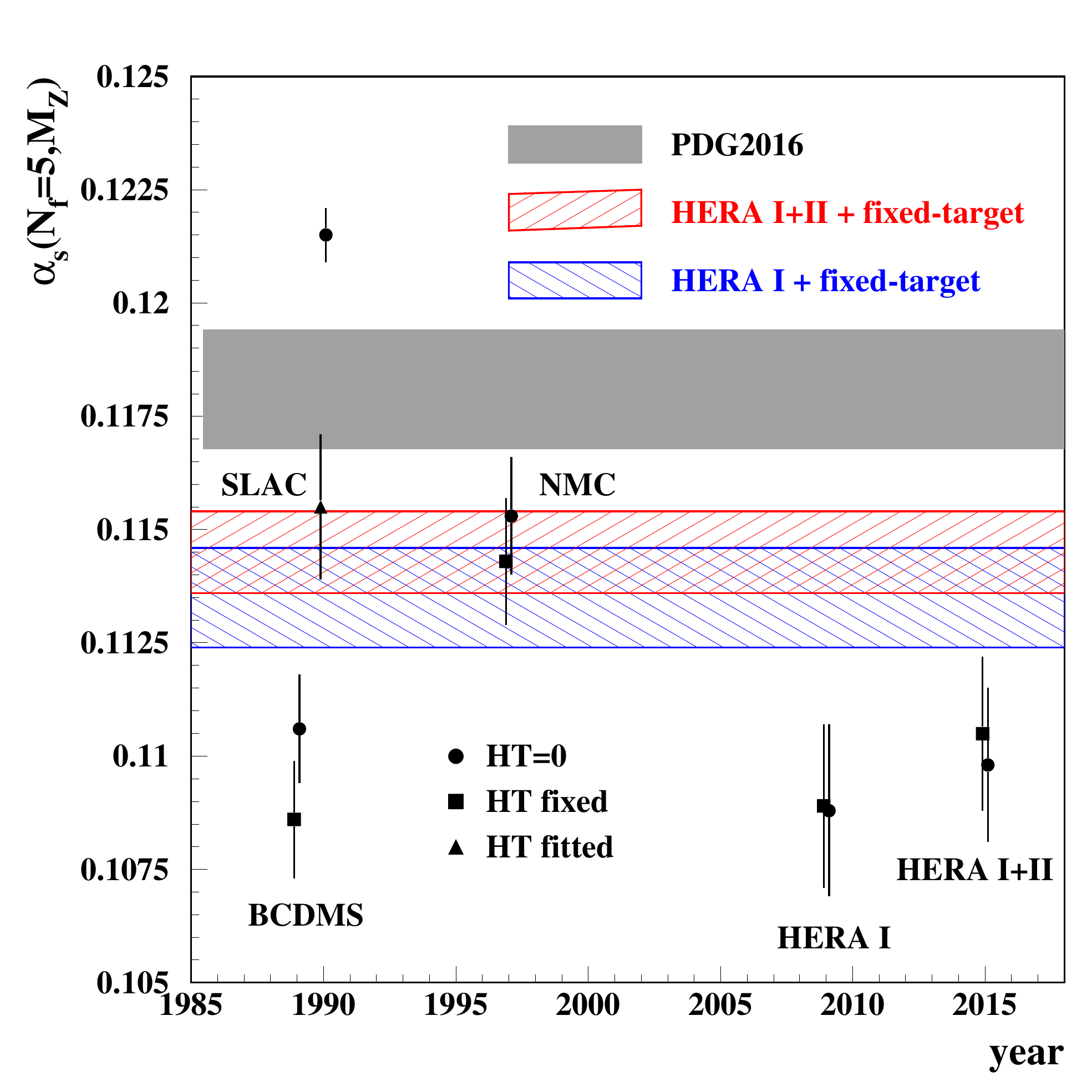}}
  \caption{\small
    \label{fig:alphas-history}
    The value of $\alpha_s^{(n_f=5)}(M_Z)$ in the \msbar scheme for $n_f=5$ at NNLO in QCD 
    determined by the individual data sets as a function of the year of their publication.
    Data from SLAC~\cite{Bodek:1979rx,Mestayer:1982ba,Whitlow:1990gk} (proton),
    BCDMS~\cite{Benvenuti:1989rh},
    NMC~\cite{Arneodo:1996qe} (proton), 
    the HERA run I~\cite{Aaron:2009aa} as well as 
    the HERA run I+II combination~\cite{Abramowicz:2015mha}
    are considered in three variants for the treatment of the higher twist
    terms defined in Eq.~(\ref{eq:htwist}):
    {\it (i)} the higher twist terms are set to zero (circles);
    {\it (ii)} the higher twist terms are fixed to the values obtained in the ABMP16
    fit from considering all data sets (squares);
    {\it (iii)} the higher twist terms are fitted to the individual data set
    under study (triangles).
    The bands for $\alpha_s^{(n_f=5)}(M_Z) $ obtained by using the combination 
    of the SLAC, BCDMS and NMC samples together with those from the HERA run I 
    (left-tilted hatches) and the run I+II combination (right-tilted hatches)
    as well as the 2016 PDG 
    average~\cite{Olive:2016xmw} (shaded area) are presented for comparison.  
  }
\end{figure}

In addition we consider for all cases also the variants when the HT terms are set to zero 
in order to check the sensitivity of the extracted value of $\alpha_s$ to this contribution. 
The values of $\alpha_s^{(n_f=5)}(M_Z)$ which have been obtained in this way are displayed in Fig.~\ref{fig:alphas-history} 
in chronological order of the experiment. 
The most recent HERA run I+II data prefer the value of
$\alpha_s^{(n_f=5)}(M_Z)= 0.1105 \pm 0.0017$.
This is somewhat larger than the value preferred by the HERA run I data. 
However, it is significantly smaller than the 2016 PDG average~\cite{Olive:2016xmw}.
It is worth noting that the HERA run I+II data are somewhat sensitive to the 
HT contribution, in contrast to those from HERA run I only. 
This is in line with the comparisons given in Fig.~\ref{fig:ht} and the discussion of 
Sec.~\ref{sec:ht}. 
As a result of this trend the value of $\alpha_s$ extracted from the
combination of the world DIS data used in the ABMP16 analysis 
increases by $\sim 1\sigma$ compared to our earlier determination 
in ABM12, which was based on the HERA run I data.
However, the 2016 PDG average~\cite{Olive:2016xmw} is still by $\sim 2\sigma$ larger than 
our present determination based on the DIS data in Eq.~(\ref{eq:asdis}).
Only the $\alpha_s^{(n_f=5)}(M_Z)$ value derived from the SLAC data overlaps with the PDG error band.
All other values are lower, in particular those from the BCDMS and HERA data.  

The sensitivity of the DIS data to the treatment of the HT terms documented in Sec.~\ref{sec:ht}
also has an impact on the $\alpha_s$ determination. 
Indeed, for the variant of our analysis when the HT terms are not taken into 
account and the kinematic cuts in Eq.~(\ref{eq:cut1}) are applied, 
the value 
\begin{eqnarray}
  \label{eq:ascut1}
  \alpha_s^{(n_f=5)}(M_Z)&=& 0.1167 \pm 0.0005 \, 
\end{eqnarray}
is obtained at NNLO. 
This is larger than our nominal value in Eq.~(\ref{eq:asdis}) by $\sim 3\sigma$. 
Imposing the more stringent cuts of Eq.~(\ref{eq:cut2}) leads to the NNLO value of 
\begin{eqnarray}
  \label{eq:ascut2}
  \alpha_s^{(n_f=5)}(M_Z)&=& 0.1140 \pm 0.0009 \,
\end{eqnarray}
which restores the agreement with our nominal fit. 
This study demonstrates again the importance of the HT terms 
in the kinematic region $Q^2=2.5\div 10~{\rm GeV}^2$ 
which survive after applying the cuts of Eq.~(\ref{eq:cut1}). 

When the top-quark data listed in Tabs.~\ref{tab:data-tt} and~\ref{tab:data-inp} 
are included, we find the value 
\begin{eqnarray}
\label{eq:asmt}
\alpha_s^{(n_f=5)}(M_Z)&=& 0.1147 \pm 0.0008 \, , 
\end{eqnarray}
in the $n_f=5$ flavor scheme at NNLO, which is not very different from the DIS one in 
Eq.~(\ref{eq:asdis}), but has a slightly smaller statistical error. 
It is worth noting that the value of $\alpha_s$ extracted from the 
$t$-quark data strongly depends on the $t$-quark mass setting and 
the result in Eq.~(\ref{eq:asmt}) is obtained by fitting $m_t$  
simultaneously with the PDFs and $\alpha_s$, cf. Sec.~\ref{sec:hq}.

Quite comparable values to the one we have obtained in the present NNLO
analysis were obtained by the JR14~\cite{Jimenez-Delgado:2014twa} with $
\alpha_s^{(n_f=5)}(M_Z) = 0.1136 \pm 0.0004~({\rm stat.})$ 
or $\alpha_s^{(n_f=5)}(M_Z) = 0.1162 \pm 0.0006~({\rm stat.})$, depending on the PDF shape employed.  
Our value is also comparable to the one determined in earlier N$^3$LO analysis
for non-singlet DIS data~\cite{Blumlein:2006be}, quoting $\alpha_s^{(n_f=5)}(M_Z) = 0.1141^{+0.0020}_{- 0.0022}$.

Other groups, CTEQ, MSTW (MMHT), and NNPDF, have also determined the value of $\alpha_s$ 
in the PDF fits, see Ref.~\cite{Alekhin:2016evh} for a recent survey. 
The NNLO values   $\alpha_s^{(n_f=5)}(M_Z)=0.1172 \pm 0.0013$ and $\alpha_s^{(n_f=5)}(M_Z)=0.1174 \pm 0.0007$, obtained 
by the MSTW and NNPDF groups, respectively, despite looking similar, have a quite different origin, 
as has been shown in detail in Ref.~\cite{Alekhin:2016evh}. 
The most essential issues appearing in the context of a comparison with the MSTW and NNPDF results
concern the treatment of the higher twist terms discussed 
in Sec.~\ref{sec:ht}, the nuclear corrections to the data on 
heavy-nuclei targets and the impact of missing NNLO corrections. 
A lower value $\alpha_s^{(n_f=5)}(M_Z)=0.1150^{+0.0060}_{- 0.0040}$
has been obtained by the CTEQ, yet with large errors, but well compatible with ours.

Finally, one may also compare lattice simulations.  
The Alpha collaboration quotes $\alpha_s^{(n_f=5)}(M_Z) = 0.1179 \pm 0.0011$ 
as a recent result~\cite{Bruno:2017lta} while 
other lattice results are summarized in Ref.~\cite{Aoki:2016frl}. 

\subsection{Heavy-quark masses}
\label{sec:hq}

In addition to the PDF parameters listed in Tab.~\ref{tab:fitvalues} 
and to the strong coupling $\alpha_s$ the ABMP16 fit also determines the heavy-quark masses.
In the \msbar scheme we obtain at NNLO in QCD the values 
\begin{eqnarray}
\label{eq:mcmbmt}
m_c(m_c) &=& 1.252 \pm 0.018\phantom{0} \GeV \, ,\nonumber\\
m_b(m_b) &=& 3.84\phantom{0} \pm 0.12\phantom{00} \GeV \, ,\nonumber\\
m_t(m_t) &=& 160.9 \pm 1.1\phantom{000} \GeV \, ,
\end{eqnarray}
where the scale $\mu_r$ has always been chosen identical to the numerical value of the masses.
For the charm-quark mass often also the scale choice $\mu_r= 3~\GeV$ is used,
for which one obtains 
\begin{eqnarray}
\label{eq:mc3GeV}
m_c(3~\GeV) &=& 1.007 \pm 0.018~\GeV \, .
\end{eqnarray}

The uncertainties quoted in Eq.~(\ref{eq:mcmbmt}) denote the $1\sigma$ 
confidence level as determined from the data listed in Sec.~\ref{sec:data}.
Largely, these are the HERA data on DIS charm-~\cite{Abramowicz:1900rp} 
and bottom-quark production~\cite{Aaron:2009af,Abramowicz:2014zub}, cf. Tab.~\ref{dis_dy_table}.
As the theory predictions for these DIS cross sections are not complete to NNLO yet, 
as described in Sec.~\ref{sec:theory}, 
the charm- and bottom-quark masses in Eq.~(\ref{eq:mcmbmt}) carry 
an additional systematic model uncertainty due to the incomplete NNLO Wilson coefficients 
estimated as $\Delta m_c(m_c) \simeq  0.01~\GeV$, cf. Fig.~\ref{fig:aQj30}.
Compared to previous analyses~\cite{Alekhin:2012vu,Alekhin:2013nda} 
this model uncertainty has been reduced by a factor of four thanks to the
new theory improvements presented in Sec.~\ref{sec:theory}. 
Thus, the current accuracy in the determination of $m_c$ from DIS data 
is becoming competitive with other methods entering the world average~\cite{Olive:2016xmw}.
For the bottom mass, the use of approximate NNLO Wilson coefficients 
adds a model uncertainty $\Delta m_b(m_b) \simeq 0.1~\GeV$ to Eq.~(\ref{eq:mcmbmt}).
On the other hand, the residual theory uncertainties from scale variations at NNLO are small, though. 
For the DIS heavy-quark cross sections considered in the present analysis, 
it amounts to $\Delta m_c(m_c) = \pm 0.025 \GeV$ for charm~\cite{Alekhin:2010sv,Alekhin:2012vu}
and to $\Delta m_b(m_b)  = \pm 0.09 \GeV$ for bottom~\cite{Alekhin:2010sv}.

For the top-quark mass we rely predominantly on the inclusive cross section 
of the top-quark pair production, cf. Tab.~\ref{tab:data-tt}, which is known
completely to NNLO accuracy in QCD~\cite{Baernreuther:2012ws,Czakon:2012zr,Czakon:2012pz,Czakon:2013goa}.
Data on single-top quark are used as well, cf. Tab.~\ref{tab:data-inp}, 
but have a much larger uncertainty, see also~\cite{Alekhin:2016jjz}.
The high precision data on top-quark pair production from the LHC taken at 
the different center-of-mass energies $\sqrt{s}=5, 7, 8,$ and 13~TeV 
lead to the small experimental uncertainty of $\Delta m_t(m_t) = \pm 1.1 \GeV$ in Eq.~(\ref{eq:mcmbmt}).
This has to be compared to the effect of the scale variation on the extracted top-quark mass 
in the \msbar scheme in the NNLO predictions of the total cross section, 
which can be quantified as $\Delta m_t(m_t) \simeq \pm 0.7 \GeV$, see, e.g.~\cite{Alekhin:2013nda}. 

\begin{figure}[th!]
\centerline{
  \includegraphics[width=16.0cm]{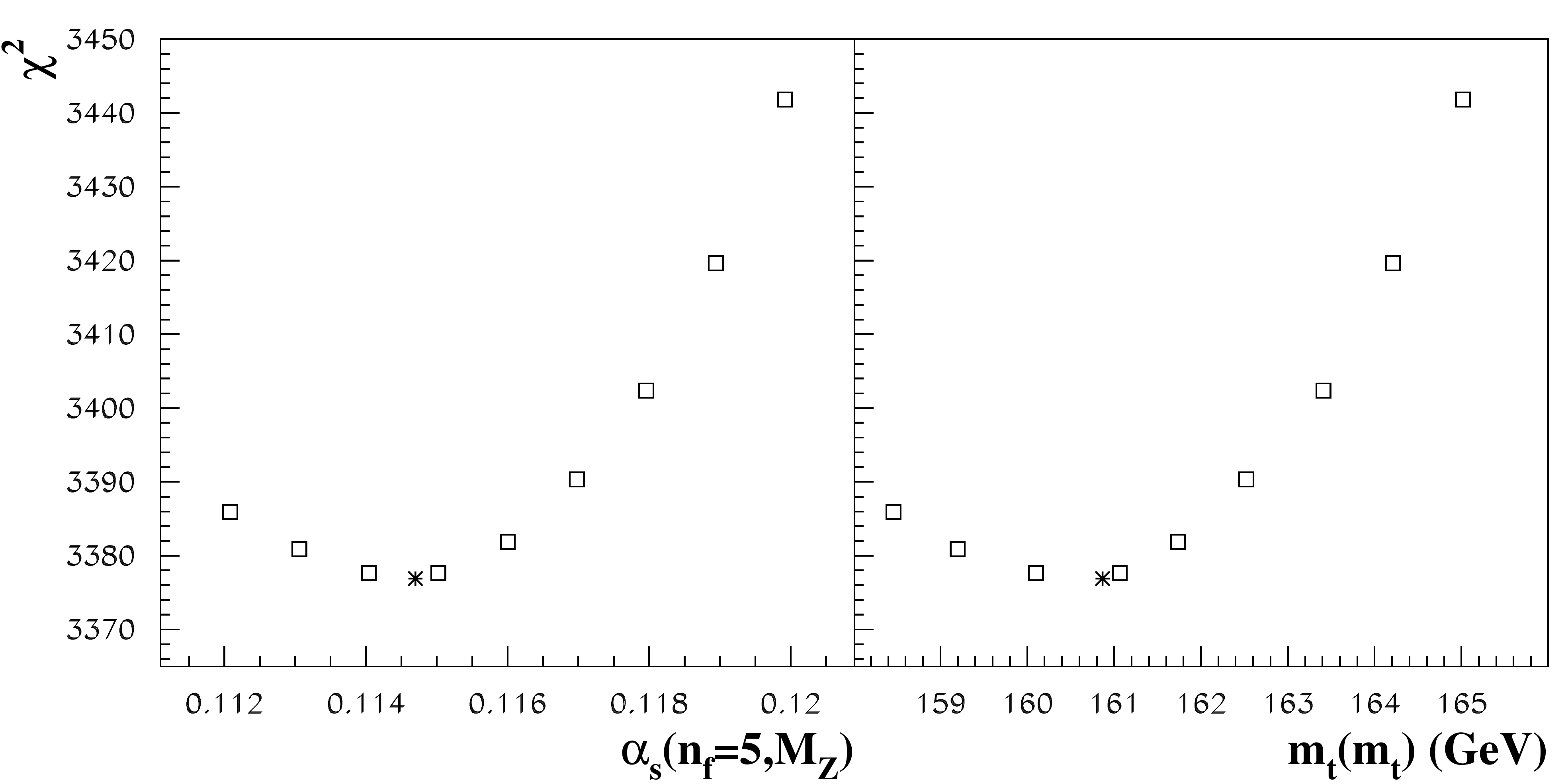}}
  \caption{\small
    \label{fig:chi-mt-alphas}
    The values of $\chi^2$ obtained in the variants of present analysis
    with $\alpha_s^{(n_f=5)}(M_z)$ fixed (squares)
    versus $\alpha_s^{(n_f=5)}(M_z)$ (left) 
    and $m_t(m_t)$ (right) in comparison with the best fit results (stars). 
  }
\end{figure}

\begin{figure}[th!]
\centerline{
  \includegraphics[width=8.75cm]{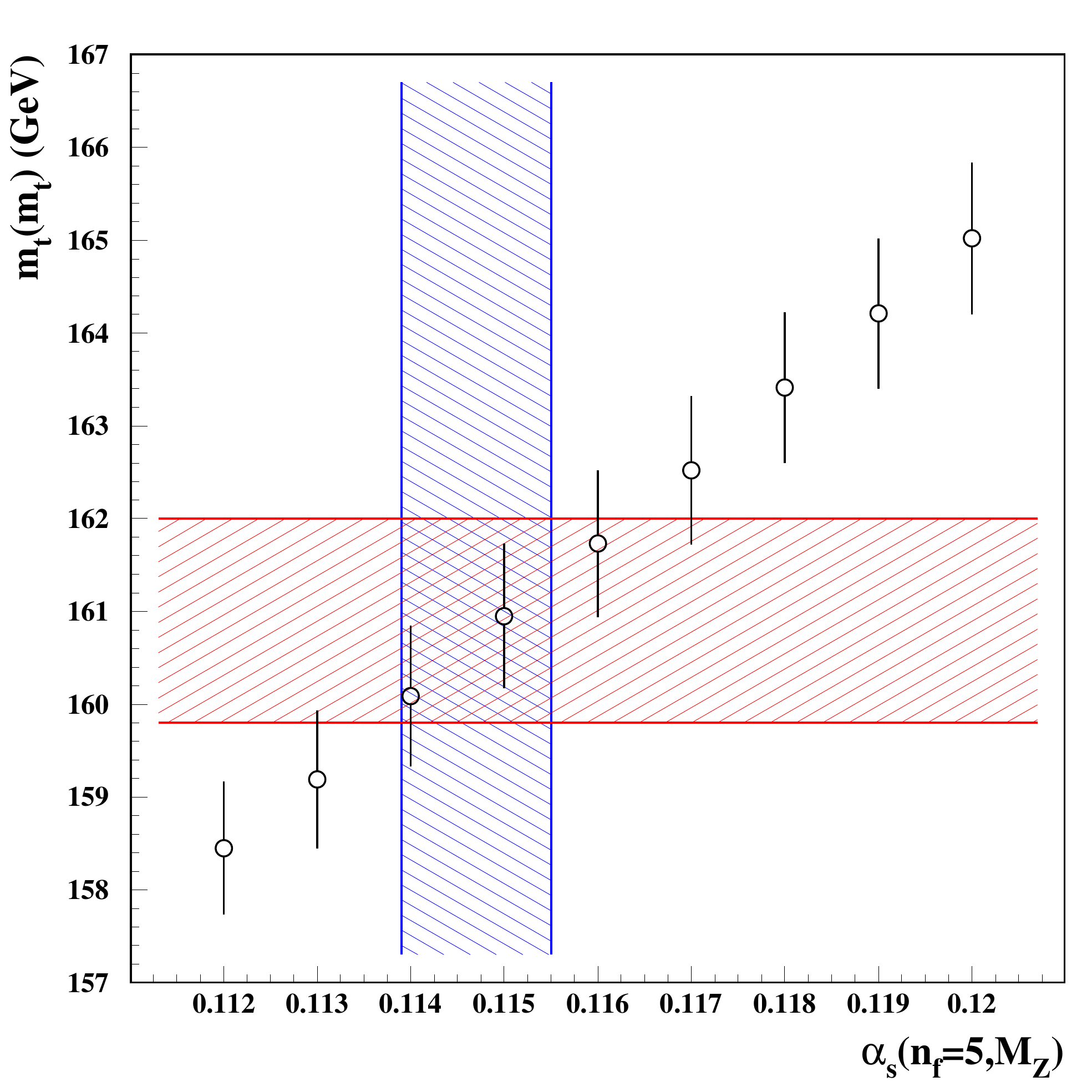}}
  \caption{\small
    \label{fig:scanalp}
    The \msbar value of the $t$-quark mass $m_t(m_t)$ obtained in the variants of 
    present analysis with the value of $\alpha_s^{(n_f=5)}(M_Z)$ fixed in comparison 
    with the $1\sigma$ bands for $m_t(m_t)$ and $\alpha_s^{(n_f=5)}(M_Z)$ 
    obtained in our nominal fit (left-tilted and right-tilted hatch, respectively). 
  }
\end{figure}

The cross section data on top-quark hadro-production in Tabs.~\ref{tab:data-tt} and \ref{tab:data-inp}
are, in fact, very sensitive to value of the mass $m_t$.
In order to illustrate this sensitivity of the data we present 
in Fig.~\ref{fig:chi-mt-alphas} the $\chi^2$ profiles versus 
the \msbar mass $m_t(m_t)$ and the strong coupling $\alpha_s^{(n_f=5)}(M_Z)$ 
obtained in the scan fit with the value of $\alpha_s^{(n_f=5)}(M_Z)$ 
spanning the range of $0.112 \div 0.120$. The profiles 
nicely demonstrate a good consistency of the top-quark mass determination
and also explain a correlation of the extracted value
of $m_t(m_t)$ with $\alpha_s$.
This correlation is explicitly displayed 
in Fig.~\ref{fig:scanalp} with the 
overlap region of the two bands for $m_t(m_t)$ and $\alpha_s$ indicating the $1\sigma$ interval.
Of course, the clear correlation shown is a direct consequence of the
parametric dependence of the total cross section for the top-quark pair
production on $\alpha_s$ and $m_t$. 
Of particular importance for $m_t$ are also the correlations with other fitted parameters, 
mostly for the gluon distribution -- 
a fact that has been addressed already in previous analyses~\cite{Alekhin:2013nda}. 
With improved precision of data on single-top production in the $t$-channel, 
the impact of $\alpha_s$ on the $m_t$ determination can be leveled, 
since that process is predominantly mediated by electroweak interactions 
and therefore sensitive to the light-flavor PDFs and the ratio
of $d/u$~\cite{Alekhin:2016jjz}.

The measured values of the charm and bottom masses in Eq.~(\ref{eq:mcmbmt}) 
can be confronted with the PDG values of 2016~\cite{Olive:2016xmw}
\begin{eqnarray}
\label{eq:PDGmcmb}
m_c(m_c)\bigr|_{\rm{PDG}} &=& 1.27 \pm 0.03 \GeV \, ,\nonumber\\
m_b(m_b)\bigr|_{\rm{PDG}} &=& 4.18\,^{\, + \,0.04}_{\, - \,0.03}\phantom{0} \GeV \, .
\end{eqnarray}
For charm, there is perfect agreement of the PDG value 
with our fit result in Eq.~(\ref{eq:mcmbmt}) which has a comparable uncertainty.
In the case of bottom, though, the fitted value in Eq.~(\ref{eq:mcmbmt}) 
carries a significantly larger error and comes out slightly lower than the PDG 
result in Eq.~(\ref{eq:PDGmcmb}) and only agrees at the level of 1 to 2$\sigma$.

For the top-quark mass in the \msbar scheme the PDG quotes 
\begin{eqnarray}
\label{eq:PDGmt}
m_t(m_t)\bigr|_{\rm{PDG}} &=& 160.0\,^{\, + \,4.8}_{\, - \,4.3} \GeV \, ,
\end{eqnarray}
which is compatible with Eq.~(\ref{eq:mcmbmt}), though 
it is subject to very large uncertainties as Eq.~(\ref{eq:PDGmt}) 
is only based on a single measurement performed at the Tevatron. 
Many other so-called direct mass measurements listed by the PDG cannot be
interpreted in quantum field theory, as they lack a well defined
renormalization scheme for the mass and need additional calibration 
of the extracted Monte-Carlo mass~\cite{Kieseler:2015jzh}. 

The on-shell scheme represents an alternative renormalization scheme 
for heavy-quark masses which is often used.
For the purpose of comparison with that scheme
 we convert the bottom- and top-quark masses in Eq.~(\ref{eq:mcmbmt}) 
to the pole masses. 
At NNLO we obtain  
\begin{eqnarray}
\label{eq:mbmt-pole}
m_b^{\rm pole} &=& 4.54\phantom{0}  \pm 0.13\phantom{0} \GeV \, ,\nonumber\\
m_t^{\rm pole} &=& 170.4 \pm 1.2\phantom{00} \GeV\, ,
\end{eqnarray}
using {\tt RunDec}~\cite{Chetyrkin:2000yt}.
For the running charm quark mass the conversion of $m_c(\mu_r)$ to the pole
mass definition at low scales $\mu_r \simeq 1.3 \GeV$ does not converge with
the known relations up to four-loop order in perturbative QCD~\cite{Marquard:2015qpa}.
Using $\alpha_s^{(n_f=5)}(M_Z)=0.1184$, for example, one obtains 
from the PDG central value in Eq.~(\ref{eq:PDGmcmb}) 
$m_c^{\rm pole}=1.47~\GeV$ at one loop,  
$m_c^{\rm pole}=1.67~\GeV$ at two loops,  
$m_c^{\rm pole}=1.93~\GeV$ at three loops,
and 
$m_c^{\rm pole}=2.39~\GeV$ at four loops~\cite{Marquard:2015qpa}.

The pole mass of the top-quark
in Eq.~(\ref{eq:mbmt-pole}) can also be compared with the PDG 
average, which quotes 
\begin{eqnarray}
\label{eq:PDGmtpole}
m_t^{\rm pole}\bigr|_{\rm{PDG}} &=& 174.2 \pm 1.4\phantom{0} \GeV\, .
\end{eqnarray}
So there is a clear tension between the values of $m_t^{\rm pole}$ in
Eqs.~(\ref{eq:mbmt-pole}) and (\ref{eq:PDGmtpole}).
The PDG average in Eq.~(\ref{eq:PDGmt}) is based on three experimental analyses of LHC data on the inclusive cross section. 
However, these analyses disregard the correlations 
of the top-quark mass with $\alpha_s$ and the PDF parameters, especially the gluon PDF,
in the theory predictions for the total cross section, as mentioned above and
illustrated in Fig.~\ref{fig:scanalp}.
We remark, that the PDG value given in Eq.~(\ref{eq:PDGmt})
for the \msbar scheme result $m_t(m_t)$ is consistent with $m_t^{\rm pole}$ in Eq.~(\ref{eq:PDGmtpole}) 
only within the large uncertainties of the former.

We briefly comment here on related studies, that have appeared in the literature. 
A determination of $m_c(m_c)$ from HERA data has been performed 
in Ref.~\cite{Bertone:2016ywq}
yielding $m_c(m_c) = 1.32 \pm 0.06 \GeV$ when fitting to DIS $c{\bar c}$-cross
section predictions at NLO in QCD in a fixed flavor-scheme. 
Within the reported uncertainties, this is compatible with Eq.~(\ref{eq:mcmbmt}).
The ZEUS collaboration has used its data~\cite{Abramowicz:2014zub} 
on DIS bottom-quark production to measure the bottom-quark mass at NLO in QCD 
and has reported $m_b(m_b) = 4.07 \pm 0.16 \GeV$ which is well compatible with
Eq.~(\ref{eq:mcmbmt}) within uncertainties. 
LHC data on heavy-flavor hadro-production cross sections measured by LHCb 
have been shown to additionally constrain PDF parameters \cite{Zenaiev:2015rfa} 
and those data can provide input for future improvements of the $m_b$ determination.

Finally, we comment on the treatment of heavy-quark masses in published PDF fits of other groups. 
With the exception of the JR14~\cite{Jimenez-Delgado:2014twa} fit, 
which implements the \msbar scheme for $m_c$, 
the available NNLO fits of other groups all use the on-shell scheme.
In addition, the value of $m_c$ is not fitted, but fixed beforehand, thereby 
disregarding any essential correlation, e.g., of $m_c$ with the gluon PDF.
In detail, the available NNLO fits use 
$m_c^{\rm pole}=1.3~\GeV$ in case of CT14~\cite{Dulat:2015mca},
$m_c^{\rm pole}=1.43~\GeV$ in case of HERAPDF2.0~\cite{Abramowicz:2015mha},
$m_c^{\rm pole}=1.4~\GeV$ in case of MMHT14~\cite{Harland-Lang:2014zoa},
and
$m_c^{\rm pole}=1.275~\GeV$ in case of NNPDF3.0~\cite{Ball:2014uwa},
covering a range of values significantly larger than the uncertainties 
obtained, e.g., in Eq.~(\ref{eq:mcmbmt}).
Some groups have performed dedicated studies of the charm and, sometimes, also
bottom-quark mass dependence in their analysis, although 
not always within their latest fit, see e.g., 
NNPDF2.1~\cite{Ball:2011mu}, CT10~\cite{Gao:2013wwa} or MMHT14~\cite{Harland-Lang:2015qea}.

Such low values for the pole mass of the charm quark as used by 
CT14~\cite{Dulat:2015mca}, MMHT14~\cite{Harland-Lang:2014zoa}
or NNPDF3.0~\cite{Ball:2014uwa} are indicators of post-truth science 
as they are not compatible with precision
determinations based on the rigorous application of quantum field theory 
as in Eq.~(\ref{eq:mcmbmt}) and with the world average in Eq.~(\ref{eq:PDGmcmb}).
That anomalously low values of $m_c^{\rm pole}$ lead to significant bias in phenomenology 
predictions even at LHC scales, in particular,
for the benchmark processes such as Higgs boson production in gluon-gluon 
fusion~\cite{Accardi:2016ndt}.

\section{Applications}
\label{sec:applications}

\subsection{Moments of PDFs and lattice results}
\label{sec:pdfmoms}

The values for the second moment of the 
 quark PDFs at the factorization 
scale $\mu^2 = 4$~GeV$^2$ for ABM11~\cite{Alekhin:2012ig}, ABM12~\cite{Alekhin:2013nda} and ABMP16 
as well as for CT14~\cite{Dulat:2015mca}, MMHT14~\cite{Harland-Lang:2014zoa} and NNPDF3.0~\cite{Ball:2014uwa} are provided in Tab.~\ref{tab:2ndmom}.
The quantity  $\langle x u_v(x)\rangle$ for the up-quark valence PDF 
is rather stable as it is mostly influenced by the data normalization.
The moments of distributions involving down-quark PDFs show a spread 
of the central values, though, 
which is much larger than the bands covering the $1\sigma$ uncertainty 
of the respective analysis.
In particular iso-spin asymmetries such as $x [u_v-d_v](x)$ or $xV(x)$ 
defined as 
\begin{eqnarray}
  \label{eq:V2ndmom}
  \langle xV(\mu^2) \rangle &=& 
  \int\limits_0^1 dx\, 
  x\left\{\left[u(x,\mu^2) + \bar{u}_{s}(x,\mu^2)\right]  
  -       
  \left[d(x,\mu^2) + \bar{d}_{s}(x,\mu^2)\right]\right\}
  \, , 
\end{eqnarray}
with $q \equiv q_{v} + q_{s}$ and $q=u,d$ are quite different among the
various analyses.

\input{table-moms}

\begin{figure}[t!]
  \centerline{
    \includegraphics[width=10.5cm]{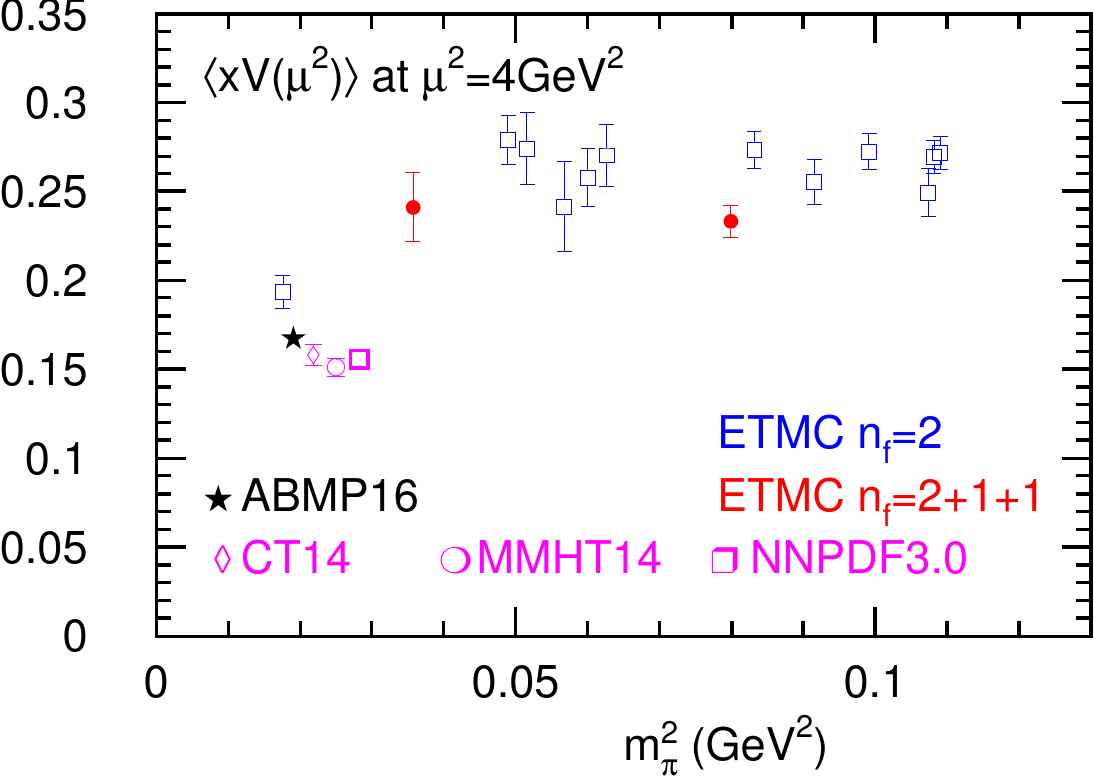}}
  \caption{\small
    \label{fig:2ndmom}
     Lattice computations of Refs.~\cite{Alexandrou:2011nr,Alexandrou:2013joa,Abdel-Rehim:2015owa,Alexandrou:2016tuo}
     for the second moment of the non-singlet distribution 
     $xV(\mu^2)$ at the scale $\mu^2=4$~GeV$^2$ 
     as a function of the pion mass squared, $m_\pi^2$, 
     including the uncertainties of the respective measurement
     compared to the results of ABMP16, CT14, MMHT14 and NNPDF3.0 PDFs given in Tab.~\ref{tab:2ndmom}. 
     The results for CT14, MMHT14 and NNPDF3.0 have been shifted in 
     $\Delta m_\pi$ with respect to ABMP16 for display purposes. 
  }
\end{figure}

At this point, we emphasize that the ABMP16 fit uses the widest set of 
Drell-Yan measurements from the LHC and Tevatron, which provide unique
constraints in the low-$x$ region and allow for a model-independent shape 
of the iso-spin asymmetry $x[\bar d - \bar u](x)$.
Comparing the ABM12~\cite{Alekhin:2013nda} and ABMP16 fits, the shifts in
the moments proportional to down-quark PDFs in Tab.~\ref{tab:2ndmom}
are reflected in a smaller value of ABMP16 for the cross section ratio $\sigma_{W^+}/\sigma_{W^-}$
for the production of $W^\pm$-bosons in the fiducial volume, shown in Fig.~\ref{fig:WandZ-ratio}.
In addition, also single-top quark production constrains the ratio $d/u$ and
the nice agreement of the 
ABMP16 fit with those data is demonstrated in Fig.~\ref{fig:single-top-pulls}.

The non-singlet distribution $xV(x)$ is also particularly suited for comparison to lattice simulations.
The results on $\langle xV(\mu^2) \rangle$ at the scale $\mu^2=4$~GeV$^2$ 
obtained with the ABMP16, CT14~\cite{Dulat:2015mca}, MMHT14~\cite{Harland-Lang:2014zoa} and NNPDF3.0~\cite{Ball:2014uwa} PDFs 
are compared with recent lattice computations as a function of the pion mass squared, $m_\pi^2$, in Fig.~\ref{fig:2ndmom} .
The lattice measurements of Refs.~\cite{Alexandrou:2011nr,Alexandrou:2013joa,Abdel-Rehim:2015owa,Alexandrou:2016tuo}
use set-ups with ($n_f=2$) and ($n_f=2+1+1$) for the number of flavors 
and low pion masses in the range $m_\pi = 133$~MeV to $329$~MeV, very close to
the physical point of $m_\pi = 138$~MeV.
There is a clear trend in the lattice results towards smaller values of $\langle xV(\mu^2) \rangle$
as the pion mass approaches the physical value.
For the lattice result taken at the lowest value in $m_\pi$ the compatibility 
with the determination from experimental data by ABMP16 is at the level of 2$\sigma$.
The PDFs of CT14, MMHT14 and NNPDF3.0, however, give significantly lower values in Tab.~\ref{tab:2ndmom}, 
so that the differences are at the level of 3 to 4$\sigma$, and compatibility is marginal.
As stressed above, experimental data clearly favors larger values of $\langle xV(\mu^2) \rangle$.

In summary, Fig.~\ref{fig:2ndmom} demonstrates nicely, that future high precision lattice
measurements in QCD with ($n_f=2+1+1$) flavors performed around the physical pion mass 
can potentially provide valuable constraints on the down-quark PDF and, most
importantly, on the the iso-spin asymmetry $x[\bar d - \bar u](x)$.

\subsection{Benchmark cross sections at the LHC}
\label{sec:bench}

Next, we proceed with cross section predictions for the 
Higgs boson production in
the gluon-gluon-fusion and hadro-production of the top-quark pairs at the LHC 
in order to benchmark the ABMP16 PDFs.

\input{table-higgs}

The gluon-gluon fusion process is the dominant production mechanism for the SM Higgs boson at the LHC
and the QCD radiative corrections to the inclusive cross section are particularly large at NLO, see e.g. Ref.~\cite{Spira:1995rr}. 
This has motivated systematic theory improvements in the effective theory 
to NNLO~\cite{Harlander:2002wh,Anastasiou:2002yz,Ravindran:2003um} 
and even to N$^3$LO accuracy~\cite{Anastasiou:2015ema,Anastasiou:2016cez}, 
by taking the limit of a large top-quark mass ($m_t \to \infty$) and integrating out the top-quark loop, 
while keeping the full $m_t$ dependence in the Born cross section.
The recent N$^3$LO results~\cite{Anastasiou:2015ema,Anastasiou:2016cez} 
demonstrate an apparent, if slow, convergence of the perturbative expansion.
The sensitivity to the choice of the renormalization and factorization scales $\mu_r$ and $\mu_f$ 
is greatly reduced and amounts to 3\% at N$^3$LO, which is also 
supported by estimates of the four-loop corrections~\cite{deFlorian:2014vta}.

In this situation, with perturbative QCD predictions of unprecedented accuracy 
for the hard scattering process being available, 
the largest remaining sources of uncertainties are the input value for the strong coupling constant $\alpha_s$ and the PDFs.
Indeed, the spread in the inclusive cross sections $\sigma(H)^{\rm NNLO}$ of
SM Higgs boson production at the LHC at NNLO~\cite{Harlander:2002wh,Anastasiou:2002yz,Ravindran:2003um} 
predicted by the PDF sets 
CJ15~\cite{Accardi:2016qay}, 
CT14~\cite{Dulat:2015mca},
HERAPDF2.0~\cite{Abramowicz:2015mha},
JR14~\cite{Jimenez-Delgado:2014twa},
MMHT14~\cite{Harland-Lang:2014zoa},
and 
NNPDF3.0~\cite{Ball:2014uwa} 
amounts to 13\% as documented in Ref.~\cite{Accardi:2016ndt}.
This is significantly larger than the PDF uncertainties 
quoted by the
individual sets and dominates by far over the residual scale uncertainty of the N$^3$LO QCD corrections.
A detailed comparison of those PDFs
has illustrated how these differences arise as a consequence of specific theory assumptions 
such as tuned values of $m_c$ used in the individual fits, cf. Sec.~\ref{sec:hq}.

In Tab.~\ref{tab:higgs} we present the results on $\sigma(H)^{\rm NNLO}$ 
for the LHC at $\sqrt{s}=13$~TeV in the effective theory (i.e., in the limit of $m_t \gg m_H$) 
with parameter choices $m_H=125.0$~GeV for the SM Higgs boson mass, 
scales $\mu_r = \mu_f = m_H$, and using the PDF sets ABM12~\cite{Alekhin:2013nda}, ABMP15~\cite{Alekhin:2015cza} and ABMP16.
The value $m_t^{\rm pole}=172.5$~GeV in Tab.~\ref{tab:higgs} has been chosen
for compatibility with the recent benchmark study in \cite{Accardi:2016ndt}. 
The quoted uncertainties are given at the 1$\sigma$ confidence level and the results for $\sigma(H)^{\rm NNLO}$ 
employ either the nominal value of the strong coupling constant $\alpha_s^{(n_f=5)}(M_Z)$
at NNLO, or fixed values of $\alpha_s^{(n_f=5)}(M_Z)=0.115$ and $\alpha_s^{(n_f=5)}(M_Z)=0.118$,
while keeping the correlation with the PDF parameters.

We also quote $\sigma(H)^{\rm NNLO}$ with uncertainties obtained with the set PDF4LHC15~\cite{Butterworth:2015oua}
which has been obtained as some average of CT14, MMHT14 and NNPDF3.0 
using the same fixed value of $\alpha_s^{(n_f=5)}(M_Z)=0.1180$ at NLO and at NNLO independent 
of the order of perturbation theory and, thereby, disregarding correlations.
This set has been employed in the recent study~\cite{deFlorian:2016spz} of
the LHC Higgs Cross Section Working Group which quotes 
a combined PDF and $\alpha_s$ uncertainty in the inclusive cross section
as small as $3.2\%$.
In view of the bias incurred with a fixed value of $\alpha_s^{(n_f=5)}(M_Z)=0.118$ 
and the benchmark studies of~\cite{Accardi:2016ndt} this
uncertainty is underestimated.

\bigskip

\input{table-ttbar}

The cross section for hadro-production of top-quark pairs at the LHC is another benchmark process.
As described in Sec.~\ref{sec:data}, the present analysis is 
based on the 
large sample of the top-quark pair and for single-top quark production data 
(cf. Tabs.~\ref{tab:data-tt} and \ref{tab:data-inp} and 
 Figs.~\ref{fig:ttbar-pulls} and \ref{fig:single-top-pulls}).

In Tab.~\ref{tab:ttbar} we present results for the inclusive cross section 
$\sigma(t{\bar t})^{\rm NNLO}$ for top-quark pair production with the theory description detailed
in Sec.~\ref{sec:theory}, i.e., NNLO accuracy in QCD~\cite{Baernreuther:2012ws,Czakon:2012zr,Czakon:2012pz,Czakon:2013goa}
using {\tt Hathor}~\cite{Aliev:2010zk}. 
We apply the PDF sets ABM12~\cite{Alekhin:2013nda}, ABMP15~\cite{Alekhin:2015cza} and ABMP16
with the top-quark mass in the \msbar scheme, $m_t(m_t)=160.9$~GeV, and scales
$\mu_r = \mu_f = m_t(m_t)$ for various center-of-mass energies of the LHC, 
$\sqrt{s}=5, 7, 8,$ and 13~TeV.
The uncertainties quoted represent the combined symmetric 1$\sigma$ uncertainty 
$\Delta \sigma({\rm PDF}+\alpha_s)$ arising from the variation 
of the PDF parameters and of $\alpha_s$ in 28 PDF sets for ABM12 or ABMP15, 
and $\Delta \sigma({\rm PDF}+\alpha_s+m_t)$ originating from 29 sets including
also the variation of $m_t$ in case of the ABMP16 PDFs.

The ABMP16 value of $\alpha_s$ has shifted upwards by about 1$\sigma$ as
compared to the ABM12 and ABMP15 one, i.e.,  $\alpha_s^{(n_f=5)}(M_Z)=0.1147$ 
for the former and $\alpha_s^{(n_f=5)}(M_Z)=0.1132$ for the later, 
cf. the discussion in Sec.~\ref{sec:alps}.
Correspondingly, the cross section predictions in Tab.~\ref{tab:ttbar}
display a systematic trend in upward shifts at the level of 1 to 2$\sigma$
in the associated uncertainties.
Note also, that the overall cross section uncertainty $\Delta \sigma({\rm PDF}+\alpha_s+m_t)$ for
ABMP16 is significantly reduced compared to previous fits thanks 
to the high precision data in Tab.~\ref{tab:data-tt}.

\subsection{Stability of the electroweak vacuum}
\label{sec:vacuum}

Recent high-precision measurements of the Higgs boson mass
$m_H$ quote 
\begin{equation} 
\label{eq:mhpdg}
m_H \,=\, 125.09 \pm 0.24 \GeV
\, 
\end{equation} 
as a very accurate average~\cite{Aad:2015zhl}.
Due to an intriguing coincidence of the value for $m_H$ in Eq.~(\ref{eq:mhpdg})
and the measured masses of all other SM particles, 
the Higgs potential can possibly develop a second minimum 
at field values as large as the Planck scale $M_{Pl} \simeq 10^{19} \GeV$ 
in addition to the one we live in, which corresponds to the
vacuum expectation value $v=246~\GeV$.
If realized in Nature, this would imply stability of the SM up to 
the high scales where unification with gravity is expected.
On the contrary, the occurrence of an instability of the electroweak vacuum at some large scales 
above the terascale but below $M_{Pl}$ indicates the breakdown of the SM and
invokes the necessity for new physics. 
Thus, it is important to test the running of the Higgs boson self-coupling
$\lambda$ in the SM up to large scales with the help of renormalization group analyses currently available  
to three-loop accuracy~\cite{Mihaila:2012fm} supplemented by the necessary
matching conditions at the two-loop level~\cite{Bezrukov:2012sa,Degrassi:2012ry}.

The evolution of the renormalization group equations from 
scales ${\cal O}(100)\GeV$ to $M_{Pl}$ critically depends on the 
input values for the SM parameters, in particular the top-quark mass and the
strong coupling constant $\alpha_s$.
Having determined both of them simultaneously including correlations, 
cf. Sec.~\ref{sec:hq}, 
we are in a position to update previous work~\cite{Alekhin:2012py} 
by investigating consequences of such a correlation 
for the study of the Higgs potential at large scales.
The condition for the vacuum stability to hold at $M_{Pl}$ 
can be formulated as a lower bound on the mass of the Higgs boson
as follows~\cite{Buttazzo:2013uya}
\begin{equation} 
\label{eq:higgs}
m_H \,=\, 
129.6 \GeV
+ 1.8 \times \left( \frac{m_t^{\rm pole} - 173.34~\GeV}{0.9} \right) 
- 0.5 \times \left( \frac{ \alpha_s^{(n_f=5)}(M_Z) - 0.1184}{0.0007} \right) \GeV
\pm 0.3~\GeV
\, ,
\end{equation} 
where $m_t$ and $\alpha_s$ are to be taken in the on-shell and \msbar schemes,
respectively, and the uncertainty of $\pm 0.3 \GeV$ 
appears due to missing higher-order corrections.
A similar condition based on a manifestly gauge-independent approach 
including two-loop matching and three-loop renormalization group evolution
has been reported in Ref.~\cite{Bednyakov:2015sca}.

\begin{figure}[t!]
\centerline{
  \includegraphics[width=8.75cm]{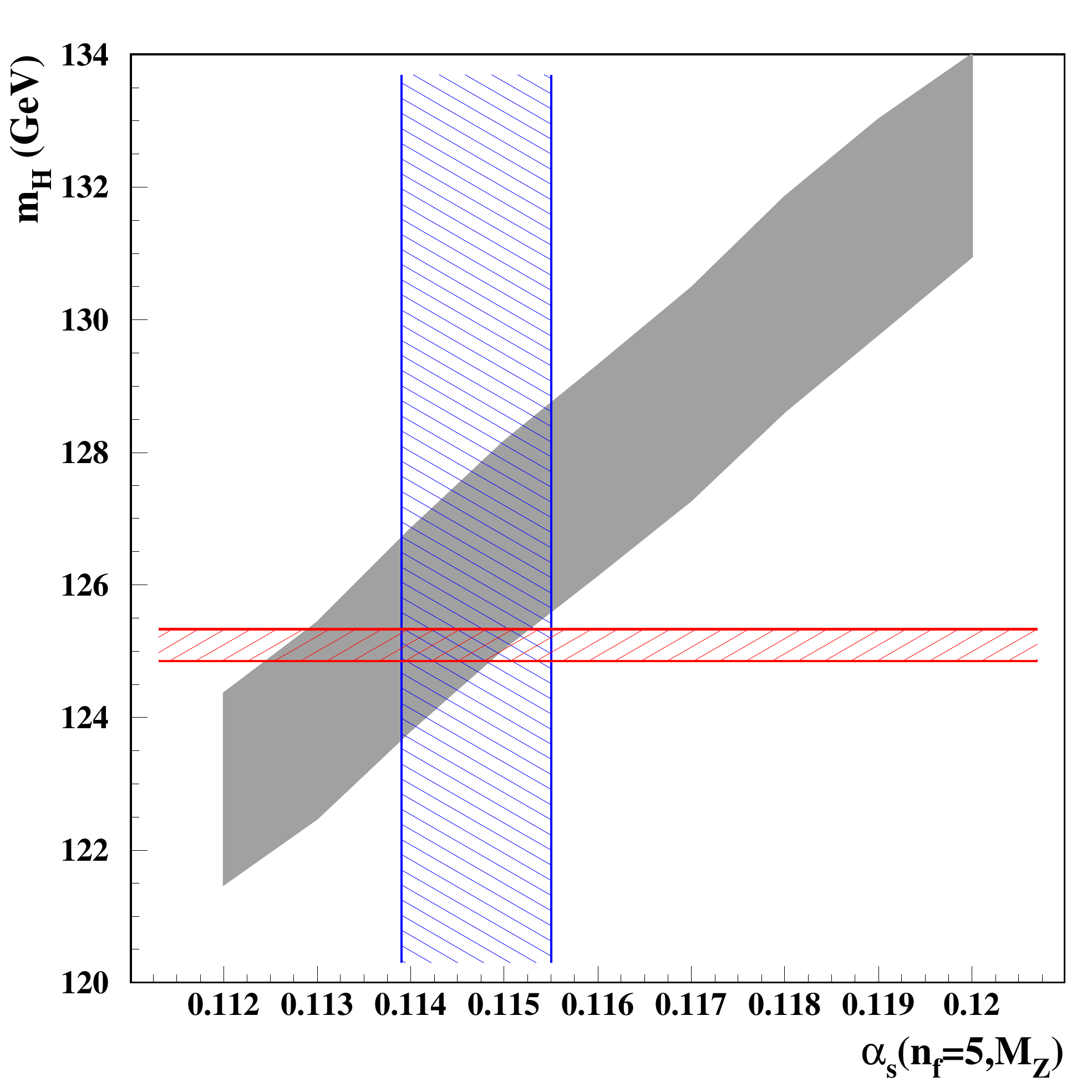}}
  \caption{\small
    \label{fig:stab}
    The value of Higgs-boson mass $m_H$ computed according to the 
    condition in Eq.~(\ref{eq:higgs}) for vacuum stability at $M_{Pl}$ 
    using the values of $m_t$ with their uncertainties 
    obtained  in the present analysis by scanning 
    $\alpha_s^{(n_f=5)}(M_Z)$ in the range of $0.112 \div 0.120$
    and disregarding the theoretical uncertainty of $\pm 0.3 \GeV$ (gray band). 
    For comparison the $1\sigma$ bands for $\alpha_s^{(n_f=5)}(M_Z)$ 
    from the nominal fit and the value of $m_H$ in Eq.~(\ref{eq:mhpdg})
    (left-tilted and right-tilted hatch, respectively) are shown.
}
%
\vspace*{10mm}
\centerline{
  \includegraphics[width=10.5cm]{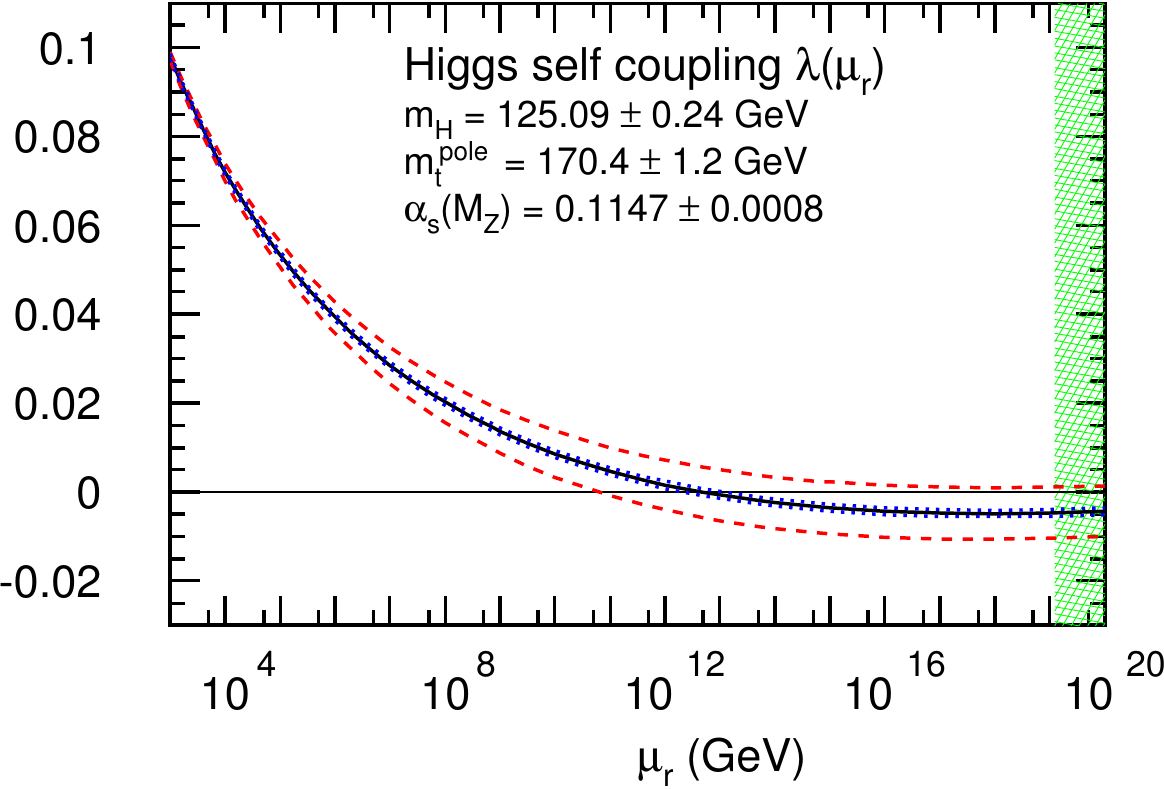}}
  \caption{\small
    \label{fig:lambda-rge}
    The renormalization group evolution of the Higgs boson self-coupling $\lambda$ as a function of scale $\mu_r$. 
    The dashed (red) lines denote the combined $1\sigma$ uncertainty for
    $\alpha_s^{(n_f=5)}(M_Z)$ and $m_t^{\rm pole}$ and  
    the dotted (blue) lines  the $1\sigma$ uncertainty in the value of $m_H$ in Eq.~(\ref{eq:mhpdg}).
    The range of scales $\mu_r \ge M_{Pl}$ is indicated by the hatched (green)
    band on the right.  
  }
\end{figure}

In Fig.~\ref{fig:stab} we display the value of $m_H$ according to Eq.~(\ref{eq:higgs}) 
evaluated as a function of $\alpha_s^{(n_f=5)}(M_Z)$.
The corresponding values of the pole mass for the top-quark 
$m_t^{\rm pole}$ are derived from the values for $m_t(m_t)$ in the \msbar scheme 
which have been obtained in the variants of our analysis 
with the values of $\alpha_s^{(n_f=5)}(M_Z)$ 
scanned in the range of $0.112 \div 0.120$. 
In doing so, the uncertainties in the fitted values for $m_t$ and 
the correlation between $m_t$ and $\alpha_s$ have been taken into account, cf. Fig.~\ref{fig:scanalp}. 
The error band on $m_H$ derived 
in this way overlaps with the one obtained from a direct measurement of $m_H$
for the value of $\alpha_s^{(n_f=5)}(M_Z)$ close to our determination in Eq.~(\ref{eq:asmt}).
This implies, that the values of $m_t$ and $\alpha_s$ obtained in our analysis 
are consistent with the condition of vacuum stability at $M_{Pl}$.

In a complementary way this is illustrated
in Fig.~\ref{fig:lambda-rge} showing the running of the Higgs boson self-coupling
$\lambda(\mu_r)$  in full three-loop accuracy and with
 $\alpha_s$ and $m_t$ obtained in our analysis as the input 
parameters as well as the 
2016 PDG values~\cite{Olive:2016xmw} for the other SM masses and couplings.
The computation has been performed
 with the code {\tt mr}, which implements matching and running of the
SM parameters~\cite{Kniehl:2016enc}.
Clearly, a vanishing Higgs self-coupling $\lambda=0$ at $M_{Pl}$ remains a scenario
which is compatible with the current values of $\alpha_s$, $m_t$ and $m_H$
within their 1$\sigma$ uncertainties. 
In addition, as follows from our analysis the value of 
$\lambda(\mu_r)$ remains strictly positive up to renormalization scales 
$\mu_r \sim {\cal O}(10^{12}\GeV)$, so that no new physics needs to be invoked
in order to stabilize the electroweak vacuum.

\subsection{LHAPDF library}
\label{sec:lhapdf}

\input{table-asmz-mq-parameters}

The PDFs obtained in the present analysis  are provided in the form of grids
accessible 
with the {\tt LHAPDF} library (version 6)~\cite{Buckley:2014ana}
and available for download under {\tt http://projects.hepforge.org/lhapdf}.
The PDFs for a fixed number of flavors, $n_f=3, 4$ and $5$, at NNLO 
\begin{verbatim}
      ABMP16_3_nnlo (0+29),
      ABMP16_4_nnlo (0+29),
      ABMP16_5_nnlo (0+29),
\end{verbatim}
consist of the central fit (set 0) and additional 29 sets for the combined symmetric uncertainties
on the PDF parameters and on the values of $\alpha_s$ and the heavy-quark masses.\footnote{
Corresponding data grids for the {\tt LHAPDF} library (version 5)~\cite{Whalley:2005nh} are available from the authors upon request.}
The quoted PDF uncertainties are calculated in the standard manner and correspond to the $\pm 1\sigma$-variation.
The PDF uncertainty $\Delta \sigma_{PDF}$ for a given cross section $\sigma_{0}$ is 
then computed according to 
\begin{equation}
  \label{eq:pdferr}
  \Delta \sigma_{PDF} \,=\, 
  \sqrt {\sum\limits_{k=1}^{n_{PDF}} \, (\sigma_{0} - \sigma_{k} )^2}
  \, ,
\end{equation}
where $\sigma_{k}$ is obtained by using the $k$-th PDF set and $n_{PDF}=29$.

The PDF set for $n_f=3$, {\tt ABMP16\_3\_nnlo}, with three light-quark
flavors is valid at all perturbative scales $\mu^2 \gtrsim 1$~GeV$^2$. 
In contrast, PDFs with a fixed number of flavors, $n_f=4$ or $5$,
are only meaningful at scales $\mu^2 \gg m_c^2$ and $\mu^2 \gg m_b^2$, respectively.
Therefore, minimal cuts in $\mu^2 \ge 3$~GeV$^2$ for the grid {\tt ABMP16\_4\_nnlo}
and $\mu^2 \ge 20$~GeV$^2$ for {\tt ABMP16\_5\_nnlo} have been imposed
 and the PDF grids are not available below these cuts.
Note, however, that by default the {\tt LHAPDF} library (version 6)~\cite{Buckley:2014ana}
extrapolates the PDFs also to kinematics outside those covered by the grid in $x$ and $\mu^2$.
Therefore, the grids {\tt ABMP16\_4\_nnlo} and {\tt ABMP16\_5\_nnlo}
at low values of $\mu^2$ should be used with a particular care.

We also remark, that the PDF sets for $n_f=3, 4$ and $5$ flavors at NNLO use the strong coupling
$\alpha_s$ correspondingly, i.e., the couplings are taken in the scheme 
$\alpha_s^{(n_f=3)}$, $\alpha_s^{(n_f=4)}$ and $\alpha_s^{(n_f=5)}$ and need to be
related by the standard decoupling relations in QCD.
Since the heavy-quark masses $m_c(m_c)$, $m_b(m_b)$ and $m_t(m_t)$ 
determined in our analysis are correlated with the PDF parameters 
they are different for each of the 29 PDF sets.
Therefore, a self-consistent prediction of PDF uncertainties
on the cross sections involving heavy quarks should be computed 
by varying the respective heavy-quark masses simultaneously with the PDFs in the loop 
Eq.~(\ref{eq:pdferr}). The corresponding heavy-quark mass values can be easily
retrieved within the {\tt LHAPDF} library framework.  
The values of $\alpha_s^{(n_f=5)}(M_Z)$ 
and the heavy-quark masses encoded in the ABMP16 grids are listed 
in Tab.~\ref{tab:parameters} for reference.
In addition, we also provide the bottom- and the top-quark pole masses, 
$m_b^{\rm pole}$ and $m_t^{\rm pole}$, 
obtained using {\tt RunDec}~\cite{Chetyrkin:2000yt} to be employed in 
corresponding computations with the on-shell scheme.
Specifically, for the central ABMP16 set the values 
$m_b^{\rm pole} = 4.537~\GeV$ and $m_t^{\rm pole} = 170.37~\GeV$ should be used.

Finally, for detailed studies of the parametric dependence of LHC observables on the strong
coupling constant $\alpha_s$ we provide the $n_f=5$ flavor NNLO PDF grids 
with the central value of $\alpha_s^{(n_f=5)}(M_Z)$ fixed. 
In total, there are 9 sets covering the range $\alpha_s^{(n_f=5)}(M_Z)=0.112 
\div 0.120$ with a spacing of 0.001.
These sets are denoted
\begin{verbatim}
      ABMP16als112_5_nnlo (0+29),
      ABMP16als113_5_nnlo (0+29),
      ABMP16als114_5_nnlo (0+29),
      ABMP16als115_5_nnlo (0+29),
      ABMP16als116_5_nnlo (0+29),
      ABMP16als117_5_nnlo (0+29),
      ABMP16als118_5_nnlo (0+29),
      ABMP16als119_5_nnlo (0+29),
      ABMP16als120_5_nnlo (0+29),
\end{verbatim}
with the value of $\alpha_s^{(n_f=5)}(M_Z)$ fixed as indicated in the file 
names and additional 29 sets for the combined symmetric uncertainties.\footnote{For the purpose of technical consistency 
the value of $\alpha_s$ in these grids is still considered as a formal parameter, however, 
with greatly suppressed uncertainty.}

\section{Conclusions}
\label{sec:concl}

The new ABMP16 PDFs presented have been determined from a global fit to the most recent experimental data 
on the basis of theory predictions at NNLO in perturbative QCD.
Essential ingredients of the ABMP16 analysis are the inclusion of 
the final HERA DIS combination data from run I+II, 
new data sets from the fixed-target DIS experiments CHORUS and NOMAD, 
the recent LHC and Tevatron DY production data 
as well as an exhaustive sample of data for the top-quark hadro-production. 
The new combined DIS data from run I+II at HERA and the fixed-target
experiments CHORUS and NOMAD in combination with the precise DY data from
Tevatron and the LHC allow for a very accurate determination of the valence quark distributions
and the flavor separation of the up- and down-quarks in a wide range of parton
momentum fractions, $x \simeq 10^{-4}$ to $0.9$. 
The accuracy of the down-quark distribution from these data is found to be comparable with the one 
from the DIS deuteron data, used previously for the ABM12 PDFs.
Therefore the latter have been discarded from the present
analysis avoiding additional uncertainties related to 
modeling of nuclear effects in deuterium targets.
Moreover, the addition of the recent charm di-muon production data from NOMAD and CHORUS
leads to an improved accuracy of the strange quark content in the proton
compared to the earlier ABM12 analysis. 
The moments of light-quark PDFs obtained with all these improvements 
are found to be in very good agreement with recent lattice measurements at the physical pion mass.

\smallskip

The DIS heavy-quark production data and the Tevatron and LHC data on the inclusive single-top 
and top-quark pair production have been used to determine the heavy-quark
masses which are considered as free parameters in the present analysis and 
fitted simultaneously with the PDFs and the strong coupling 
constant $\alpha_s$ to preserve all correlations between those parameters.
Specifically, we determine the heavy-quark masses 
$m_c(m_c)$, $m_b(m_b)$ and $m_t(m_t)$, 
in the \msbar scheme providing better perturbative stability both for 
the DIS and hadronic heavy-quark production. 

\smallskip

The theory predictions for the hard-scattering processes in the ABMP16 analysis 
maintain NNLO accuracy in QCD and employ the fixed-flavor number scheme, which 
has been shown to provide an excellent description of the existing DIS data.
In addition, the theory framework features a number of new improvements.
It contains new approximations for the NNLO Wilson coefficients
in the description of the DIS heavy-quark production 
as well as advances in predictions of single-top production in the $s$-channel
to approximate NNLO accuracy.
The ABMP16 analysis is also based on a refined treatment of higher twist contributions 
to the power corrections in DIS which extends to small values of $x$.
In particular, it has been clearly demonstrated that higher twist terms in DIS
are required for the kinematic coverage of the data analyzed.
Moreover, higher twist terms cannot be entirely eliminated by the cut 
on hadronic invariant mass $W^2\gtrsim 12~{\rm GeV}^2$ 
in DIS proposed in the literature.

As a result of the new theory improvements and newly added data in the current analysis, 
an updated value of the strong coupling constant $\alpha_s$ has been determined
in the \msbar scheme for $n_f = 5$ at NNLO in QCD accuracy. 
It yields a value of $\alpha_s^{(n_f=5)}(M_Z) =  0.1147 \pm 0.0008$, which 
represents a shift upwards in the central value by $1\sigma$ compared to the
previous ABM12 analysis due to the impact of the new combined HERA run I+II data.
For the heavy-quark masses we find at NNLO in the \msbar scheme 
$m_c(m_c) = 1.252 \pm 0.018 \GeV$, $m_b(m_b) = 3.84 \pm 0.12 \GeV$ and $m_t(m_t) = 160.9 \pm 1.1 \GeV$, respectively.
This corresponds to a pole mass $m_t^{\rm pole} = 170.4 \pm 1.2 \GeV$ for the top-quark.
Furthermore, a renormalization group analysis of the SM couplings accurate to
three-loop order shows, that the obtained values of $\alpha_s$ and $m_t^{\rm pole}$ employed with account of their correlations 
are compatible with the requirement of a stable electroweak vacuum 
up to the Planck scale ${\cal O}(10^{19}\GeV)$, 
thereby diminishing the need to introduce new physics.

\smallskip

In summary, the new ABMP16 analysis and the corresponding PDFs pave the way
for precision predictions at the LHC in run II, so that 
future precision measurements can be confronted with theory computations at highest accuracy. 
With no doubt, this will offer the chance for further improvements in the description of
the parton content of the proton. 
At the same time, benchmark comparisons with other PDF sets 
published in the literature will allow to test and, possibly, 
to eliminate the remaining underlying model assumptions and tunes in those fits. 
This will, finally, consolidate the understanding of PDFs and QCD at high scales.

\subsection*{Acknowledgments}
We are grateful to Daniel Britzger, Maria Vittoria Garzelli, Karl Jansen, 
Uta Klein, Mikhaylo Lisovyi, Frank Petriello, Andrey Pikelner,
Klaus Rabbertz and Andreas Vogt for useful discussions.

This work has been supported by Bundesministerium f\"ur Bildung und Forschung (contract 05H15GUCC1) 
and by the European Commission through PITN-GA-2012-316704 ({\it HIGGSTOOLS}). 

\appendix
\renewcommand{\theequation}{\ref{sec:appA}.\arabic{equation}}
\setcounter{equation}{0}
\renewcommand{\thefigure}{\ref{sec:appA}.\arabic{figure}}
\setcounter{figure}{0}
\renewcommand{\thetable}{\ref{sec:appA}.\arabic{table}}
\setcounter{table}{0}
\section{Correlations}
\label{sec:appA}
In Tabs.~\ref{tab:cor1}--\ref{tab:cor3} 
we present the covariance matrix for the correlations of the fit parameters of ABMP16 
discussed in Sec.~\ref{sec:pdfs}, cf. Tab.~\ref{tab:fitvalues} and
Eqs.~(\ref{eq:asmt}) and (\ref{eq:mcmbmt}) for the strong coupling $\alpha_s$ and the heavy-quark masses.
Note, that in Eq.~(\ref{eq:asmt}) we quote $\alpha_s^{(n_f=5)}(M_Z)$, whereas 
below the correlations are given for $\alpha_s^{(n_f=3)}(\mu_0)$ with $\mu_0=1.5~\GeV$.

\input{table-cor}

\newpage
\cleardoublepage


\end{document}

%% file: table-slac-data1.tex
\begin{table}[t!]
\begin{center}
\begin{tabular}{|c|c|c|c|c|}
\hline
Experiment&Process &Beam energy & Reference & Normalization  \\
&&(GeV)&&   \\
\hline
SLAC-49a&$ep \rightarrow e X$&$7\div 20$&~\cite{Bodek:1979rx,Whitlow:1990gk}  & $1.001(11)$   \\
\hline
SLAC-49b&$ep \rightarrow e X$&$4.5\div 18$&~\cite{Bodek:1979rx,Whitlow:1990gk}  & $1.010(15)$   \\
\hline
SLAC-87&$ep \rightarrow e X$&$8.7\div 20$&~\cite{Bodek:1979rx,Whitlow:1990gk}   & $1.012(11)$  \\
\hline
SLAC-89b&$ep \rightarrow e X$&$6.5\div 19.5$&~\cite{Mestayer:1982ba,Whitlow:1990gk}  & $1.000(11)$  \\
\hline
BCDMS&$\mu p \rightarrow \mu X$&$100\div 280$&~\cite{Benvenuti:1989rh} & 0.976(7)  \\
\hline
NMC&$\mu p \rightarrow \mu X$&90&~\cite{Arneodo:1996qe} & 0.993(13)  \\
&&120&  &  1.011(12) \\
&&200&  & 1.022(12)  \\
&&280&  & 1.012(12)  \\
\hline
\end{tabular}
\caption{
\label{tab:SLAC}
\small
The values of fitted normalization factors for the fixed-target DIS 
data sets used in the present analysis with the uncertainties quoted in 
parentheses. 
}
\end{center}
\end{table}

%% file: table-dis-dy.tex
\begin{table}[th!]
 \footnotesize
 \centering
  \begin{tabular}{|l|c|c|c|l|c|}
    \hline
    Experiment & Beam ($E_b$) or center-    &  $\mathcal{L}$  & Process  & \multicolumn{1}{c|}{Kinematic cuts used in the present analysis} & Ref. \\
               & of-mass energy $(\sqrt{s})$ &  (1/fb)         &          &  \multicolumn{1}{c|}{(cf. orginal references for notations)}                                      & \\ \cline{1-6} 
    \hline
    \multicolumn{1}{l}{\bf DIS} & \multicolumn{1}{c}{} & \multicolumn{1}{c}{} &\multicolumn{1}{c}{} & \multicolumn{1}{c}{} & \multicolumn{1}{c}{} \\ \cline{1-6}
    \hline
\hline
    HERA I+II  & $\sqrt s =0.225\div0.32$ & 0.5 & $e^{\pm}p \rightarrow e^{\pm}X   $ & $2.5 \le Q^2 \le 50000$ GeV$^2$, $2.5 \cdot 10^{-5} \le x \le 0.65$ & \cite{Abramowicz:2015mha} \\
               &   TeV &     & $e^{\pm}p \rightarrow \overset{(-)}{\nu} X$ & $200 \le Q^2 \le 50000$ GeV$^2$, $1.3 \cdot 10^{-2} \le x \le 0.40$ & \\
    \hline
\hline
    BCDMS      & $E_b=$100$\div$280 GeV & & $\mu^+ p \rightarrow \mu^+ X$ & $7 < Q^2 < 230$ GeV$^2$, $0.07 \leq x \leq 0.75$ & \cite{Benvenuti:1989rh} \\ 
\hline
    NMC        & $E_b=$90$\div$280 GeV & & $\mu^+ p \rightarrow \mu^+ X$ & $2.5 \le Q^2 < 65$ GeV$^2$, $0.009 \le x < 0.5$ & \cite{Arneodo:1996qe} \\ 
    \hline
     SLAC-49a & $E_b=$7$\div$20 GeV & & $e^-p \rightarrow e^-X$ & $2.5 \le Q^2 < 8$ GeV$^2$, $0.1 < x < 0.8$, $W \ge $1.8 GeV& \cite{Bodek:1979rx} \\

     & & & & & \cite{Whitlow:1990gk} \\
\hline
    SLAC-49b & $E_b=$4.5$\div$18 GeV & & $e^-p \rightarrow e^-X$ & $2.5 \le Q^2 < 20$ GeV$^2$, $0.1 < x < 0.9$, $W \ge $1.8 GeV & \cite{Bodek:1979rx} \\
     & & & &  & \cite{Whitlow:1990gk} \\
\hline
    SLAC-87  & $E_b=$8.7$\div$20 GeV & & $e^-p \rightarrow e^-X$ &$2.5 \le Q^2 < 20$ GeV$^2$, $0.3 < x < 0.9$, $W \ge $1.8 GeV & \cite{Bodek:1979rx}\\ 
     & & & &  & \cite{Whitlow:1990gk} \\

\hline
     SLAC-89b & $E_b=$6.5$\div$19.5 GeV & & $e^-p \rightarrow e^-X$ & $2.5 \le Q^2 \le 19$ GeV$^2$, $0.17 < x < 0.9$, $W \ge $1.8 GeV& \cite{Mestayer:1982ba} \\
     & & & &  & \cite{Whitlow:1990gk} \\
\hline 
   \multicolumn{2}{l}{\bf DIS heavy-quark production} & \multicolumn{1}{c}{} &\multicolumn{1}{c}{} & \multicolumn{1}{c}{} & \multicolumn{1}{c}{} \\ 
\hline
    \hline
    HERA I+II & $\sqrt s=$0.32 TeV&  & $e^{\pm}p\rightarrow e^{\pm} c  X  $ & $2.5 \le Q^2 \le 2000$ GeV$^2$, $2.5 \cdot 10^{-5} \le x \le 0.05$ & \cite{Abramowicz:1900rp} \\
    \hline
    H1 & $\sqrt s=$0.32 TeV& 0.189 & $e^{\pm}p \rightarrow e^{\pm} b  X   $ & $5 \le Q^2 \le 2000$ GeV$^2$, $2 \cdot 10^{-4} \le x \le 0.05$ & \cite{Aaron:2009af} \\
    \hline
    ZEUS & $\sqrt s=$0.32 TeV & 0.354 & $e^{\pm}p \rightarrow e^{\pm} b X   $ & $6.5 \le Q^2 \le 600$ GeV$^2$,  $1.5 \cdot 10^{-4} \le x \le 0.035$          & \cite{Abramowicz:2014zub} \\
    \hline
\hline
    CCFR  & 87 $\lesssim E_b \lesssim $333 GeV & & $ \overset{(-)}{\nu} p \rightarrow \mu^{\pm} cX$ & $1 \le Q^2 < 170$ GeV$^2$, $0.015 \le x \le 0.33$ &  \cite{Goncharov:2001qe} \\  \cline{1-4}
    \hline
    CHORUS & $\langle E_b\rangle \approx $27 GeV  & &  $\nu p \rightarrow \mu^+ cX$ & &  \cite{KayisTopaksu:2011mx} \\ 
    \hline
    NOMAD  & $6 \le E_b \le $300 GeV & & $\nu p \rightarrow \mu^+ cX$ & $1 \le Q^2 < 20$ GeV$^2$, $0.02 \lesssim x \le 0.75$ &  \cite{Samoylov:2013xoa} \\  \cline{1-4}
    \hline
    NuTeV  & 79 $\lesssim E_b \lesssim $245 GeV & & $\overset{(-)}{\nu} p \rightarrow \mu^{\pm} cX$ &$1 \le Q^2 < 120$ GeV$^2$, $0.015 \le x \le 0.33$  &  \cite{Goncharov:2001qe} \\  \cline{1-4}
    \hline
    \multicolumn{1}{l}{\bf DY} & \multicolumn{1}{c}{} & \multicolumn{1}{c}{} & \multicolumn{1}{c}{} & \multicolumn{1}{c}{} & \multicolumn{1}{c}{} \\
    \hline  
\hline
    \multirow{2}{*}{ATLAS} & \multirow{2}{*}{$\sqrt s=$7 TeV}& 0.035 & $ pp \rightarrow W^{\pm}X \rightarrow l^{\pm} \nu X$ & $p_{T}^l > 20$ GeV, $p_{T}^{\nu} > 25$ GeV, $m_T > 40$ GeV  & \cite{Aad:2011dm}  \\
                           &    &    & $pp \rightarrow Z X \rightarrow l^+l^- X$ & $p_{T}^l > 20$ GeV, $66<m_{ll}<116$ GeV & \\\cline{2-6}
                           & \multirow{2}{*}{$\sqrt s=$13 TeV} & 0.081 & $ pp \rightarrow W^{\pm} X \rightarrow l^{\pm} \nu X$ & $p_{T}^{\nu} > 25$ GeV, $m_T > 50$ GeV  & \cite{Aad:2016naf} \\
                           &    &    & $pp \rightarrow Z X \rightarrow l^+l^- X$ & $p_{T}^l > 25$ GeV, $66<m_{ll}<116$ GeV & \\\cline{2-6} 
    \hline 
    \multirow{2}{*}{CMS} & $\sqrt s=$7 TeV& 4.7  & $ pp \rightarrow W^{\pm} X \rightarrow \mu^{\pm} \nu X$ & $p_{T}^{\mu} > 25$ GeV & \cite{Chatrchyan:2013mza} \\
\cline{2-6}
                         & $\sqrt s=$8 TeV  & 18.8  & $ pp \rightarrow W^{\pm} X \rightarrow \mu^{\pm} \nu X$ & $p_{T}^{\mu} > 25$ GeV & \cite{Khachatryan:2016pev} \\
    \hline 
    \multirow{2}{*}{{D\O}} & \multirow{2}{*}{$\sqrt s=$1.96 TeV} & 7.3 & $\bar{p}p \rightarrow  W^{\pm} X \rightarrow \mu^{\pm} \nu X$ & $p_{T}^{\mu} > 25$ GeV, $\cancel{E}_{T} > 25$ GeV & \cite{Abazov:2013rja} \\
\cline{3-6}
                           &       & 9.7  & $ \bar{p}p \rightarrow W^{\pm} X \rightarrow e^{\pm} \nu X$ & $p_{T}^{e} > 25$ GeV, $\cancel{E}_{T} > 25$ GeV & \cite{D0:2014kma} \\
    \hline 
    \multirow{2}{*}{LHCb} & \multirow{2}{*}{$\sqrt s=$7 TeV} & 1 & $  pp \rightarrow W^{\pm} X \rightarrow \mu^{\pm} \nu X$ & $p_{T}^{\mu} > 20$ GeV                         & \cite{Aaij:2015gna} \\
                          &     &   & $pp \rightarrow Z X\rightarrow {\mu}^+{\mu}^- X$        & $p_{T}^{\mu} > 20$ GeV, $60<m_{\mu \mu}<120$ GeV & \\\cline{2-6}
                          & \multirow{3}{*}{$\sqrt s=$8 TeV} & 2  & $pp \rightarrow Z X \rightarrow e^+e^- X $ & $p_{T}^e > 20$ GeV, $60<m_{ee}<120$ GeV       & \cite{Aaij:2015vua} \\\cline{3-6}
                          &     & 2.9 & $ pp \rightarrow W^{\pm} X\rightarrow \mu^{\pm} \nu X$ & $p_{T}^{\mu} > 20$ GeV & \cite{Aaij:2015zlq} \\
                          &     &   & $pp \rightarrow Z X\rightarrow {\mu}^+{\mu}^- X$        & $p_{T}^{\mu} > 20$ GeV, $60<m_{\mu \mu}<120$ GeV & \\
\hline
\hline
    FNAL-605 & $E_b = 800$ GeV   &  & $pCu \rightarrow \mu^+ \mu^- X$ & 7$ \le M_{\mu\mu} \le $18 GeV  & \cite{Moreno:1990sf} \\ \cline{1-6}
    FNAL-866 & \multirow{2}{*}{ $E_b = 800$ GeV} &  & $pp \rightarrow \mu^+ \mu^- X$ & 4.6$ \le  M_{\mu\mu} \le $12.9 GeV& \cite{Towell:2001nh} \\
         & &  & $pD \rightarrow \mu^+ \mu^- X$ & &                      \\
    \hline
  \end{tabular}
    \caption{
      \label{dis_dy_table} 
      \small  
      The list of DIS and DY data used in the current analysis with the collider data listed first.
      The top-quark production data are detailed in Tabs.~\ref{tab:data-tt},~\ref{tab:data-inp}. 
    }
\end{table}

%% file: table-ttbar-data.tex
\begin{sidewaystable}[Ht!]
\renewcommand{\arraystretch}{1.3}
\begin{center}                   
{\small                          
\begin{tabular}{|c|c|c|c|c|c|c|c|c|c}   
\hline                                                    
\multicolumn{2}{|c|}{}&                      
\multicolumn{7}{|c|}{Cross section (pb)}                      
\\
\hline                                                    
\multicolumn{2}{|c|}{$\sqrt s$~(TeV)}                      
&\multicolumn{1}{|c|}{5}
&\multicolumn{2}{|c|}{7}
&\multicolumn{2}{|c|}{8}                         
&\multicolumn{2}{|c|}{13}                         
\\
\hline                                                    
\multicolumn{2}{|c|}{Experiment}                      
&{CMS}
&{ATLAS}

&{CMS}  
&{ATLAS}
&{CMS}  
&{ATLAS}
&{CMS}  
\\                                                        
\hline
\multirow{2}{*}{}
&dilepton + $b$-jet(s)
&
&$183\pm48$~\cite{Aad:2014kva}
&
& $243\pm8$~\cite{Aad:2014kva}
&
&$818\pm36$~\cite{Aaboud:2016pbd} 
& $792\pm43$~\cite{Khachatryan:2016kzg}
\\
\cline{2-9}
&dilepton + jets
&
& $181\pm11$~\cite{Aad:2014jra} 
&$174\pm6$~\cite{Khachatryan:2016mqs}
& 
&$245\pm9$~\cite{Khachatryan:2016mqs}
&
& $746\pm86$~\cite{Khachatryan:2015uqb}
\\
\cline{2-9}
\multirow{2}{*}{Decay mode}
&lepton + jets
&
& 
&$162\pm14$~\cite{Khachatryan:2016yzq} 
&$260\pm24$~\cite{Aad:2015pga}
&$229\pm15$~\cite{Khachatryan:2016yzq}
&
& $836\pm133$~\cite{CMS:2015toa}
\\
\cline{2-9}
\multirow{2}{4em}{}
&lepton + jets, $b~\rightarrow~\mu~\nu~X$
&
&$165\pm38$~\cite{ATLAS:2012gpa} 
&
&
&  
&
& 
\\
\cline{2-9}
\multirow{2}{4em}{}
&lepton + $\tau\rightarrow$~hadrons 
&
&$183\pm25$~\cite{Aad:2015dya}
&$143\pm26$~\cite{Chatrchyan:2012vs}
& 
&$257\pm25$~\cite{Khachatryan:2014loa}
& 
&
\\
\cline{2-9}
\multirow{2}{4em}{}
&jets + $\tau\rightarrow$~hadrons
&
&$194\pm49$~\cite{Aad:2012vip} 
& $152\pm34$~\cite{Chatrchyan:2013kff}
& 
& 
&
& 
\\
\cline{2-9}
\multirow{2}{4em}{}
&all-jets
&
&$168\pm60$~\cite{ATLAS-CONF-2012-031}
& $139\pm28$~\cite{Chatrchyan:2013ual}
&  
& $276\pm39$~\cite{Khachatryan:2015fwh}
&
& $834^{+123}_{-109}$~\cite{CMS:2016rtp}
\\
\cline{2-9}
\multirow{2}{4em}{}
&e$\mu$
& $82\pm23$~\cite{CMS:2016pqu} 
& 
&
& 
&
&
& 
\\
\cline{2-9}                           
\hline
\end{tabular}
}
\caption{
\label{tab:data-tt}
\small 
The data on the $t\bar{t}$-production cross section
from the LHC used in the present analysis. The errors given 
are combinations of the statistical and systematic ones. 
An additional error of 1.4, 3.3, 4.2 and 12~pb due to the beam energy uncertainty 
applies to all entries for the collision energy of $\sqrt{s}=5$, 7, 8 and 13 TeV, respectively.
The quoted values are rounded for the purpose of a compact presentation. 
}
\end{center}
\end{sidewaystable}

%% file: table-single-top-data.tex
\begin{table}[b!]
\renewcommand{\arraystretch}{1.3}
\begin{center}                   
{\small                          
\begin{tabular}{|c|c|c|c|c|c|c|c|}   
\hline                           
Experiment                      
&\multicolumn{3}{c}{ATLAS}
&\multicolumn{3}{|c|}{CMS}  
&CDF\&D{\O}
\\
\hline                                                    
{$\sqrt s$~(TeV)}                      
&{7}                         
&{8}                         
&{13}                         
&{7}                         
&{8}                         
&{13}                         
&{1.96}
\\                                                        
\hline
{Final states} 
& $tq$
& $tq$
& $tq$
& $tq$
& $tq$
& $tq$
&$tq, t\bar{b}$
\\
\hline                           
{Reference}                      
&\cite{Aad:2014fwa}                         
&\cite{Tepel:2014kna}
&\cite{Aaboud:2016ymp}
&\cite{Chatrchyan:2012ep}
&\cite{Khachatryan:2014iya}
&\cite{Sirunyan:2016cdg}
&\cite{Aaltonen:2015cra}
\\
\hline                                                    
{Luminosity (1/fb)}                      
&4.59                         
&20.3
&3.2
&2.73
&19.7
&2.3
&9.7x2
\\                                                        
\hline
{Cross section (pb)}                      
& $68 \pm 8$  
& $82.6 \pm 12.1$
&$247 \pm 46$
& $67.2 \pm 6.1$
& $83.6 \pm 7.7$
& $232 \pm 30.9$
& $3.30^{+0.52}_{-0.40}$ (sum)
\\                                                        
\hline
\end{tabular}
}
\caption{
\label{tab:data-inp}
\small 
The data on single-top production 
in association with a light quark $q$ or $\bar{b}$-quark
from the LHC and Tevatron used in the present analysis. The errors given 
are combinations of the statistical, systematic, and  
luminosity ones.} 
\end{center}
\end{table}

%% file: table-heavyquark1.tex
\begin{table}[t!]
\renewcommand{\arraystretch}{1.3}
\begin{center}                   
  \begin{tabular}{|l|c|c|c|c|}
    \hline
Experiment&Process& Reference & $NDP$&$\chi^2$
\\
\hline
\multicolumn{1}{l}{\bf DIS} & \multicolumn{1}{c}{} & \multicolumn{1}{c}{} &\multicolumn{1}{c}{} & \multicolumn{1}{c}{}  \\ 
\hline
HERA~I+II &$e^{\pm}p \rightarrow e^{\pm} X$ & \cite{Abramowicz:2015mha} & 1168 & 1510 
\\
   &$e^{\pm}p \rightarrow \overset{(-)}{\nu} X$ &  &  &  
\\
\hline
\hline
BCDMS &$\mu^+ p \rightarrow \mu^+ X$ & \cite{Benvenuti:1989rh}& 351 & 411
\\
\hline
NMC &$\mu^+ p \rightarrow \mu^+ X$ & \cite{Arneodo:1996qe}& 245 & 343
\\
\hline
SLAC-49a &$e^-p \rightarrow e^- X$ & \cite{Bodek:1979rx,Whitlow:1990gk} & 38 & 59
\\
\hline
SLAC-49b &$e^-p \rightarrow e^- X$ & \cite{Bodek:1979rx,Whitlow:1990gk} & 154 & 171
\\
\hline
SLAC-87 &$e^-p \rightarrow e^- X$ & \cite{Bodek:1979rx,Whitlow:1990gk} & 109 & 103
\\
\hline
SLAC-89b &$e^-p \rightarrow e^- X$ & \cite{Mestayer:1982ba,Whitlow:1990gk}& 90 & 79
\\
\hline
\multicolumn{1}{l}{\bf DIS heavy-quark production} & \multicolumn{1}{l}{} & \multicolumn{1}{c}{} &\multicolumn{1}{c}{} & \multicolumn{1}{c}{}  \\ 
\hline
HERA~I+II &$e^{\pm}p \rightarrow e^{\pm}c X$ & \cite{Abramowicz:1900rp} & 52 & 62 
\\
\hline
H1  &$e^{\pm}p \rightarrow e^{\pm}b X$ & \cite{Aaron:2009af} & 12 & 5 
\\
\hline
ZEUS &  $e^{\pm}p \rightarrow e^{\pm}b X$         & \cite{Abramowicz:2014zub} & 17 & 16 
\\
\hline
\hline
CCFR & $\overset{(-)}{\nu} p \rightarrow \mu^{\pm} c X$   & \cite{Goncharov:2001qe} & 89 & 62   
\\
\hline
CHORUS &$\nu p \rightarrow \mu^+ c X$  & \cite{KayisTopaksu:2011mx} & 6 & 7.6 
\\
\hline
NOMAD &$\nu p \rightarrow \mu^+ c X$  & \cite{Samoylov:2013xoa} & 48 & 59 
\\
\hline
NuTeV & $\overset{(-)}{\nu} \rightarrow \mu^{\pm} c X$    & \cite{Goncharov:2001qe} & 89 & 49 
\\
\hline
\multicolumn{1}{l}{\bf DY} & \multicolumn{1}{c}{} & \multicolumn{1}{c}{} &\multicolumn{1}{c}{} & \multicolumn{1}{c}{}  \\ 
\hline

FNAL-605 &$p Cu \rightarrow \mu^+ \mu^- X$ & \cite{Moreno:1990sf}& 119 & 165
\\
\hline
FNAL-866 &$p p \rightarrow \mu^+ \mu^- X$ & \cite{Towell:2001nh}& 39 & 53
\\
 &$p D \rightarrow \mu^+ \mu^- X$ & & &
\\
\hline
\multicolumn{1}{l}{\bf Top-quark production} & \multicolumn{1}{l}{} & \multicolumn{1}{c}{} &\multicolumn{1}{c}{} & \multicolumn{1}{c}{}  \\ 
\hline
ATLAS, CMS & $pp \rightarrow tqX $ & 
~\cite{Aad:2014fwa,Tepel:2014kna,Aaboud:2016ymp,Chatrchyan:2012ep,Khachatryan:2014iya,Sirunyan:2016cdg}  & 10 & 2.3 
\\
\hline
CDF\&D{\O} &$\bar{p}p \rightarrow tb X$ & \cite{Aaltonen:2015cra} & 2 & 1.1  
\\
& $\bar{p}p \rightarrow tqX $  &  &  &  
\\
\hline
\hline
ATLAS, CMS & $pp \rightarrow t\bar{t}X $ & 
~\cite{Aad:2014jra,Khachatryan:2016mqs,Khachatryan:2015uqb,
    Aad:2014kva,Aaboud:2016pbd,Khachatryan:2016kzg,Khachatryan:2016yzq,Aad:2015pga,CMS:2015toa,ATLAS:2012gpa,
    Aad:2015dya,Chatrchyan:2012vs,Khachatryan:2015fwh,Aad:2012vip,Chatrchyan:2013kff,ATLAS-CONF-2012-031,
    Chatrchyan:2013ual,CMS:2016rtp,Khachatryan:2014loa,CMS:2016pqu} 
  & 23 & 13
\\
\hline
CDF\&D{\O} &$\bar{p}p \rightarrow t\bar{t}X $ &
~\cite{Aaltonen:2015cra}  & 1 & 0.2
\\
\hline
  \end{tabular}
\caption{\small 
\label{tab:datahq}
{The values of $\chi^2$ obtained in the present analysis for the data on 
inclusive DIS, the fixed-target DY process, and on heavy-quark production.
The collider DY data are listed in Tab.~\ref{tab:dydata}.
}
}
\end{center}
\end{table}

%% file: table-WandZ-data.tex
\begin{sidewaystable}[H!]
\renewcommand{\arraystretch}{1.3}
\fontsize{10.25}{11.25}\selectfont
\begin{center}
\begin{tabular}{|c|c||c|c|c|c|c|c|c|c|c|}   
\hline                           
\multicolumn{2}{|c|}{Experiment}                      
&\multicolumn{2}{c|}{ATLAS}  
&\multicolumn{2}{c|}{CMS}  

&\multicolumn{2}{c|}{D\O}
&\multicolumn{3}{c|}{LHCb}
\\
\hline                                                    
\multicolumn{2}{|c|}{$\sqrt s$~(TeV)}                      
&\multicolumn{1}{c|}{7}
&\multicolumn{1}{c|}{13}
&\multicolumn{1}{c|}{7}
&\multicolumn{1}{c|}{8}

&\multicolumn{2}{c|}{1.96}
&7
&\multicolumn{2}{c|}{8}
\\                                                        
\hline
\multicolumn{2}{|c|}{Final states} 
& $W^+\rightarrow l^+\nu$
& $W^+\rightarrow l^+\nu$
& $W^+\rightarrow \mu^+\nu$

& $W^+\rightarrow \mu^+\nu$
&$W^+\rightarrow \mu^+\nu$
& $W^+\rightarrow e^+\nu$
&$W^+\rightarrow \mu^+\nu$
& $Z\rightarrow e^+e^-$                                                        
&$W^+\rightarrow \mu^+\nu$
\\
\multicolumn{2}{|c|}{ }                                          
& $W^-\rightarrow l^-\nu$
& $W^-\rightarrow l^-\nu$
&$W^-\rightarrow \mu^-\nu$
&$W^-\rightarrow \mu^-\nu$
&$W^-\rightarrow \mu^-\nu$
&$W^-\rightarrow e^-\nu$
&$W^-\rightarrow \mu^-\nu$
&                                                         
&$W^-\rightarrow \mu^-\nu$
\\                                                        
\multicolumn{2}{|c|}{ }                                          
& $Z\rightarrow l^+l^-$
& $Z\rightarrow l^+l^-$
& (asym)                                                        
&
& (asym)
& (asym)                                                        
& $Z\rightarrow \mu^+\mu^-$
&                                                         
& $Z\rightarrow \mu^+\mu^-$
\\
\hline                                                    
\multicolumn{2}{|c|}{Cut on the lepton $P_T$ }                      
&$P_T^l>20~{\rm GeV}$      
&$P_T^e>25~{\rm GeV}$      
&$P_T^{\mu}>25~{\rm GeV}$   
&$P_T^{\mu}>25~{\rm GeV}$   
&$P_T^{\mu}>25~{\rm GeV}$
&$P_T^{e}>25~{\rm GeV}$
&$P_T^{\mu}>20~{\rm GeV}$
&$P_T^{e}>20~{\rm GeV}$                       
&$P_T^{\mu}>20~{\rm GeV}$                       
\\                                                        
\hline                                                    
\multicolumn{2}{|c|}{Luminosity (1/fb)}                      
&0.035
&0.081
&4.7
&18.8
&7.3
&9.7
&1
&2                        
&2.9
\\
\hline                           
\multicolumn{2}{|c|}{Reference}                      
&\cite{Aad:2011dm}                         
&\cite{Aad:2016naf}
&\cite{Chatrchyan:2013mza}
&\cite{Khachatryan:2016pev}
&\cite{Abazov:2013rja}
&\cite{D0:2014kma}
&\cite{Aaij:2015gna}
&\cite{Aaij:2015vua}
&\cite{Aaij:2015zlq}
\\                                                        
\hline                                                    
\multicolumn{2}{|c|}{$NDP$}
&30
&6
&11  
&22
&10
&13
&31
&17                            
&32                            
\\                                                        
\hline
\multirow{2}{4em}{ }
& present analysis
    \footnote{
      The ABM12~\cite{Alekhin:2013nda} analysis has used 
  older data sets from CMS and LHCb.}
& 31.0
& 9.2
 &22.4
 &16.5
 &17.6
 &19.0
&45.1
&21.7
 &40.0
\\
\cline{2-11}
 &CJ15~\cite{Accardi:2016qay}
 &--
 &--
&--
 &--
 &20
&29
 &--
&--
 &--
\\
\cline{2-11}
 &CT14~\cite{Dulat:2015mca}
&42
 &--
 &--~\footnote{ 
For the statistically less significant data with the cut of 
$P_T^{\mu}>35~{\rm GeV}$ the  value of $\chi^2=12.1$ was obtained.}
&--
&--
&34.7
 &--
&--
 &--
\\
\cline{2-11}
$\chi^2$ &JR14~\cite{Jimenez-Delgado:2014twa} 
 &--
 &--
 &--
&--
 &--
&--
 &--
&--
 &--
\\
\cline{2-11}
&{\tt HERAFitter}~\cite{Camarda:2015zba}
 &--
 &--
 &--
&--
 &13
&19
 &--
&--
 &--
\\
\cline{2-11}
&MMHT14~\cite{Harland-Lang:2014zoa}
 &39
 &--
 &--
&--
 &21
&--
 &--
&--
 &--
\\
\cline{2-11}
&NNPDF3.0~\cite{Ball:2014uwa}
 &35.4
 &--
 &18.9
 &--
 &--
&--
 &--
&--
 &--
\\
\hline                                          
\end{tabular}
\caption{\small 
  \label{tab:dydata}
  Compilation of precise data on $W$- and $Z$-boson production in $pp$ and $p\bar{p}$ collisions and the $\chi^2$ values  obtained for these data sets
  in different PDF analyses using their individual definitions of $\chi^2$.
  The NNLO fit results are quoted as a default, while the NLO values are 
  given for the CJ15~\cite{Accardi:2016qay} and 
  {\tt HERAFitter}~\cite{Camarda:2015zba} PDFs. 
  Missing table entries indicate that the respective data sets have not been used in the analysis.
}
\end{center}
\end{sidewaystable}

%% file: table-pdf-parameters.tex
\begin{table}[h!]
\renewcommand{\arraystretch}{1.5}
\begin{center} 
\footnotesize
\begin{tabular}{|c|c|c|c|c|c|c|c|} 
\hline 
\multicolumn{1}{|c|}{ } & 
\multicolumn{1}{c|}{$a$} & 
\multicolumn{1}{c|}{$b$} & 
\multicolumn{1}{c|}{$\gamma_{-1}$} & 
\multicolumn{1}{c|}{$\gamma_1$} & 
\multicolumn{1}{c|}{$\gamma_2$} & 
\multicolumn{1}{c|}{$\gamma_3$} & 
\multicolumn{1}{c|}{$A$} 
\\ \hline 
$u_v$
 &  0.623 $\pm$ 0.033
 &  3.443  $\pm$ 0.064
 &
 &  -0.22  $\pm$ 0.33
 &  -2.88  $\pm$ 0.46
 &  2.67  $\pm$ 0.80
 &
\\
$d_v$
 &  0.372  $\pm$ 0.068
 &  4.47  $\pm$ 0.55
 &
 &  -3.20  $\pm$ 0.77
 &  -0.61  $\pm$ 1.96  
 &  0$\pm$ 0.001~\footnote{This parameter is poorly determined from the fit.
   Therefore its variation is constrained within the range $\gamma_{3,d} \in [-0.001,0.001]$.}  
 &
\\
$u_s$
 & -0.415 $\pm$ 0.031
 &  7.75  $\pm$ 0.39
 &  0.0373 $\pm$ 0.0032
 &  4.44  $\pm$ 0.95
 &
 & 
 & 0.0703 $\pm$ 0.0081
\\
$d_s$
 &  -0.17  $\pm$ 0.011
 &  8.41  $\pm$ 0.34
 &
 &  13.3  $\pm$ 1.7
 &
 &
 & 0.1408$\pm$ 0.0076
\\
$s_s$ 
 & -0.344 $\pm$ 0.019
 &  6.52  $\pm$ 0.27
 &
 &
 &
 &
 & 0.0594 $\pm$ 0.0042
\\
$g$ 
 & -0.1534 $\pm$ 0.0094
 &  6.42  $\pm$ 0.83
 &
 &  -11.8  $\pm$ 3.7
 &
 &
 &
\\
\hline 
\end{tabular} 
\caption{ \small
  \label{tab:fitvalues} 
  The fitted PDF parameters Eqs.~(\ref{eq:val})--(\ref{eq:poly}) 
  and their $1\sigma$ errors due to statistical and systematic uncertainties
  in the data.}
\end{center} 
\end{table}

%% file: table-hts.tex
\begin{table}[b!]
\renewcommand{\arraystretch}{1.5}
\begin{center} 
\begin{tabular}{|c|c|c|} 
\hline 
\multicolumn{1}{|c|}{ } & 
\multicolumn{1}{c|}{$H_2^{\tau=4}(x)/{\rm GeV}^2$} & 
\multicolumn{1}{c|}{$H_T^{\tau=4}(x)/{\rm GeV}^2$} 
\\ \hline 
$x=0.0$
 &  0.023  $\pm$ 0.019
 &  -0.319  $\pm$ 0.126
\\ 
$x=0.1$
 &  -0.032  $\pm$ 0.013
 &  -0.134  $\pm$ 0.040
\\ 
$x=0.3$
 &  -0.005  $\pm$ 0.009
 &  -0.052  $\pm$ 0.030
\\
$x=0.5$
 &  0.025  $\pm$ 0.006
 &  0.071 $\pm$ 0.025
\\
$x=0.7$
 &  0.051  $\pm$ 0.005
 &  0.030  $\pm$ 0.012
\\
$x=0.9$
 &  0.003  $\pm$ 0.004
 & 0.003  $\pm$ 0.007
\\
$x=1$
 &  0
 & 0
\\
\hline 
\end{tabular} 
\caption{ \small
  \label{tab:hts} 
  The knots of the twist-4 splines $S_{2,T}(x)$ in Eq.~(\ref{eq:htspline}) obtained in the present
  analysis.
} 
\end{center} 
\end{table}

%% file: table-moms.tex
\begin{table}[t!]
\renewcommand{\arraystretch}{1.3}
\begin{center}
{\small
\begin{tabular}{|l|l|l|l|l|}
\hline
\multicolumn{1}{|c|}{ } &
\multicolumn{1}{c|}{$\langle x u_v(x)\rangle$} &
\multicolumn{1}{c|}{$\langle x d_v(x)\rangle$} &
\multicolumn{1}{c|}{$\langle x [u_v-d_v](x)\rangle$} &
\multicolumn{1}{c|}{$\langle x V(x)\rangle$} \\
\hline
ABM11~\cite{Alekhin:2012ig}
& $0.2966 \pm 0.0039$
& $0.1172 \pm 0.0050$
& $0.1794 \pm 0.0041$
& $0.1652 \pm 0.0039$
\\
ABM12~\cite{Alekhin:2013nda} 
& $0.2950 \pm 0.0029$ 
& $0.1212 \pm 0.0016$  
& $0.1738 \pm 0.0025$  
& $0.1617 \pm 0.0031$
\\
ABMP16 (this work)
& $0.2911 \pm 0.0024$ 
& $0.1100 \pm 0.0031$  
& $0.1811 \pm 0.0032$  
& $0.1674 \pm 0.0037$  
\\
\hline
CT14~\cite{Dulat:2015mca} \footnote{
  The PDF uncertainties of CT14 denote the 90\% confidence level and need rescaling
  with a factor of 1.645 for comparison with the 68\% confidence level
  uncertainties quoted for all other results.}
& $0.2887~^{+~0.0074}_{-~0.0073}$
& $0.1180~^{+~0.0053}_{-~0.0041}$
& $0.1707~^{+~0.0078}_{-~0.0092}$
& $0.1579~^{+~0.0095}_{-~0.0117}$
\\
\hline
MMHT14~\cite{Harland-Lang:2014zoa} 
& $0.2852~^{+~0.0052}_{-~0.0034}$
& $0.1202~^{+~0.0030}_{-~0.0031}$
& $0.1650~^{+~0.0047}_{-~0.0034}$
& $0.1509~^{+~0.0053}_{-~0.0039}$
\\
\hline
NNPDF3.0~\cite{Ball:2014uwa}
& $0.2833 \pm 0.0042$ 
& $0.1183 \pm 0.0049$  
& $0.1650 \pm 0.0054$  
& $0.1553 \pm 0.0037$  
\\
\hline
\end{tabular}
}
\caption{\small 
  \label{tab:2ndmom}
  Second moment of valence quark distributions at NNLO at $\mu^2 = 4$~GeV$^2$ 
  with their uncertainties. 
  For ABM11, ABM12 and ABMP16 the sets with $n_f=3$ flavors have been used.
}
\end{center}
\end{table}

%% file: table-higgs.tex
\begin{table}[b!]
\begin{center}
\renewcommand{\arraystretch}{1.3}
\begin{tabular}{|l|l|l|l|l|}
\hline
PDF sets &
$\alpha_s^{(n_f=5)}(M_Z)$
  & \multirow{2}{8em}{$\sigma(H)^{\rm NNLO}$~[pb] nominal $\alpha_s$}
  & \multirow{2}{8em}{$\sigma(H)^{\rm NNLO}$~[pb] $\alpha_s^{(n_f=5)}(M_Z)=0.115$}
  & \multirow{2}{8em}{$\sigma(H)^{\rm NNLO}$~[pb] $\alpha_s^{(n_f=5)}(M_Z)=0.118$}
\\[3.5ex]
\hline
ABM12~\cite{Alekhin:2013nda} &
$0.1132 \pm 0.0011$
  & $ 39.80 \pm 0.84 $
  & $ 41.62 \pm 0.87 $
  & $ 44.70 \pm 0.91 $
\\[0.5ex]
\hline
ABMP15~\cite{Alekhin:2015cza} &
$0.1132 \pm 0.0011$
  & $ 39.46 \pm 0.77 $
  & $ 41.30 \pm 0.79 $
  & $ 44.36 \pm 0.83 $
\\[0.5ex]
\hline
ABMP16 (this work) &
$0.1147 \pm 0.0009$
  & $ 40.20 \pm 0.63 $
  & $ 40.50 \pm 0.64 $
  & $ 43.50 \pm 0.67 $
\\[0.5ex]
\hline
PDF4LHC15~\cite{Butterworth:2015oua}&
$0.1180$ (fixed)
  & $ 42.42 \pm 0.78 $ 
  & $ 39.49 \pm 0.73 $
  & $ 42.42 \pm 0.78 $ 
\\[0.5ex]
\hline
\end{tabular}
  \caption{\small 
  \label{tab:higgs}
  Cross section for the Higgs boson production from the gluon fusion 
  at NNLO in QCD (computed in the effective theory) 
  with the PDF and $\alpha_s$ uncertainties
  at $\sqrt{s}=13$~TeV for $m_H=125.0$~GeV for $\mu_r=\mu_f=m_H$. 
  The columns list the value of 
  $\alpha_s$ for each PDF set 
  and the cross sections values obtained with both the nominal PDF set
  and the choices for $\alpha_s$ indicated. 
  }
\end{center}
\end{table}

%% file: table-ttbar.tex
\begin{table}[t!]
\begin{center}
\renewcommand{\arraystretch}{1.3}
\begin{tabular}{|l|l|l|l|l|}
\hline
PDF sets
  & \multirow{2}{7em}{$\sigma(t{\bar t})^{\rm NNLO}$~[pb] at $\sqrt{s}=5$~TeV}
  & \multirow{2}{7em}{$\sigma(t{\bar t})^{\rm NNLO}$~[pb] at $\sqrt{s}=7$~TeV}
  & \multirow{2}{7em}{$\sigma(t{\bar t})^{\rm NNLO}$~[pb] at $\sqrt{s}=8$~TeV}
  & \multirow{2}{7em}{$\sigma(t{\bar t})^{\rm NNLO}$~[pb] at $\sqrt{s}=13$~TeV}
\\[3.5ex]
\hline
ABM12~\cite{Alekhin:2013nda} 
  & $ 56.03 \pm 2.76 $  
  & $ 156.8 \pm 6.4 $  
  & $ 228.6 \pm 8.6 $  
  & $ 793.9 \pm 22.1 $  
\\[0.5ex]
\hline
ABMP15~\cite{Alekhin:2015cza} 
  & $ 56.79 \pm 2.98 $  
  & $ 157.4 \pm 6.9 $  
  & $ 229.0 \pm 9.3 $   
  & $ 791.0 \pm 23.7 $   
\\[0.5ex]
\hline
ABMP16 (this work)
  & $ 63.66 \pm 1.60 $  
  & $ 171.8 \pm 3.4 $  
  & $ 247.5 \pm 4.6 $  
  & $ 831.4 \pm 14.5 $  
\\[0.5ex]
\hline
\end{tabular}
\caption{\small 
  \label{tab:ttbar}
  Cross section for the top-quark pair production at NNLO in QCD 
  with the PDF uncertainties for the top-quark mass $m_t(m_t)=160.9$~GeV 
  in the \msbar scheme and $\mu_r=\mu_f=m_t(m_t)$ 
  at various center-of-mass energies of the LHC.
}
\end{center}
\end{table}

%% file: table-asmz-mq-parameters.tex
\begin{table}[ht!]
\begin{center}
\renewcommand{\arraystretch}{1.3}
\begin{tabular}{|l|c|c|c|c|c|c|}
\hline
PDF set
  & $\alpha_s^{(n_f=5)}(M_Z)$
  & $m_c(m_c)$~[GeV]
  & $m_b(m_b)$~[GeV]
  & $m_b^{\rm pole}$~[GeV]
  & $m_t(m_t)$~[GeV]
  & $m_t^{\rm pole}$~[GeV]
\\[0.5ex]
\hline
  {\bf \phantom{0}0} &   {\bf 0.11471} &   {\bf 1.252} &   {\bf 3.838} &   {\bf 4.537} &   {\bf 160.86} &   {\bf 170.37} 
\\[0.5ex]
\hline
  \phantom{0}1 & 0.11471 & 1.252 & 3.839 & 4.538 & 160.86 & 170.37 
\\
  \phantom{0}2 & 0.11472 & 1.251 & 3.838 & 4.537 & 160.86 & 170.37 
\\
  \phantom{0}3 & 0.11471 & 1.252 & 3.839 & 4.538 & 160.86 & 170.37 
\\
  \phantom{0}4 & 0.11468 & 1.252 & 3.839 & 4.537 & 160.86 & 170.37 
\\
  \phantom{0}5 & 0.11463 & 1.252 & 3.839 & 4.536 & 160.86 & 170.36 
\\
  \phantom{0}6 & 0.11468 & 1.252 & 3.839 & 4.537 & 160.86 & 170.37 
\\
  \phantom{0}7 & 0.11471 & 1.251 & 3.839 & 4.538 & 160.86 & 170.37       
\\
  \phantom{0}8 & 0.11456 & 1.252 & 3.839 & 4.535 & 160.86 & 170.35 
\\
  \phantom{0}9 & 0.11510 & 1.252 & 3.838 & 4.545 & 160.86 & 170.41
\\
 10 & 0.11453 & 1.251 & 3.839 & 4.534 & 160.86 & 170.35 
\\
 11 & 0.11472 & 1.250 & 3.838 & 4.537 & 160.86 & 170.37 
\\
 12 & 0.11468 & 1.250 & 3.839 & 4.537 & 160.86 & 170.37 
\\
 13 & 0.11469 & 1.267 & 3.838 & 4.536 & 160.86 & 170.37 
\\
 14 & 0.11478 & 1.250 & 3.838 & 4.538 & 160.86 & 170.38 
\\
 15 & 0.11487 & 1.249 & 3.838 & 4.540 & 160.86 & 170.38 
\\
 16 & 0.11453 & 1.254 & 3.840 & 4.535 & 160.86 & 170.35 
\\
 17 & 0.11477 & 1.252 & 3.961 & 4.671 & 160.86 & 170.37 
\\
 18 & 0.11469 & 1.252 & 3.861 & 4.561 & 160.86 & 170.37 
\\
 19 & 0.11460 & 1.252 & 3.846 & 4.543 & 160.87 & 170.37 
\\
 20 & 0.11481 & 1.251 & 3.835 & 4.536 & 160.85 & 170.37 
\\
 21 & 0.11471 & 1.252 & 3.848 & 4.548 & 160.89 & 170.40 
\\
 22 & 0.11483 & 1.254 & 3.827 & 4.527 & 160.80 & 170.32 
\\
 23 & 0.11467 & 1.252 & 3.837 & 4.535 & 160.81 & 170.31 
\\
 24 & 0.11492 & 1.250 & 3.835 & 4.538 & 161.34 & 170.89 
\\
 25 & 0.11461 & 1.252 & 3.844 & 4.541 & 160.57 & 170.05 
\\
 26 & 0.11522 & 1.252 & 3.836 & 4.545 & 161.51 & 171.10 
\\
 27 & 0.11486 & 1.251 & 3.831 & 4.532 & 161.04 & 170.57
\\
 28 & 0.11466 & 1.252 & 3.840 & 4.538 & 160.86 & 170.36
\\
 29 & 0.11497 & 1.248 & 3.827 & 4.530 & 161.47 & 171.03 
\\[0.5ex]
\hline
\end{tabular}
\caption{\small 
  \label{tab:parameters}
  Values of $\alpha_s^{(n_f=5)}(M_Z)$ and heavy-quark masses 
  $m_c(m_c)$, $m_b(m_b)$ and $m_t(m_t)$ 
  in the \msbar scheme obtained for the individual PDF sets of ABMP16. 
  For bottom and top, also the values for pole masses $m_b^{\rm pole}$ and
  $m_t^{\rm pole}$ in the on-shell scheme are given.
}
\end{center}
\end{table}

%% file: table-cor.tex
\begin{center}
\begin{table}
\renewcommand{\arraystretch}{1.3}
\begin{center}
\small
\begin{tabular}{|c|r|r|r|r|r|r|r|r|r|r|}
\hline
&$a_u$
&$b_u$
&$\gamma_{1,u}$
&$\gamma_{2,u}$
&$\gamma_{3,u}$
&$a_d$
&$b_d$
&$\gamma_{1,d}$
&$\gamma_{2,d}$
&$\gamma_{3,d}$
\\[1.0ex]
\hline
%
$a_u$
& 1.0
& 0.7617
& 0.9372
&- 0.5078
& 0.4839
& 0.4069
& 0.3591
& 0.4344
&- 0.3475
& 0.0001
\\ 
$b_u$
& 0.7617
& 1.0
& 0.6124
&- 0.1533
&- 0.0346
& 0.3596
& 0.2958
& 0.3748
&- 0.2748
& 0.0001
\\
$\gamma_{1,u}$
& 0.9372
& 0.6124
& 1.0
&- 0.7526
& 0.7154
& 0.2231
& 0.2441
& 0.2812
&- 0.2606
& 0.0001
\\ 
$\gamma_{2,u}$
&- 0.5078
&- 0.1533
&- 0.7526
& 1.0
&- 0.9409
& 0.2779
& 0.2276
& 0.2266
&- 0.1860
& 0.0
\\ 
$\gamma_{3,u}$
& 0.4839
&- 0.0346
& 0.7154
&- 0.9409
& 1.0
&- 0.1738
&- 0.1829
&- 0.1327
& 0.1488
& 0.0
\\ 
$a_d$
& 0.4069
& 0.3596
& 0.2231
& 0.2779
&- 0.1738
& 1.0
& 0.7209
& 0.9697
&- 0.6529
& 0.0001
\\ 
$b_d$
& 0.3591
& 0.2958
& 0.2441
& 0.2276
&- 0.1829
& 0.7209
& 1.0
& 0.7681
&- 0.9786
&- 0.0001
\\ 
$\gamma_{1,d}$
& 0.4344
& 0.3748
& 0.2812
& 0.2266
&- 0.1327
& 0.9697
& 0.7681
& 1.0
&- 0.7454
& 0.0002
\\
$\gamma_{2,d}$
&- 0.3475
&- 0.2748
&- 0.2606
&- 0.1860
& 0.1488
&- 0.6529
&- 0.9786
&- 0.7454
& 1.0
&- 0.0002
\\ 
$\gamma_{3,d}$
& 0.0001
& 0.0001
& 0.0001
& 0.0
& 0.0
& 0.0001
&- 0.0001
& 0.0002
&- 0.0002
& 1.0
\\ 
$a_{us}$
&- 0.0683
&- 0.0081
&- 0.2094
& 0.3881
&- 0.3206
& 0.2266
& 0.1502
& 0.2000
&- 0.1293
& 0.0
\\ 
$b_{us}$
&- 0.3508
&- 0.3089
&- 0.3462
& 0.0906
&- 0.0537
&- 0.1045
&- 0.2000
&- 0.2241
& 0.2798
& 0.0
\\ 
$\gamma_{-1,us}$
& 0.2296
& 0.1387
& 0.3367
&- 0.4043
& 0.3474
&- 0.1171
&- 0.1127
&- 0.0810
& 0.0767
& 0.0
\\ 
$\gamma_{1,us}$
&- 0.4853
&- 0.4119
&- 0.3844
&- 0.0365
& 0.0064
&- 0.4380
&- 0.3592
&- 0.4957
& 0.3771
&- 0.0001
\\
$A_{us}$
& 0.0506
& 0.0807
&- 0.0949
& 0.3198
&- 0.2560
& 0.2527
& 0.1648
& 0.2350
&- 0.1509
& 0.0
\\
$a_{ds}$
&- 0.0759
&- 0.0443
&- 0.0951
& 0.0263
&- 0.0382
&- 0.2565
&- 0.2541
&- 0.2666
& 0.2380
& 0.0
\\
$b_{bs}$
& 0.0452
&- 0.0197
& 0.0345
&- 0.0589
& 0.0683
&- 0.2084
& 0.0190
&- 0.1841
&- 0.0522
& 0.0
\\ 
$\gamma_{1,ds}$
&- 0.0492
&- 0.0809
& 0.0101
&- 0.1791
& 0.1309
&- 0.5576
&- 0.2029
&- 0.4584
& 0.0946
& 0.0
\\ 
$A_{ds}$
&- 0.1980
&- 0.1262
&- 0.2349
& 0.1526
&- 0.1428
&- 0.1113
&- 0.2167
&- 0.1739
& 0.2407
& 0.0
\\ 
$a_{ss}$
&- 0.2034
&- 0.1285
&- 0.2362
& 0.2328
&- 0.2080
& 0.0960
& 0.1596
& 0.0661
&- 0.1054
& 0.0
\\ 
$b_{ss}$
&- 0.1186
&- 0.0480
&- 0.1532
& 0.1549
&- 0.1536
& 0.0486
& 0.1508
& 0.0267
&- 0.1161
& 0.0
\\ 
$A_{ss}$
&- 0.1013
&- 0.0411
&- 0.1458
& 0.1802
&- 0.1625
& 0.1216
& 0.1678
& 0.0924
&- 0.1196
& 0.0
\\ 
$a_g$
& 0.0046
&- 0.0374
& 0.1109
&- 0.1934
& 0.1653
&- 0.0288
&- 0.0122
& 0.0053
& 0.0059
& 0.0
\\ 
$b_g$
& 0.2662
& 0.3141
& 0.1579
&- 0.0050
&- 0.0207
& 0.0973
& 0.0870
& 0.0646
&- 0.0666
& 0.0
\\
$\gamma_{1,g}$
& 0.2008
& 0.2274
& 0.0706
& 0.0876
&- 0.0835
& 0.0919
& 0.0574
& 0.0493
&- 0.0364
& 0.0
\\ 
$\alpha_s^{(n_f=3)}(\mu_0)$
& 0.1083
&- 0.0607
& 0.0848
&- 0.0250
& 0.0765
& 0.0763
&- 0.0306
& 0.0725
& 0.0243
& 0.0
\\ 
$m_c(m_c)$
&- 0.0006
& 0.0170
&- 0.0104
& 0.0206
&- 0.0201
&- 0.0123
&- 0.0161
&- 0.0114
& 0.0108
& 0.0
\\ 
$m_b(m_b)$
& 0.0661
& 0.0554
& 0.0605
&- 0.0367
& 0.0287
&- 0.0116
& 0.0029
&- 0.0074
&- 0.0051
& 0.0
\\ 
$m_t(m_t)$
&- 0.1339
&- 0.2170
&- 0.0816
& 0.0081
& 0.0250
&- 0.0616
&- 0.0813
&- 0.0491
& 0.0736
& 0.0
\\
\hline
\end{tabular}
\end{center}
\caption{\small 
  \label{tab:cor1}
  Correlation matrix of the fitted parameters for the PDFs, the strong coupling and the heavy-quark masses.
  Note, that $\alpha_s^{(n_f=3)}(\mu_0)$ is evaluated at the scale $\mu_0=1.5~\GeV$.
}
\end{table}
\end{center}

\begin{center}
\begin{table}
\renewcommand{\arraystretch}{1.3}
\begin{center}
\small
\begin{tabular}{|c|r|r|r|r|r|r|r|r|r|r|}
\hline
&$a_{us}$
&$b_{us}$
&$\gamma_{-1,us}$
&$\gamma_{1,us}$
&$A_{us}$
&$a_{ds}$
&$b_{bs}$
&$\gamma_{1,ds}$
&$A_{ds}$
&$a_{ss}$
\\[1.0ex]
\hline
%
$a_u$
&- 0.0683
&- 0.3508
& 0.2296
&- 0.4853
& 0.0506
&- 0.0759
& 0.0452
&- 0.0492
&- 0.1980
&- 0.2034
\\ 
$b_u$
&- 0.0081
&- 0.3089
& 0.1387
&- 0.4119
& 0.0807
&- 0.0443
&- 0.0197
&- 0.0809
&- 0.1262
&- 0.1285
\\
$\gamma_{1,u}$
&- 0.2094
&- 0.3462
& 0.3367
&- 0.3844
&- 0.0949
&- 0.0951
& 0.0345
& 0.0101
&- 0.2349
&- 0.2362
\\ 
$\gamma_{2,u}$
& 0.3881
& 0.0906
&- 0.4043
&- 0.0365
& 0.3198
& 0.0263
&- 0.0589
&- 0.1791
& 0.1526
& 0.2328
\\ 
$\gamma_{3,u}$
&- 0.3206
&- 0.0537
& 0.3474
& 0.0064
&- 0.2560
&- 0.0382
& 0.0683
& 0.1309
&- 0.1428
&- 0.2080
\\ 
$a_d$
& 0.2266
&- 0.1045
&- 0.1171
&- 0.4380
& 0.2527
&- 0.2565
&- 0.2084
&- 0.5576
&- 0.1113
& 0.0960
\\ 
$b_d$
& 0.1502
&- 0.2000
&- 0.1127
&- 0.3592
& 0.1648
&- 0.2541
& 0.0190
&- 0.2029
&- 0.2167
& 0.1596
\\ 
$\gamma_{1,d}$
& 0.2000
&- 0.2241
&- 0.0810
&- 0.4957
& 0.2350
&- 0.2666
&- 0.1841
&- 0.4584
&- 0.1739
& 0.0661
\\
$\gamma_{2,d}$
&- 0.1293
& 0.2798
& 0.0767
& 0.3771
&- 0.1509
& 0.2380
&- 0.0522
& 0.0946
& 0.2407
&- 0.1054
\\ 
$\gamma_{3,d}$
& 0.0
& 0.0
& 0.0
&- 0.0001
& 0.0
& 0.0
& 0.0
& 0.0
& 0.0
& 0.0
\\ 
$a_{us}$
& 1.0
&- 0.3156
&- 0.8947
&- 0.5310
& 0.9719
& 0.2849
& 0.0241
&- 0.0470
& 0.2983
& 0.4131
\\ 
$b_{us}$
&- 0.3156
& 1.0
& 0.1372
& 0.8258
&- 0.3995
& 0.0467
&- 0.0221
&- 0.1190
& 0.1856
& 0.0291
\\ 
$\gamma_{-1,us}$
&- 0.8947
& 0.1372
& 1.0
& 0.2611
&- 0.7829
&- 0.1695
& 0.0156
& 0.0501
&- 0.2117
&- 0.7191
\\ 
$\gamma_{1,us}$
&- 0.5310
& 0.8258
& 0.2611
& 1.0
&- 0.6479
& 0.0086
& 0.0076
& 0.1460
& 0.0781
&- 0.0010
\\
$A_{us}$
& 0.9719
&- 0.3995
&- 0.7829
&- 0.6479
& 1.0
& 0.2983
& 0.0515
&- 0.0404
& 0.3055
& 0.2811
\\
$a_{ds}$
& 0.2849
& 0.0467
&- 0.1695
& 0.0086
& 0.2983
& 1.0
&- 0.1608
& 0.0719
& 0.9152
&- 0.2941
\\
$b_{bs}$
& 0.0241
&- 0.0221
& 0.0156
& 0.0076
& 0.0515
&- 0.1608
& 1.0
& 0.7834
&- 0.3022
&- 0.0390
\\ 
$\gamma_{1,ds}$
&- 0.0470
&- 0.1190
& 0.0501
& 0.1460
&- 0.0404
& 0.0719
& 0.7834
& 1.0
&- 0.1838
&- 0.1373
\\ 
$A_{ds}$
& 0.2983
& 0.1856
&- 0.2117
& 0.0781
& 0.3055
& 0.9152
&- 0.3022
&- 0.1838
& 1.0
&- 0.1833
\\ 
$a_{ss}$
& 0.4131
& 0.0291
&- 0.7191
&- 0.0010
& 0.2811
&- 0.2941
&- 0.0390
&- 0.1373
&- 0.1833
& 1.0
\\ 
$b_{ss}$
& 0.2197
& 0.0643
&- 0.4479
& 0.1286
& 0.1193
&- 0.1579
&- 0.0260
& 0.0169
&- 0.0896
& 0.6522
\\ 
$A_{ss}$
& 0.3627
& 0.0261
&- 0.6319
& 0.0102
& 0.2412
&- 0.2688
&- 0.0180
&- 0.0960
&- 0.1797
& 0.9280
\\ 
$a_g$
&- 0.2570
& 0.0001
& 0.2196
& 0.0039
&- 0.2493
&- 0.2190
&- 0.0454
&- 0.1031
&- 0.2571
& 0.0626
\\ 
$b_g$
&- 0.1419
& 0.1266
& 0.0694
& 0.2648
&- 0.1715
&- 0.0515
& 0.0917
& 0.2130
&- 0.0469
&- 0.0092
\\
$\gamma_{1,g}$
&- 0.0241
& 0.0332
&- 0.0226
& 0.1296
&- 0.0489
&- 0.0137
& 0.0503
& 0.1409
& 0.0022
&- 0.0279
\\ 
$\alpha_s^{(n_f=3)}(\mu_0)$
& 0.0954
&- 0.2866
&- 0.0341
&- 0.3493
& 0.1110
&- 0.0604
&- 0.1265
&- 0.1811
&- 0.1330
&- 0.0841
\\ 
$m_c(m_c)$
& 0.0704
&- 0.0093
&- 0.0033
&- 0.0462
& 0.1182
& 0.0849
& 0.0547
& 0.0413
& 0.1193
&- 0.0728
\\ 
$m_b(m_b)$
&- 0.0183
&- 0.0132
& 0.0044
& 0.0209
&- 0.0298
&- 0.0006
& 0.0332
& 0.0695
&- 0.0432
&- 0.0159
\\ 
$m_t(m_t)$
& 0.0641
&- 0.1841
&- 0.0408
&- 0.2635
& 0.0755
&- 0.0573
&- 0.1067
&- 0.2003
&- 0.0869
& 0.0169
\\
\hline
\end{tabular}
\end{center}
\caption{\small 
  \label{tab:cor2}
  Tab.~\ref{tab:cor1} continued.
}
\end{table}
\end{center}

\begin{center}
\begin{table}
\renewcommand{\arraystretch}{1.3}
\begin{center}
\small
\begin{tabular}{|c|r|r|r|r|r|r|r|r|r|}
\hline
&$b_{ss}$
&$A_{ss}$
&$a_g$
&$b_g$
&$\gamma_{1,g}$
&$\alpha_s^{(n_f=3)}(\mu_0)$
&$m_c(m_c)$
&$m_b(m_b)$
&$m_t(m_t)$
\\[1.0ex]
\hline
%
$a_u$
&- 0.1186
&- 0.1013
& 0.0046
& 0.2662
& 0.2008
& 0.1083
&- 0.0006
& 0.0661
&- 0.1339
\\ 
$b_u$
&- 0.0480
&- 0.0411
&- 0.0374
& 0.3141
& 0.2274
&- 0.0607
& 0.0170
& 0.0554
&- 0.2170
\\
$\gamma_{1,u}$
&- 0.1532
&- 0.1458
& 0.1109
& 0.1579
& 0.0706
& 0.0848
&- 0.0104
& 0.0605
&- 0.0816
\\ 
$\gamma_{2,u}$
& 0.1549
& 0.1802
&- 0.1934
&- 0.0050
& 0.0876
&- 0.0250
& 0.0206
&- 0.0367
& 0.0081
\\ 
$\gamma_{3,u}$
&- 0.1536
&- 0.1625
& 0.1653
&- 0.0207
&- 0.0835
& 0.0765
&- 0.0201
& 0.0287
& 0.0250
\\ 
$a_d$
& 0.0486
& 0.1216
&- 0.0288
& 0.0973
& 0.0919
& 0.0763
&- 0.0123
&- 0.0116
&- 0.0616
\\ 
$b_d$
& 0.1508
& 0.1678
&- 0.0122
& 0.0870
& 0.0574
&- 0.0306
&- 0.0161
& 0.0029
&- 0.0813
\\ 
$\gamma_{1,d}$
& 0.0267
& 0.0924
& 0.0053
& 0.0646
& 0.0493
& 0.0725
&- 0.0114
&- 0.0074
&- 0.0491
\\
$\gamma_{2,d}$
&- 0.1161
&- 0.1196
& 0.0059
&- 0.0666
&- 0.0364
& 0.0243
& 0.0108
&- 0.0051
& 0.0736
\\ 
$\gamma_{3,d}$
& 0.0
& 0.0
& 0.0
& 0.0
& 0.0
& 0.0
& 0.0
& 0.0
& 0.0
\\ 
$a_{us}$
& 0.2197
& 0.3627
&- 0.2570
&- 0.1419
&- 0.0241
& 0.0954
& 0.0704
&- 0.0183
& 0.0641
\\ 
$b_{us}$
& 0.0643
& 0.0261
& 0.0001
& 0.1266
& 0.0332
&- 0.2866
&- 0.0093
&- 0.0132
&- 0.1841
\\ 
$\gamma_{-1,us}$
&- 0.4479
&- 0.6319
& 0.2197
& 0.0694
&- 0.0226
&- 0.0341
&- 0.0034
& 0.0044
&- 0.0408
\\ 
$\gamma_{1,us}$
& 0.1286
& 0.0102
& 0.0039
& 0.2648
& 0.1296
&- 0.3493
&- 0.0462
& 0.0209
&- 0.2635
\\
$A_{us}$
& 0.1193
& 0.2412
&- 0.2493
&- 0.1715
&- 0.0489
& 0.1110
& 0.1182
&- 0.0298
& 0.0755
\\
$a_{ds}$
&- 0.1579
&- 0.2688
&- 0.2190
&- 0.0515
&- 0.0137
&- 0.0604
& 0.0849
&- 0.0006
&- 0.0573
\\
$b_{bs}$
&- 0.0260
&- 0.0180
&- 0.0454
& 0.0917
& 0.0503
&- 0.1265
& 0.0547
& 0.0332
&- 0.1067
\\ 
$\gamma_{1,ds}$
& 0.0169
&- 0.0960
&- 0.1031
& 0.2130
& 0.1409
&- 0.1811
& 0.0413
& 0.0695
&- 0.2003
\\ 
$A_{ds}$
&- 0.0896
&- 0.1797
&- 0.2571
&- 0.0469
& 0.0022
&- 0.1330
& 0.1193
&- 0.0432
&- 0.0869
\\ 
$a_{ss}$
& 0.6522
& 0.9280
& 0.0626
&- 0.0092
&- 0.0279
&- 0.0841
&- 0.0728
&- 0.0159
& 0.0169
\\ 
$b_{ss}$
& 1.0
& 0.6427
&- 0.0179
& 0.1967
& 0.1164
&- 0.2390
&- 0.0965
& 0.0169
&- 0.1675
\\ 
$A_{ss}$
& 0.6427
& 1.0
&- 0.0211
& 0.1403
& 0.0997
&- 0.1385
& 0.0216
& 0.0072
&- 0.1109
\\ 
$a_g$
&- 0.0179
&- 0.0211
& 1.0
&- 0.5279
&- 0.8046
& 0.1838
&- 0.2829
& 0.0076
& 0.3310
\\ 
$b_g$
& 0.1967
& 0.1403
&- 0.5279
& 1.0
& 0.8837
&- 0.5124
& 0.1438
& 0.1255
&- 0.7275
\\
$\gamma_{1,g}$
& 0.1164
& 0.0997
&- 0.8046
& 0.8837
& 1.0
&- 0.2511
& 0.1829
& 0.0814
&- 0.5180
\\ 
$\alpha_s^{(n_f=3)}(\mu_0)$
&- 0.2390
&- 0.1385
& 0.1838
&- 0.5124
&- 0.2511
& 1.0
&- 0.1048
& 0.0423
& 0.6924
\\ 
$m_c(m_c)$
&- 0.0965
& 0.0216
&- 0.2829
& 0.1438
& 0.1829
&- 0.1048
& 1.0
& 0.0328
&- 0.1577
\\ 
$m_b(m_b)$
& 0.0169
& 0.0072
& 0.0076
& 0.1255
& 0.0814
& 0.0423
& 0.0328
& 1.0
&- 0.0900
\\ 
$m_t(m_t)$
&- 0.1675
&- 0.1109
& 0.3310
&- 0.7275
&- 0.5180
& 0.6924
&- 0.1577
&- 0.0900
& 1.0
\\
\hline
\end{tabular}
\end{center}
\caption{\small 
  \label{tab:cor3}
  Tab.~\ref{tab:cor1} continued.
}
\end{table}
\end{center}